\documentclass{article}

\usepackage{arxiv}      

\usepackage[utf8]{inputenc} 
\usepackage[T1]{fontenc}    

\usepackage{amsmath, amssymb, amsfonts, amsthm} 
\usepackage{mathtools}     
\usepackage{bm}            
\usepackage{mathrsfs}      

\usepackage{graphicx}      
\usepackage{subfig}        
\usepackage{float}         

\usepackage{booktabs}      
\usepackage{multirow}      
\usepackage{makecell}      

\usepackage{algorithm}
\usepackage{algorithmicx}
\usepackage{algpseudocode}

\usepackage[numbers]{natbib} 
\usepackage{doi}            

\usepackage{hyperref}
\usepackage{url}

\usepackage{xcolor}         
\usepackage{textcomp}       
\usepackage{manyfoot}       
\usepackage{lipsum}         
\usepackage{microtype}      
\usepackage{nicefrac}       

\usepackage[title]{appendix}

\usepackage{lineno}         

\title{Attention mechanism for scalable mesh-based neural surrogates of free-surface fluids}

\author{Federico Lanteri$^{1,*}$, Massimiliano Cremonesi$^{1}$\\
	$^{1}$Department of Civil and Environmental Engineering, Politecnico di Milano\\
    Piazza Leonardo da Vinci, 32, 20133, Milano, Italy\\
	$^{*}$\texttt{federico.lanteri@polimi.it} \\
}

\begin{document}
\maketitle

\begin{abstract}
High-fidelity simulations of free-surface flows using Lagrangian methods such as the Particle Finite Element Method (PFEM) are computationally intensive due to the need to continuously update the computational domain and solve the underlying governing equations. This challenge is further amplified in the presence of non-Newtonian rheologies, where material nonlinearities can significantly increase computational cost. These limitations motivate the development of efficient surrogate models capable of approximating the evolution of PFEM simulations at a reduced computational expense. While data-driven deep learning approaches are promising, a key challenge lies in designing models that can effectively operate on arbitrary and evolving geometries.

In this work, we propose a self-attention-based neural surrogate for PFEM simulations of free-surface flows. The proposed architecture leverages attention mechanism to model interactions between nodes to capture complex spatial dependencies. At the same time, it preserves the underlying PFEM discretization, which provides a geometric and topological framework for remeshing and node redistribution. This helps in preserving good quality spatial discretization and geometrical representation during rollouts, enhancing long-term stability, and enables the reconstruction of derived mechanical quantities through standard finite element operators. Two attention formulations are investigated: a standard self-attention variant and a linear attention variant designed to alleviate the computational limitations of standard attention and improve scalability to larger problem sizes.

The proposed models are evaluated on representative two- and three-dimensional free-surface flow benchmarks involving evolving geometries, varying material parameters, and non-Newtonian fluids. The results demonstrate accurate prediction of both transient dynamics and final free-surface configurations, while significantly improving scalability to large problem sizes. Furthermore, the mesh-based formulation enables the reconstruction of derived mechanical quantities, such as stress fields, directly from the predicted solution. Overall, the proposed framework provides an accurate and scalable surrogate modeling strategy for PFEM simulations of free-surface flows, enabling the application of data-driven Lagrangian surrogates to engineering-scale problems involving evolving geometries.
\end{abstract}


\section{Introduction}\label{sec:intro}
Free-surface flows are characterized by the presence of a moving interface between a fluid and the surrounding air. More generally, many fluid mechanics problems involve interfaces whose position evolves in time, including multiphase flows, where the interface separates immiscible fluids, and fluid–structure interaction (FSI) problems, where the fluid interacts with deformable solid bodies. In these problems, the interface evolves dynamically as a result of the underlying fluid motion and its interaction with the surrounding media. Since the interface position is part of the unknown solution, the computational domain changes continuously in time, giving rise to highly transient configurations with complex evolving geometries. Such flows arise in a wide range of applications, including industrial processes, geotechnical engineering and natural hazards.

The Particle Finite Element Method (PFEM) \cite{idelsohn2004,cremonesi2020} is a Lagrangian mesh-based technique specifically designed to address these challenges. The Lagrangian description offers several advantages, including the natural tracking of free surfaces and the elimination of nonlinear convective terms from the governing equations. However, in a Lagrangian approach, large deformations or displacements progressively deteriorate the computational mesh, compromising accuracy and stability. PFEM overcomes this limitation by equipping a standard Lagrangian finite element framework with an efficient re-meshing technique. Over the years, PFEM has been successfully applied to a broad range of problems, including landslides \cite{franci2020}, geotechnical applications \cite{carbonell2022}, non-Newtonian fluids \cite{rizzieri2024a, rizzieri2025}, multiphase flows \cite{idelsohn2009}, and fluid–structure interaction \cite{meduri2018, cerquaglia2019, fu2024}, as well as industrial processes such as glass forming \cite{ryzhakov2016} and 3D concrete printing \cite{rizzieri2024}. 

Despite continuous advances in numerical algorithms, meshing strategies \cite{leyssens2024}, and computational power, high-fidelity PFEM simulations remain extremely demanding. This is further complicated when modeling complex non-Newtonian fluid behavior, where material nonlinearities significantly complicate the convergence and increase computational cost. This has motivated the development of surrogate models that approximate the input–output behavior of the physical system, bypassing the need to solve computationally expensive Full-Order Models (FOMs). Projection-based Reduced-Order Models (ROMs), such as Proper Orthogonal Decomposition (POD) or Proper Generalized Decomposition (PGD), have shown remarkable results in many applications \cite{quarteroni2016}. However, these methods typically assume a fixed computational domain throughout the parametric manifold of interest, and their extension to problems involving moving boundaries or evolving geometries is complex. The reduced basis must be continuously adapted to account for the domain deformation and topological changes, often requiring sophisticated mapping or domain-tracking strategies. In \cite{beckermann2025}, a first attempt in this direction was presented, but it remains limited to relatively simple cases.

Deep learning surrogate models provide an attractive alternative in this context, as modern architectures have demonstrated strong capabilities in handling arbitrary and time-evolving geometries \cite{brivio2025}. Several approaches have been proposed to learn the dynamics of physical systems directly from simulation data. Among these, Graph Neural Networks (GNNs) have emerged as a powerful framework for learning the dynamics of physical systems \cite{tierz2025, sharma2026}, particularly in the context of meshfree, particle-based simulations \cite{sanchez-gonzalez2020, li2022, zhao2025}. In these formulations, particles are typically represented as nodes carrying physical quantities (e.g., position, velocity, pressure), while edges, defined through fixed-radius neighborhoods, encode pairwise interactions through message passing operations. These approaches learn the time evolution of the system in an autoregressive manner and have demonstrated strong predictive capabilities across a wide range of applications, including granular flows in geotechnics \cite{choi2024} and molecular dynamics \cite{li2022a}.

While effective, particle-based GNN surrogates incur significant computational and memory costs. A key limitation stems from the local nature of message passing: information is exchanged only between neighboring nodes at each layer, requiring multiple message-passing steps to propagate information across the computational domain. In \cite{tesan2026}, it is shown that for parabolic and elliptic systems, a sufficient number of message passing steps is required to propagate information across the entire domain, with the required depth depending on mesh topology and the geometry of the domain. This requirement becomes particularly challenging in large-scale three-dimensional simulations, where each node interacts with many neighbors and the resulting graph contains a very large number of edges. Consequently, both memory consumption and computational cost increase rapidly with problem size. In addition, autoregressive graph-based surrogates are known to be sensitive to error accumulation, which can progressively degrade predictive accuracy during long-term rollouts.

Mesh-based surrogates offer significant advantages over purely meshless approaches by directly exploiting the computational mesh as an underlying interaction structure. A representative example is NeuralPFEM \cite{lanteri2025}, which operates by exploiting the mesh connectivity naturally available in PFEM simulations both in training and prediction. At inference, on unseen cases, the mesh is reconstructed using the same efficient meshing algorithm employed in the numerical solver. This strategy enables the use of established PFEM tools for mesh quality preservation and field interpolation, helping maintain a uniform node distribution and improving robustness during long autoregressive rollouts.  Although mesh-based connectivity generally yields fewer edges than particle-based neighborhood searches, the explicit storage of edge embeddings and the repeated application of message-passing operations remain a significant memory bottleneck, limiting scalability to large engineering problems.

To address the scaling limitations of graph-based models, neural surrogate architectures based on attention mechanisms have recently emerged as a promising alternative. Originally introduced for sequence modeling in natural language processing \cite{vaswani2017}, attention mechanisms have since demonstrated strong performance across a wide range of domains, including computer vision \cite{dosovitskiy2021} and computational biology \cite{jumper2021}. By leveraging attention, these models compute interactions in a data-adaptive manner, without requiring the explicit construction, storage, or update of a fixed graph structure. Attention-based architectures have been investigated for surrogate modeling \cite{li2023, wu2024, adams2025, iparraguirre2026}. However, most existing studies focus on fixed-domain or weakly deforming fluid problems, while applications to free-surface flow problems \cite{wang2026, alkin2025} and non-Newtonian fluids  \cite{saberi2025} remain relatively scarce. In this work, we propose a novel strategy to integrate an attention-based neural surrogate within a mesh-based Lagrangian framework, mitigating the memory bottlenecks associated with GNN-based formulations while preserving the advantages of a dynamically evolving mesh representation.

In contrast to GNNs, where interactions are explicitly defined through graph connectivity, attention mechanisms model interactions through weights computed from dot products between node embeddings. This allows the network to learn spatial correlations adaptively, without relying on predefined neighborhood structures, explicit edge features, or multiple message-passing iterations to propagate information across the domain. As a result, the memory overhead associated with edge embeddings is eliminated. Moreover, since the model does not explicitly depend on edge lengths, it is not tied to a specific mesh resolution and can therefore better accommodate discretizations that differ from those used during training, as well as resolution changes arising from mesh adaptation. This reduces the sensitivity of the model to variations in inter-particle distances, which can become significant during autoregressive rollouts even for well-trained models.

A well-known limitation of standard attention is its quadratic computational and memory complexity with respect to the number of nodes. Recent developments, such as FlashAttention~\cite{Dao2022}, alleviate the memory bottleneck by reformulating attention computation to achieve linear memory complexity. Nevertheless, the computational cost remains quadratic, which can become prohibitive for large-scale simulations. To address this limitation, several recent works have proposed strategies to reduce the effective cost without altering the attention operator itself, enabling applications to larger problems~\cite{zhou2026, alkin2025a}. An alternative approach is provided by linear attention mechanisms \cite{katharopoulos2020, zhuoran2021}, which reformulate the attention operator so that the expensive pairwise interaction matrix is never explicitly constructed. By exploiting the associativity of matrix multiplication and removing or approximating the softmax normalization, the computation can be reorganized to achieve linear complexity with respect to the number of nodes.

We evaluate the proposed architecture through a comparative study against a GNN baseline, considering both standard and linear attention variants. The assessment focuses on predictive accuracy, computational cost, and memory footprint across representative free-surface flow benchmarks. In addition to geometric variability, these benchmarks incorporate variations in material properties, with particular emphasis on the model’s ability to accurately capture and reproduce the dynamical response of non-Newtonian fluids. The results show that attention-based NeuralPFEM models achieve competitive predictive performance while reducing computational and memory requirements and improving scalability. These characteristics make attention-based architectures a promising alternative for scalable surrogate modeling of large-scale PFEM simulations.

\section{Mathematical modeling and numerical approximation}
The numerical framework employed to perform the simulations and generate the datasets was originally presented in \cite{rizzieri2025}, where it was validated against two- and three-dimensional benchmarks involving free-surface flows of Newtonian and non-Newtonian fluids. Governing equations and modeling assumptions are briefly recalled below.

\subsection{Governing equations}\label{sec:governing}
We consider the motion of a homogeneous, incompressible fluid governed by the Navier–Stokes equations, which express the conservation of momentum and mass:

\begin{align}
\rho \frac{d \bm{v}}{d t} = \nabla_{\bm{x}} \cdot \bm{\sigma} + \rho \bm{b} \quad \text{in } \Omega_{t} \times [0, T], \label{eq:1} \\
\ \nabla_{\bm{x}} \cdot \bm{v} = 0 \quad \text{in } \Omega_{t} \times [0, T], \label{eq:2}
\end{align}
\noindent where $\rho$ is the fluid density, $\bm{v}=\bm{v}(\bm{x},t)$ is the velocity field, $p=p(\bm{x},t)$ is the pressure field, $\bm{\sigma}= \bm{\sigma}(\bm{x},t)$ is the Cauchy stress tensor, $\bm{b}$ is the vector of the external accelerations and $\frac{d (\bullet)}{dt} = \frac{\partial (\bullet)}{\partial t}\Big\vert_{\bm{x}} + \bm{c} \cdot \nabla_{\bm{x}}(\bullet)$ represents the total time derivative, with $\bm{c}$ being the convective velocity.

The Cauchy stress tensor $\bm{\sigma}$ can be decomposed into a volumetric and a deviatoric part:
\begin{equation}
\bm{\sigma} = -p \bm{I} + \bm{\tau}, \label{eq:3}
\end{equation}
where $\bm{I}$ is the identity tensor and $\bm{\tau}$ denotes the deviatoric stress tensor.
The latter is related to the deviatoric strain rate tensor $\bm{D} = \frac{1}{2}(\nabla_{\bm{x}} \bm{v} + \nabla_{\bm{x}} \bm{v}^{\mathrm{T}})$ through a rheological law of the form:
\begin{equation}
\bm{\tau} = 2 \eta_\text{eff}(\bm{D})\bm{D}, \label{eq:4}
\end{equation}
where $\eta_{\text{eff}}$ denotes the effective viscosity. For Newtonian fluids, $\eta_{\text{eff}}$ is constant, yielding a linear relationship between stress and strain rate. In contrast, non-Newtonian fluids exhibit a non-linear dependence of the deviatoric stress on the strain rate. A particularly important class of such materials is described by the Bingham model, which captures the behavior of viscoplastic fluids. Bingham materials are characterized by the presence of a yield stress $\tau_0$. The material behaves as a rigid body when the applied shear stress is below $\tau_0$, and flows as a viscous fluid once this threshold is exceeded. This behavior is representative of a wide range of natural and industrial flows, including mudflows, debris flows and fresh concrete.

The constitutive law for a Bingham fluid can be expressed as:
\begin{equation}
\begin{cases}
\bm{D} = \bm{0}, & \text{if } |\bm{\tau}| \le \tau_0, \\
\bm{\tau} = \left( \tau_0 \dfrac{\bm{D}}{|\bm{D}|} + 2 \mu \bm{D} \right), & \text{if } |\bm{\tau}| > \tau_0,
\end{cases} \label{eq:bingham}
\end{equation}
where $|\cdot|$ denotes the second invariant of the deviatoric tensor, and $\mu$ is the plastic viscosity, defining the slope of the stress–strain rate curve in the yielded regime.
The discontinuous nature of the Bingham law and the sharp fluid-to-solid transition, may lead to numerical instabilities. To facilitate numerical solution, the rheological law is regularized through an exponential viscosity model \cite{papanastasiou1987}. 

\subsection{Space and time discretizations}
In the numerical framework adopted in this work, Navier–Stokes equations are solved using the Finite Element Method (FEM). The continuum domain is discretized with a finite element mesh employing standard Galerkin shape functions. Since PFEM relies on frequent remeshing, linear shape functions are employed to simplify the interpolation and transfer of field variables between successive meshes. It is worth noting that this choice, which employs shape functions of the same order for both velocity and pressure fields, violates the Ladyzhenskaya–Babuška–Brezzi (LBB) stability condition. To address this limitation, the present framework incorporates a Pressure Stabilizing Petrov–Galerkin (PSPG) formulation \cite{Hughes1986}, enabling the stable use of equal-order velocity–pressure approximations. Time discretization is then performed by subdividing the time interval of interest into finite time steps $\Delta t$ and approximating derivatives using the implicit backward Euler scheme. This choice provides unconditional stability for the time integration scheme. Further details regarding the spatial and temporal discretization, as well as the solution of the resulting linearized algebraic system, can be found in \cite{rizzieri2025}.

\subsection{Particle Finite Element Method (PFEM)}

The Particle Finite Element Method (PFEM) \cite{onate2004,cremonesi2020} combines a standard Lagrangian finite element formulation with an efficient remeshing strategy, enabling the simulation of problems characterized by large deformations and evolving free surfaces.

The computational domain is initially discretized using a finite element mesh, on which the discretized Navier–Stokes equations are solved. The resulting velocity field is then used to update the mesh nodes according to the Lagrangian description. As the simulation progresses, large deformations may significantly deteriorate mesh quality, compromising numerical accuracy and stability. To address this issue, the mesh is periodically regenerated by preserving the nodal set while reconstructing the connectivity. The new mesh is obtained through a Delaunay triangulation of the nodes. The physical domain is subsequently reconstructed from the triangulation, typically using the $\alpha$-shape technique \cite{edelsbrunner1992}, which identifies and removes the unphysical most distorted elements.

As the nodes are advected with the flow, their spatial distribution gradually becomes non-uniform, with some regions exhibiting excessive nodal concentration and others suffering from insufficient resolution. This imbalance further degrades mesh quality and may affect numerical accuracy. To preserve an adequate nodal distribution throughout the simulation, several strategies have been proposed in the literature, including mesh smoothing techniques \cite{meduri2019}, adopted in the present framework, and mesh adaptation procedures \cite{leyssens2024}. These operations play an important role in maintaining a regular discretization of the domain and, as clarified later, will be particularly beneficial for long autoregressive surrogate simulations, where excessive node clustering may otherwise amplify prediction errors.

\newpage
\section{NeuralPFEM framework}
\begin{figure}[h]
    \centering
    \includegraphics[width=0.8\linewidth]{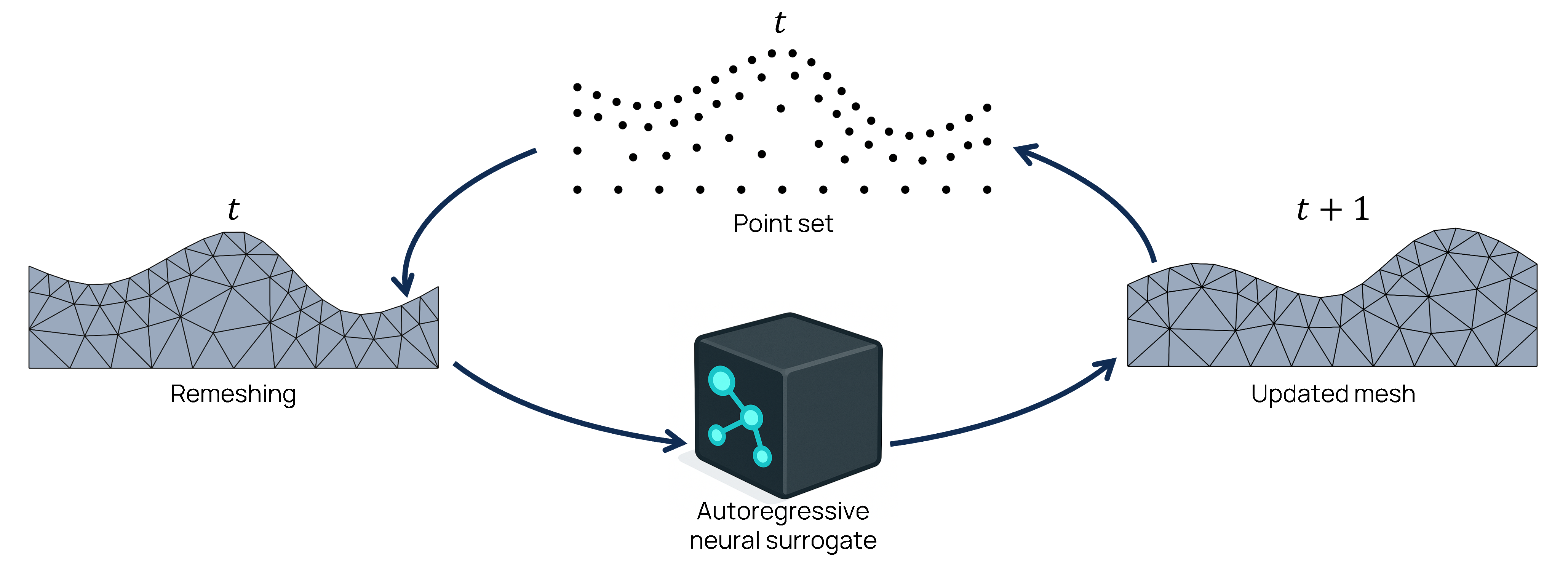}
    \caption{A scheme of NeuralPFEM framework during the prediction phase.}
    \label{fig:1}
\end{figure} 
NeuralPFEM \cite{lanteri2025} is a mesh-based autoregressive Lagrangian neural surrogate designed to simulate free-surface fluid flows. As illustrated in Figure \ref{fig:1}, the method retains the key components of the PFEM pipeline but replaces the most computationally expensive step, the solution of the discretized Navier–Stokes equations, with a learned model. More specifically, starting from a given configuration at time $t$, the fluid domain is represented by a set of nodes carrying physical quantities and connected through a mesh defining the domain geometry. The current state, including nodal positions, velocities, pressures, and material parameters, is provided as input to the deep learning architecture. The model predicts the physical state at the next time step $t+\Delta t$, effectively learning the time-advancing operator that, in PFEM, is obtained through the numerical solution of the governing equations. Once the new velocities are predicted, node positions are updated through a Lagrangian advection step, so that the mesh moves consistently with the flow. Because this motion can lead to mesh distortion, a remeshing procedure is periodically applied to restore mesh quality and maintain a suitable discretization of the domain. After remeshing, physical quantities are transferred or interpolated onto the new mesh before proceeding to the next step. The autoregressive formulation lies in the recursive use of the predicted state at $t + \Delta t$ as input for subsequent steps, enabling the simulation of arbitrary temporal horizons. Training is carried out on pairs of consecutive states extracted from high-fidelity PFEM simulations, allowing the network to approximate the underlying time-advancing operator.

A known limitation of autoregressive models is error accumulation \cite{zhou2025}: small inaccuracies compound over successive steps, gradually driving the solution away from the true physical trajectory. In a Lagrangian framework, this effect is particularly pronounced because the system state is defined not only by physical variables (e.g., velocity, pressure), but also by the geometry itself, represented by node positions and mesh connectivity. 
Consequently, errors in the predicted dynamics directly induce errors in the geometry. Nodes may progressively move far from their correct locations, leading to incorrect spatial discretization and a degraded reconstruction of the domain shape. This issue becomes especially critical for models that rely on edge-based features, such as inter-nodal distances. During training, these distances are typically close to uniform due to the regularity of the discretization. However, as errors accumulate during rollout, the geometry can become increasingly irregular. Such configurations deviate significantly from those seen during training and therefore fall outside the model’s learned data distribution. When the model encounters these out-of-distribution states, its predictions become less reliable, further amplifying the error. This creates a loop in which inaccuracies in the dynamics degrade the geometry, and the degraded geometry, in turn, leads to increasingly inaccurate predictions. 

Several strategies have been proposed to mitigate these instabilities \cite{mccabe2023}. Noise injection during training, as adopted in the Graph Neural Simulator (GNS) \cite{sanchez-gonzalez2020}, improves robustness while maintaining computational efficiency. An alternative is to exploit the stabilization mechanisms already embedded in the numerical frameworks. As shown in \cite{lanteri2025}, remeshing and mesh adaptation procedures help preserve mesh quality and maintain a well-distributed set of nodes throughout the simulation. By periodically regenerating the mesh connectivity and redistributing nodes, while consistently interpolating the field variables onto the updated discretization, these operations limit node clustering. As a result, the occurrence of out-of-distribution configurations is reduced, mitigating error accumulation and improving the stability of long autoregressive rollouts.

\newpage
\subsection{Graph Neural Network architecture}
\begin{figure}[h]
    \centering
    \includegraphics[width=0.9\linewidth]{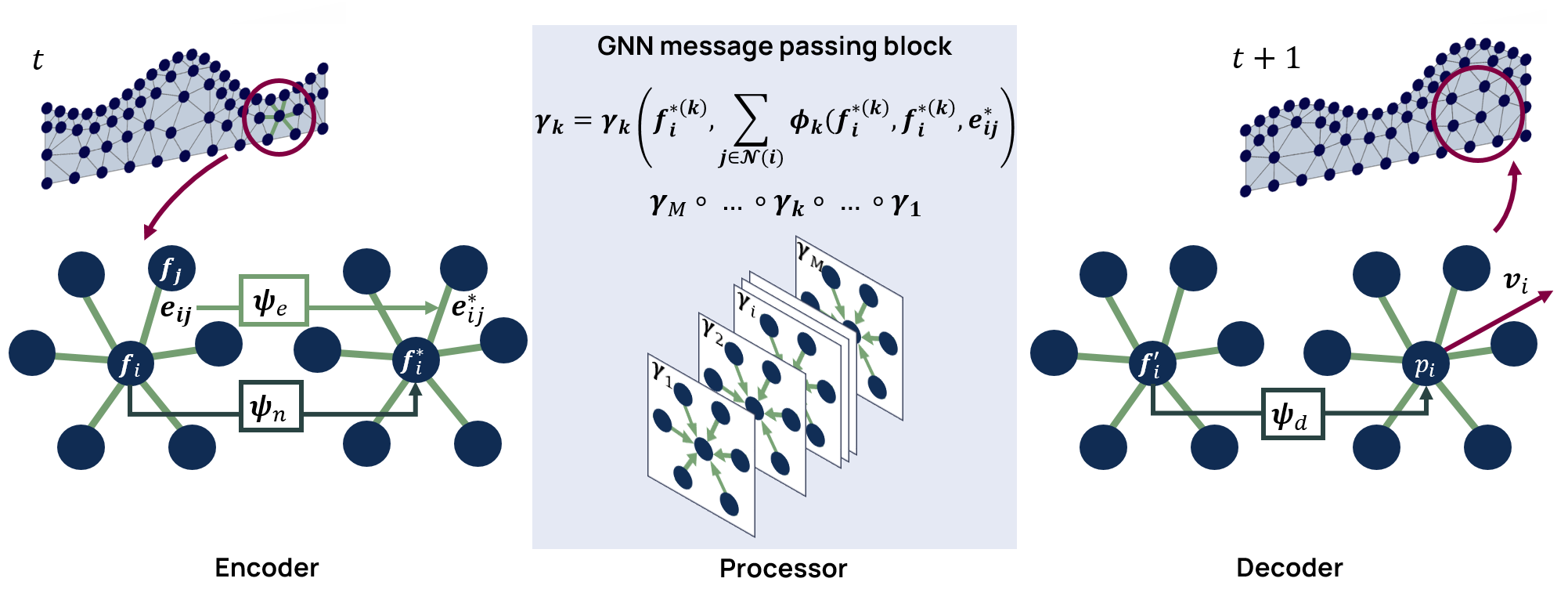}
    \caption{A schematic of the GNN architecture during the prediction phase. The model leverages mesh connectivity to define graph edges. The encoder embeds node and edge features into a latent graph representation, after which the processor performs multiple message-passing steps. Finally, the decoder outputs the predicted physical quantities, and nodes are advected according to the resulting velocity field.}
    \label{fig:2}
\end{figure} 

Mesh-based surrogates typically rely on a Graph Neural Network (GNN) processor~\cite{scarselli2009, battaglia2018}, a natural choice given that the finite element mesh inherently defines a graph over the fluid domain. Unlike purely particle-based Lagrangian surrogates, which dynamically construct graph connectivity by linking particles within a fixed interaction radius~\cite{sanchez-gonzalez2020, li2022}, the GNN-based NeuralPFEM exploits the existing mesh connectivity to define graph edges, avoiding the overhead of dynamic neighbor search and preserving the geometric structure of the discretization.

The model follows an encoder–processor–decoder architecture, illustrated in Figure~\ref{fig:2}. To provide the network with short-term temporal information, node features $\bm{f}_i \in \mathbb{R}^m$ include a history of $C$ consecutive velocity $\bm{v}$ and pressure $p$ states:
\begin{equation}
\bm{f}_i = \left[ \bm{v}_i^{(t-C+1)},\dots, \bm{v}_i^{(t)}, \dots, p_i^{(t-C+1)}, \dots  p_i^{(t)} \right].
\end{equation}
where $i \in \{1, \dots, N\}$ denotes the node index, and $N$ is the total number of nodes in the mesh. Additionally, other relevant nodal characteristics, such as material properties or boundary condition indicators, can be appended to this feature vector.
Edge features $\bm{e}_{ij} \in \mathbb{R}^q$ consist of the normalized relative distance vector between connected nodes $i$ and $j$ and its magnitude, providing the model with local geometric awareness:
\begin{equation}
\bm{e}_{ij} = \left[ \frac{\bm{x}_i - \bm{x}_j}{|\bm{x}_i - \bm{x}_j|}, |\bm{x}_i - \bm{x}_j| \right].
\end{equation}

The encoder maps node and edge features to latent representations $\bm{f}_i^* \in \mathbb{R}^d$ and $\bm{e}_{ij}^* \in \mathbb{R}^d$ via independent multi-layer perceptrons (MLPs) $\psi_n$ and $\psi_e$, respectively. $d$ denotes the dimension of the latent space \footnote{For notational simplicity, we use a single symbol, although the dimensionality may differ for node and edge features.}. The processor then applies $M$ message-passing blocks $\gamma_1, \dots, \gamma_M$, each updating node representations by aggregating transformed features from neighboring nodes and edges. Formally, at each block, the node representation is updated as
\begin{equation}
    \bm{f}_i^{*(k+1)} = \gamma_k\!\left(\bm{f}_i^{*(k)},\, \sum_{j\in \mathcal{N}(i)} \phi_k\!\left(\bm{f}_i^{*(k)},\, \bm{f}_j^{*(k)},\, \bm{e}_{ij}^*\right)\right),
\end{equation}
where $\phi_k$ is a learnable message function, and $\mathcal{N}(i)$ is the set of neighbors of node $i$. Stacking $M$ such blocks allows information to propagate across increasingly large neighborhoods, with the effective receptive field of each node growing with depth. Finally, the decoder $\psi_d$ applies a nodewise MLP to map the processed latent representations to the predicted physical quantities --- nodal velocities and pressures at the next time step --- after which nodes are advected according to the predicted velocity field.

\subsubsection{Challenges of GNN-based approaches}\label{sec:memory}

Despite its strong predictive performance, the GNN formulation carries a memory cost that scales unfavorably with problem size~\cite{duan2023}. At each message-passing layer, the network must keep in memory the intermediate processed node features, edge features, and associated gradients for every node and every edge in the graph. As node counts grow with mesh resolution, edge counts scale aggressively as $\mathcal{O}(N \cdot k)$, where $k$ is the average number of neighbors per node. In three-dimensional unstructured meshes $k$ can easily exceed ten, causing the edge count to exceed the node count by an order of magnitude. Since edge embeddings must be stored and updated at every message passing step, they constitute the dominant term in the memory budget. This cost is further amplified by the requirement of bidirectional message passing: each undirected mesh edge is represented as two directed edges, one in each direction, doubling the number of stored quantities. To perform backpropagation through $M$ message passing steps, all intermediate node and edge activations must be retained during the forward pass to enable gradient computation via the chain rule, yielding a total memory cost that scales as $\mathcal{O}(M \cdot (N + E) \cdot d)$, where $E$ is the number of directed edges and $d$ the dimension of the latent space.

Critically, the number of message passing $M$ cannot be freely reduced to lower this cost without sacrificing predictive quality. As shown in~\cite{tesan2026}, full propagation of information across the computational domain yields significantly improved performance. This behavior suggests that these data-driven operators benefit from a global receptive field, unlike numerical FEM solvers, where interactions are restricted to immediate neighbors. The value $M$ for which this condition is satisfied, however, increases substantially with mesh resolution and domain extent, making this condition highly demanding in practice and generally feasible only in simple problems. Moreover, too many message passing steps in GNNs are known to lead to the undesired phenomenon of oversmoothing \cite{rusch2023}. As the number of message-passing layers increases, node representations are repeatedly aggregated with those of their neighbors. While this promotes information propagation, it also progressively reduces the distinguishability of node features: embeddings tend to converge toward a common, low-variance representation across the graph. Beyond a certain depth, this loss of feature diversity counteracts the benefits of additional information propagation, leading to a saturation—or even degradation—of predictive performance. In practice, $M$ is typically treated as a hyperparameter, selecting it empirically at the point where performance saturates and further depth yields diminishing returns, an optimal intermediate regime, where the receptive field is sufficiently large without incurring excessive smoothing

A second limitation of GNN formulations is their dependence on a fixed mesh resolution \cite{li2023a}. In typical graph-based discretizations, edge features are constructed from normalized inter-nodal distances, which implicitly encode the characteristic element size of the training mesh. As a result, the model does not learn scale-invariant physical interactions, but rather relationships that are tied to a specific discretization length scale. When the model is applied to meshes with different resolutions, these geometric features no longer correspond to the same physical distances, leading to a mismatch between the learned representation and the underlying physics. Consequently, the mapping between edge features and physical interactions deteriorates, resulting in a loss of accuracy. This limits the transferability of the trained model and poses challenges for applications involving mesh adaptivity or multi-resolution discretizations.

\subsection{Self-attention architecture} \label{sec:attn-npfem}
The memory bottleneck of the GNN processor is fundamentally rooted in its edge representations. As analyzed above, storing and updating edge embeddings across $M$ message-passing layers constitutes the dominant term in the memory budget, and this cost cannot be relieved by reducing $M$ without sacrificing the global receptive field required for accurate simulation. This motivates the search for an architecture that achieves global information propagation without relying on explicit edge representations.

The self-attention mechanism offers a natural solution to this problem. In this section we propose a new architecture for the NeuralPFEM based on the self-attention mechanism. As established in~\cite{joshi2025}, the self-attention mechanism can be interpreted as a message-passing operating on a fully connected graph: self-attention computes pairwise interactions between all nodes simultaneously, with the attention weights playing the role of learned, dynamic edge weights. Crucially, this graph-like communication is never explicitly instantiated as a set of edges. Both GNNs and self-attention aggregate pairwise node interactions into updated representations, but while the GNN is topologically constrained to communicate along mesh edges, self-attention learns from data alone which inter-node relationships are relevant. In our context, this reinterpretation carries a concrete practical implication with respect to the depth condition of~\cite{tesan2026}. Full domain-wide information propagation is achieved by self-attention in a single layer, since every node attends to every other node regardless of distance.

To process the physical state, the proposed architecture employs the same node encoder and decoder structures detailed for the GNN architecture described in the previous Section, mapping the nodal history to latent representations and back to physical quantities. However, self-attention does not require an edge representation and therefore the edge encoder is removed, eliminating the dominant source of memory overhead. Spatial awareness is instead incorporated directly into the node representations via a dedicated positional encoding module, described in Section~\ref{sec:rope}. This modification also reduces the dependence of the learned representation on the characteristic length scales of the training mesh. In the absence of explicit geometric edge features tied to a specific element size, attention-based models have been shown to be less sensitive to the characteristic length scale of the training mesh~\cite{li2023}.

Although self-attention mechanisms do not inherently require an explicit connectivity structure between nodes, the PFEM mesh remains advantageous for several reasons. First, it provides a geometric and topological framework that enables consistent remeshing and node redistribution. In particular, the mesh defines elements with quantifiable quality metrics, which are essential for detecting mesh degradation and guiding corrective operations such as node insertion, removal, or repositioning. As demonstrated in \cite{lanteri2025}, this is beneficial in maintaining a uniform spatial resolution during long autoregressive rollouts, enhancing stability over time. Second, the mesh supports the a posteriori evaluation of spatial derivatives through finite element shape functions. This allows the reconstruction of gradients and derived physical quantities, such as stresses or internal forces, in a manner consistent with the underlying discretization. Finally, retaining the mesh structure facilitates the development of hybrid approaches that combine the classical PFEM finite element solver with the neural surrogate model. In such frameworks, the surrogate can be selectively applied in regions of the domain or during time intervals where its predictions are sufficiently reliable, while the remaining computations are delegated to the numerical solver, thereby improving robustness and overall accuracy. 

The resulting architecture combines the global communication capabilities of attention mechanisms with the geometric and numerical advantages provided by the PFEM discretization.
We consider two variants of the attention processor, differing in how pairwise interactions are computed: a standard softmax formulation and a linearized variant that reduces the quadratic complexity in the number of nodes. Both are described below.

\subsubsection{Standard attention}
\begin{figure}[h]
    \centering
    \includegraphics[width=\linewidth]{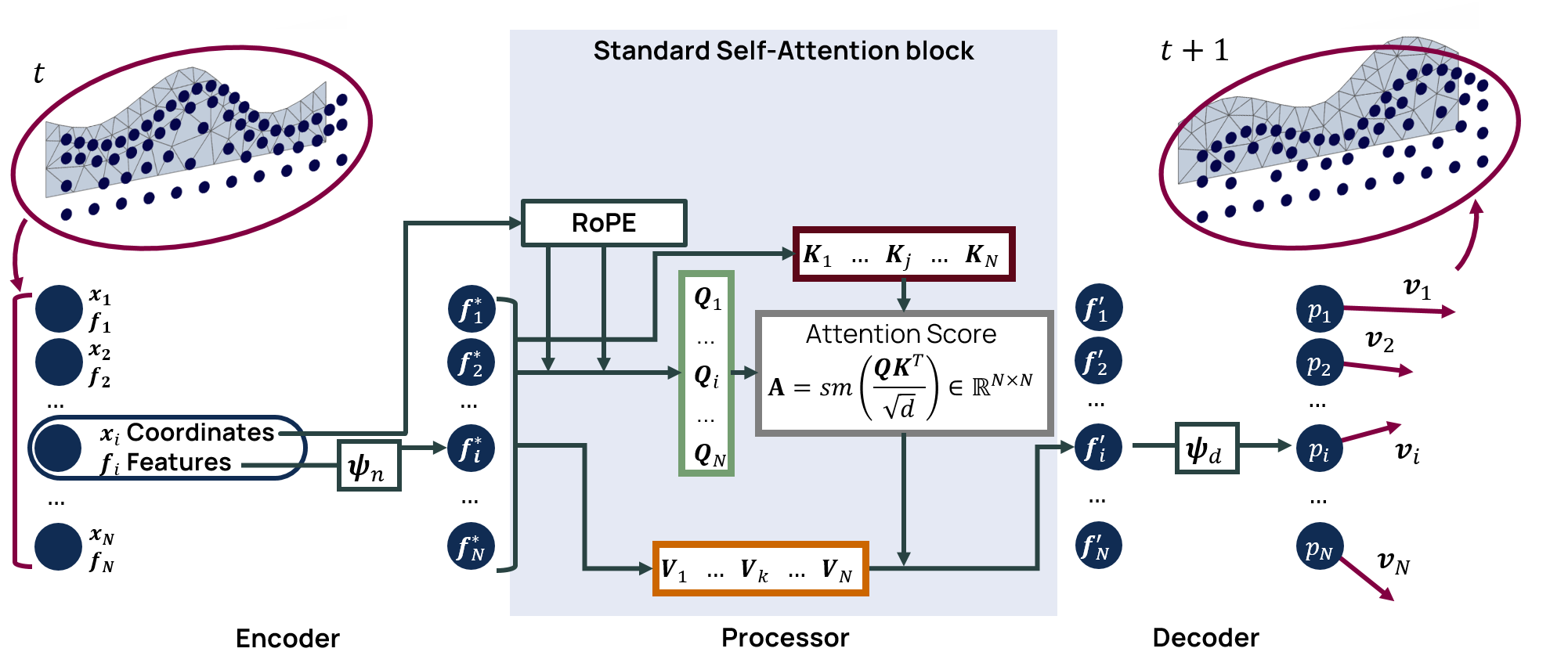}
    \caption{Standard self-attention architecture during the prediction phase. Here, $\mathrm{sm}$ denotes the softmax function. The attention mechanism removes explicit edge representations, so the model operates on an unordered set of nodes rather than a graph structure. Because this makes the architecture geometrically agnostic, node coordinates are incorporated to provide positional information.}
    \label{fig:3}
\end{figure} 

The overall structure of standard self-attention NeuralPFEM is presented in Figure \ref{fig:3}.

Given the encoded node features $\bm{f}_i^* \in \mathbb{R}^d$, the processor begins by independently projecting each node representation as
\begin{equation}
\bm{q}_i = \phi_Q(\bm{f}_i^*), \quad \bm{k}_i = \phi_K(\bm{f}_i^*), \quad \bm{\mathrm{v}}_i = \phi_V(\bm{f}_i^*),
\end{equation}
where $\phi_Q, \phi_K, \phi_V : \mathbb{R}^d \to \mathbb{R}^d$ are three independent MLPs. These individual vectors are then stacked to form the query, key and value matrices $\bm{Q}, \bm{K}, \bm{V} \in \mathbb{R}^{N \times d}$. The attention score between nodes $i$ and $j$ is computed as the scaled dot product of their respective query and key vectors. These scores are subsequently normalized across all target nodes $j$ via a softmax operation to produce the final attention weights:
\begin{equation}
s_{ij} = \frac{\bm{q}_i^\top \bm{k}_j}{\sqrt{d}}, \quad \alpha_{ij} = \mathrm{softmax}_j(s_{ij})
= \frac{\exp(s_{ij})}{\sum_{j'} \exp(s_{ij'})}.
\end{equation}
The $\sqrt{d}$ scaling factor ensures that dot products do not grow excessively large in magnitude as the embedding dimension increases; without it, the softmax function would be pushed into regions of near-zero gradients, destabilizing the training process. By definition, the softmax operation ensures that $\alpha_{ij} \ge 0$ and $\sum_j \alpha_{ij} = 1$. This nonlinearity exponentially amplifies high-scoring pairs, encouraging the model to concentrate its attention on the most physically relevant nodes rather than distributing it uniformly across the computational domain. The updated representation of node $i$ is then a weighted aggregation of all value vectors:
\begin{equation}
    \bm{f}'_i = \sum_{j=1}^N \alpha_{ij} \mathrm{v}_j.
\end{equation}
The query–key interaction acts as a data-driven adaptive kernel measuring pairwise compatibility between nodes, which is then used to weight the aggregation of the value vectors representing the nodal quantities of the physical fields.

The explicit computation of the $N \times N$ score matrix $S = [s_{ij}] \in \mathbb{R}^{N\times N}$ is the source of the $\mathcal{O}(N^2 d)$ complexity of standard attention. To avoid explicitly constructing this matrix in memory, which would introduce a quadratic memory cost and offset the gains from removing edge embeddings, we adopt the FlashAttention implementation~\cite{Dao2022}, which computes exact attention without storing the full attention matrix, yielding significant reductions in GPU memory consumption without any approximation.

Multiple attention heads are employed in parallel, each projecting node features into a distinct query-key-value subspace and computing an independent attention pattern. The outputs of all heads are concatenated and linearly projected back to the original embedding dimension. This multi-head structure allows the model to simultaneously capture different types of relationships within a single layer enhancing representational capacity.

\subsubsection{Linear attention}
\begin{figure}[h]
    \centering
    \includegraphics[width=\linewidth]{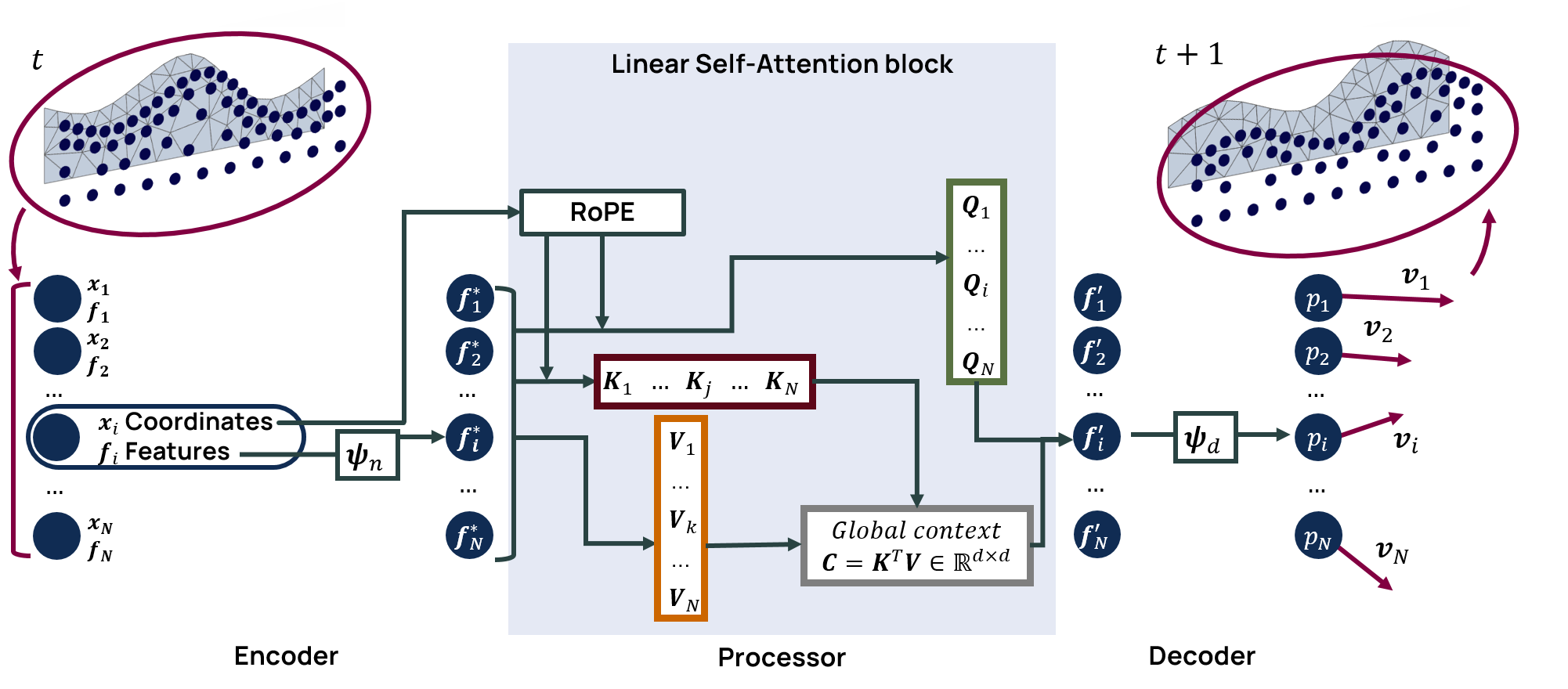}
    \caption{Linear self-attention architecture during the prediction phase. By removing the softmax nonlinearity, attention consists in two matrix multiplications. Reordering these operations avoids constructing the full pairwise attention matrix, instead computing a global context representation that is applied to all queries. This reduces computational complexity from quadratic to linear scaling with respect to the number of nodes.}
    \label{fig:4}
\end{figure}

Despite the memory savings achieved by employing FlashAttention implementation, standard attention retains $\mathcal{O}(N^2)$ computational complexity, which becomes prohibitive for large problems. A linearized alternative \cite{katharopoulos2020, zhuoran2021} can be derived by observing that, when the softmax nonlinearity is removed, the attention output reduces to two consecutive matrix multiplications:
\begin{equation}
    \bm{f}' = \left(\bm{Q}\bm{K}^\top\right)\bm{V}.
\end{equation}
Since matrix multiplication is associative, this expression can be equivalently reordered as:
\begin{equation}
    \bm{f}' = \bm{Q}\left(\bm{K}^\top\bm{V}\right).
\end{equation}
Rather than forming the $N \times N$ interaction matrix $\bm{A}=\bm{Q}\bm{K}^\top$, one first computes the $d \times d$ context matrix $\bm{C}=\bm{K}^\top\bm{V}$, which is then applied to all query vectors simultaneously. This reordering reduces the overall complexity from $\mathcal{O}(N^2 d)$ to $\mathcal{O}(N d^2)$, which scales linearly in the number of nodes. Since $d \ll N$ in practice, this term does not constitute a new computational bottleneck. However, this simplification comes at the cost of reduced expressivity. In standard attention, softmax induces a normalized interaction across all pairs of nodes, while the linearized formulation relies entirely on dot products, resulting in a more diffuse, global aggregation of features where nodes interact through a shared low-rank operator. The linearized attention block can simply replace the standard one, leaving the rest of the architecture untouched (Figure \ref{fig:4}). This linearized formulation was employed in the context of operator learning~\cite{cao2021}, where it admits a natural interpretation as a Galerkin-type approximation of an integral operator: the context matrix $\bm{C}$ acts as a global, data-driven kernel applied uniformly to all query vectors, in analogy with the projection of a solution onto a set of basis functions in classical variational methods. This theoretical grounding has motivated its adoption in surrogate modeling~\cite{li2023}.

\subsubsection{Positional encoding}\label{sec:rope}
The attention mechanism is inherently position-agnostic when no positional information is provided in the input features: nodes with identical feature vectors are treated as indistinguishable regardless of their spatial location, and the model has no means of inferring geometric structure from the input alone. Several strategies have been proposed to inject positional information  into attention-based models. The Transformer~\cite{vaswani2017} introduces fixed sinusoidal encodings added to the input token embeddings before the first attention layer, where each position is mapped to a vector of sine and cosine functions at geometrically increasing frequencies. A more direct approach, adopted in~\cite{cao2021}, consists of concatenating raw spatial coordinates to the input features. Although simple to implement, this encodes absolute positions, making the model sensitive to the global coordinate frame and potentially limiting generalization across domains of different size or location.

Given the limitations of absolute position encodings, we adopt a more effective alternative that is Rotary Positional Embeddings (RoPE) \cite{su2024}, which have been shown to lead to superior performance in computational physics applications \cite{li2023}. RoPE incorporates positional information directly into the attention mechanism by modifying the query and key vectors prior to the dot product. Specifically, each vector is rotated in latent space by an angle determined by the spatial position of the corresponding node. As a result, the dot product between two vectors implicitly depends on their relative positional difference. Consequently, attention scores reflect not only feature similarity but also the spatial relationship between nodes. This formulation is particularly well-suited to physical simulations, where interactions depend primarily on relative geometry rather than absolute coordinates.

Although originally introduced for one-dimensional sequences, RoPE extends naturally to higher-dimensional domains by applying independent rotations along each spatial coordinate, enabling the encoding of relative positional information in two- and three-dimensional settings.

\subsubsection{Material properties}

Several strategies have been proposed in the literature to encode material characteristic into node- and particle-based neural surrogates. In~\cite{sanchez-gonzalez2020}, different materials are identified by a discrete tag appended to the node features. However, this approach restricts the model to a fixed and finite set of predefined materials, preventing generalization to unseen combinations of physical parameters. A more flexible alternative, adopted for example in~\cite{choi2024}, consists of appending the material property values directly to the feature vector of each node. This formulation can in principle handle spatially varying material distributions, but in the common setting where material properties are uniform across the domain it introduces a high degree of redundancy: the same values are replicated identically across all $N$ node feature vectors, contributing no additional information while unnecessarily inflating the input dimensionality.

For spatially uniform material properties, we propose a conditioning strategy based on Feature-wise Linear Modulation (FiLM)~\cite{perez2017}. Rather than appending identical material parameters to every node feature, the material properties are isolated and processed globally. A dedicated MLP evaluates the material properties once, projecting them into a scaling vector $\bm{s}$ and bias vector $\bm{b}$. These are used to modulate the node features via an affine transformation $\bm{s} \odot \psi_n(\bm{f}_i^*) + \bm{b}$ where $\odot$ denotes elementwise multiplication. The material information is thus injected into the latent node representations in a single global operation, without any replication across nodes. This design cleanly separates the encoding of local physical state, handled by the node MLP, from the encoding of global material description, handled by the FiLM MLP, while allowing the latter to influence all node representations in a parameter-efficient, non-redundant manner.

The effectiveness of the conditioning strategy also depends on the representation adopted for the material parameters. In~\cite{rizzieri2026} we have shown that using physically informed, aggregated parameters as model inputs — rather than the raw dimensional quantities individually — can significantly benefit the learning process. Condensing correlated parameters into dimensionless groups reduces the effective input dimensionality, encodes known physical relationships between variables directly into the model inputs, enhancing generalization: even if individual dimensional parameters fall outside the training range, accurate predictions can still be obtained as long as the corresponding dimensionless groups remain within the learned domain.

For Bingham fluids, the flow dynamics are more appropriately characterized by the dimensionless yield stress, $\tau_0^* = \frac{\tau_0}{\rho g D}$, where $g$ is the gravitational acceleration and $D$ represents a characteristic length scale. This formulation condenses the absolute yield stress $\tau_0$ and fluid density $\rho$ into a single parameter that quantifies the ability of the material's microstructure to counterbalance gravity-induced stresses. Rather than providing $\tau_0$ and $\rho$ as separate inputs, only $\tau_0^*$ and $\mu$ are supplied, reducing the material feature vector from three components to two while embedding physically meaningful structure into the input representation. This principle extends naturally to other families of materials. It is important to note that, although $\tau_0^*$ is used as a model input, the training dataset is generated by sampling the original dimensional parameters. This is dictated by the requirements of the underlying PFEM solver, which operates on dimensional quantities. Since the mapping from dimensional to dimensionless parameters is not one-to-one, multiple combinations of $\tau_0$ and $\rho$ may yield the same value of $\tau_0^*$. Therefore, defining a sampling strategy directly on the dimensionless parameter would require a non-trivial inverse mapping and additional constraints to ensure physical admissibility. 

\subsubsection{Training and Evaluation}
The models are trained in a supervised manner to predict the next-step velocities and pressures. To balance the gradient contributions of different physical quantities, the input features and target variables are normalized to zero mean and unit variance using statistics computed over the training dataset prior to optimization.

For all architectures, the network parameters $\Theta$ are optimized by minimizing the Mean Squared Error (MSE) between the predicted normalized fields and the reference PFEM targets. While MSE serves as the single step training objective, the overall geometric accuracy of autoregressive rollouts is evaluated using the Chamfer distance. In Lagrangian frameworks, continuous advection eliminates the direct point-to-point correspondence required for standard Eulerian node-to-node error metrics. To address this, the domains are treated as unordered point clouds. Given predicted nodal positions $A\subset \mathbb{R}^n$ and reference positions $B\subset \mathbb{R}^n$, (where $n$ is the space dimension, $n=2$ in 2D and $n=3$ in 3D), the Chamfer distance $M_v$ is defined as:

\begin{equation}
    M_v=\mathit{Ch}(A,B) = \sum_{a \in A} \min_{b \in B} d_X(a,b)
\end{equation}
where $d_X$ is the Euclidean distance. This metric provides a measure of the geometric discrepancy between the predicted and reference configurations independently of node ordering.

\section{Results}\label{sec:results}
In this section, the proposed attention-based NeuralPFEM architectures are evaluated on three benchmark problems involving free-surface fluid flows of increasing complexity. The benchmarks are designed to assess predictive accuracy, memory consumption, and computational cost, while also investigating the ability of the models to handle evolving geometries and varying material properties, including non-Newtonian rheologies. Both standard and linear attention formulations are considered and compared against the original GNN-based NeuralPFEM. The test cases span two- and three-dimensional configurations, ranging from moderate-scale problems to large-scale simulations. In addition to the prediction of the primary variables, the analysis also considers the reconstruction of derived quantities of engineering interest obtained from spatial derivatives of the predicted fields. For all benchmarks, predictions are compared against high-fidelity PFEM reference solutions on configurations not seen during training.

\subsection{2D Bingham flow on inclined plane}
\begin{figure}[h]
    \centering
    \subfloat[\label{fig:5a}]{\includegraphics[width=0.5\linewidth]{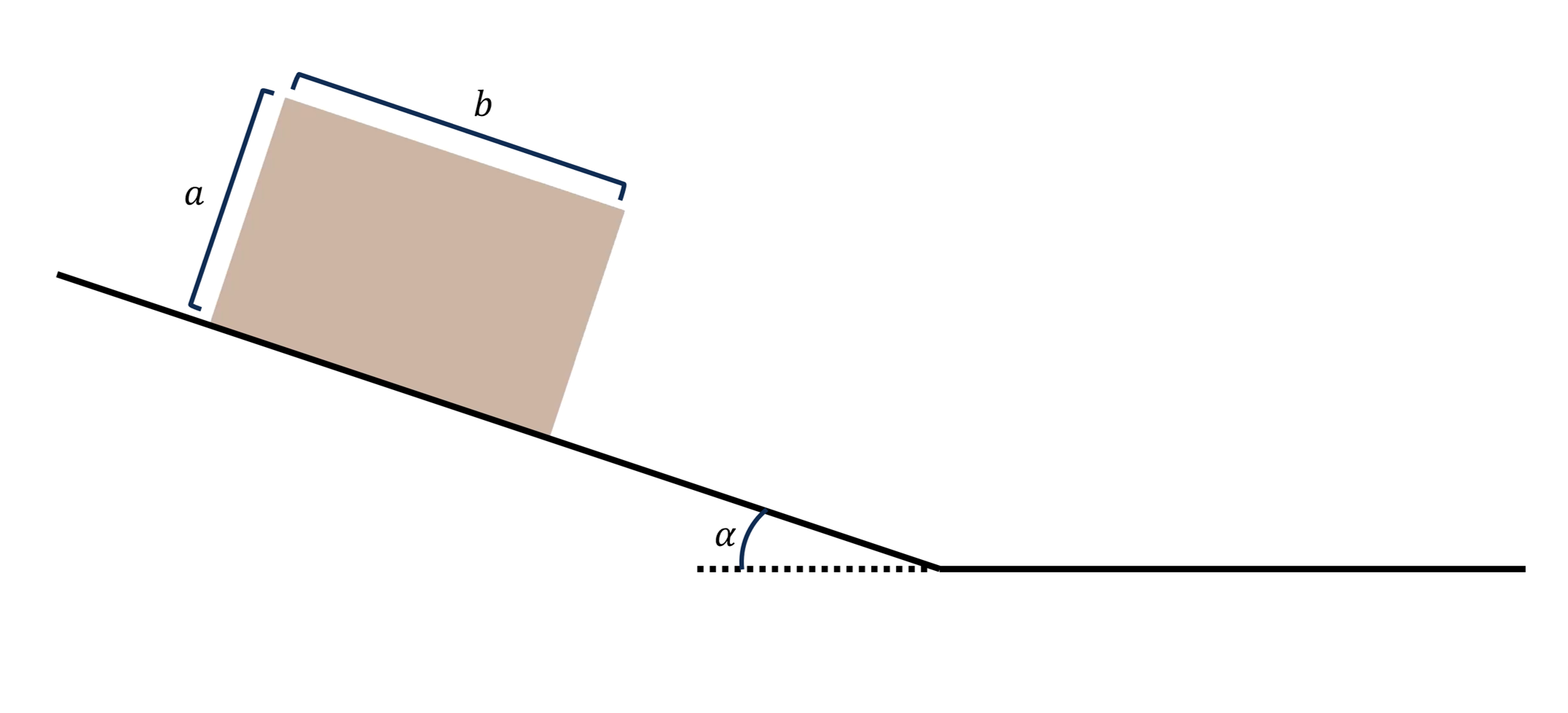}} 
    \subfloat[\label{fig:5b}]{\includegraphics[width=0.5\linewidth]{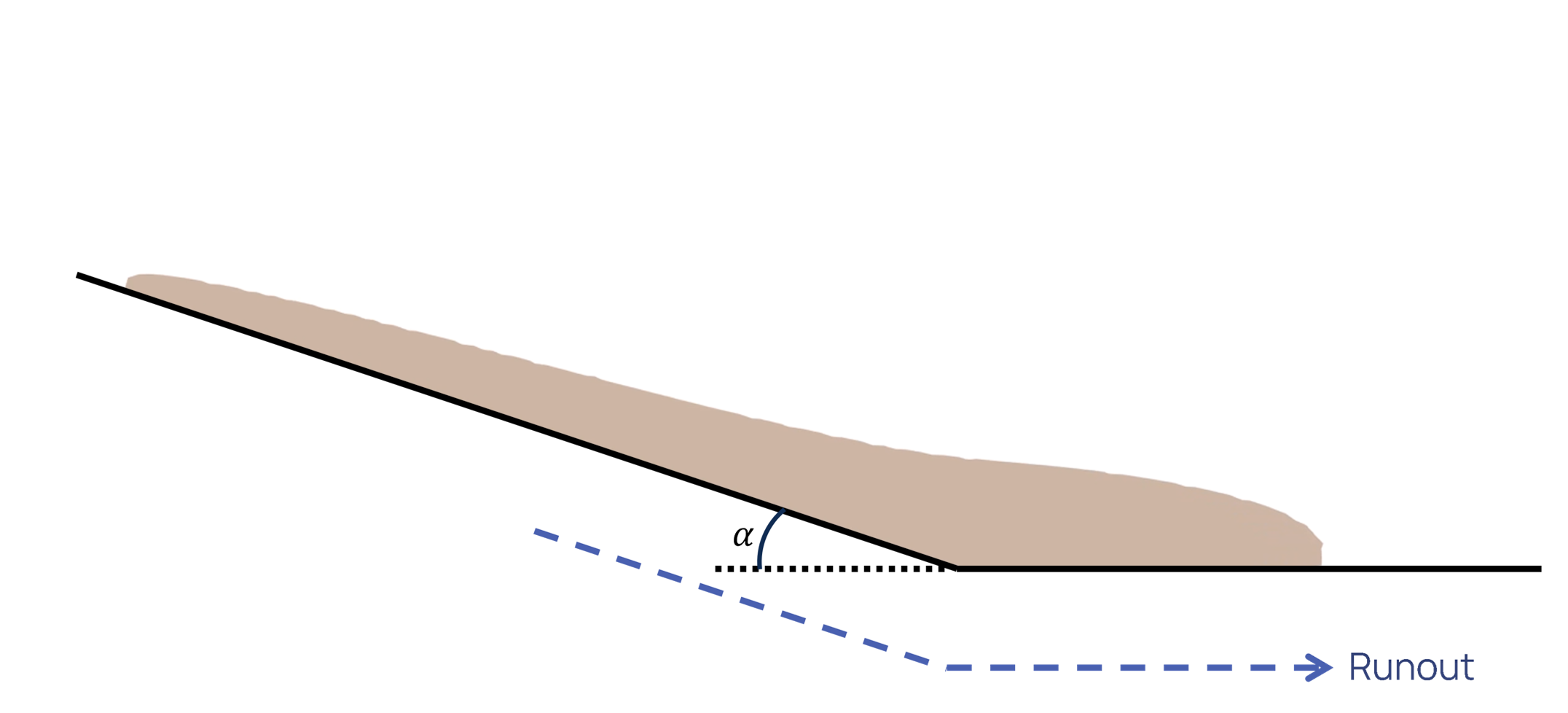}}
    \caption{2D inclined plane benchmark: initial configuration (left) and final equilibrium state (right).}
    \label{fig:5}
\end{figure}

The first benchmark considers a two-dimensional visco-plastic Bingham fluid released from rest on an inclined plane. The initial configuration consists of a rectangular block of dimensions $a \times b$, placed on a surface inclined at an angle $\alpha$ with respect to the horizontal (Figure~\ref{fig:5a}). Upon release, the fluid spreads under the action of gravity. Unlike Newtonian fluids, which would continue to flow indefinitely, Bingham fluids exhibit a different behavior: as stated in Section~\ref{sec:governing}, flow occurs only where the local deviatoric stress exceeds the threshold set by the yield stress value $\tau_0$, while material in sub-yield regions behaves as a rigid body. As the fluid spreads and the free surface flattens, the gravitational driving stress progressively decreases until it falls below $\tau_0$ everywhere in the domain, at which point the material arrests and a static equilibrium is reached. Depending on the material properties and inclination angle, two qualitatively distinct outcomes are possible: the fluid either reaches the end of the inclined section and continues spreading onto the horizontal plane (as in Figure~\ref{fig:5b}), or comes to rest on the slope before reaching the transition point. All simulations are carried out until this equilibrium state is achieved, making the final deposit geometry, in addition to the transient dynamics, a meaningful validation target. Since Bingham fluids provide a natural model for geomaterials such as debris flows and mudslides, this configuration is widely used as a benchmark in geotechnical and civil engineering applications.

A series of PFEM simulations were carried out to generate the training, validation, and test datasets. In total, 60 simulations were used for training, 10 for validation, and 20 for testing. The initial rectangular geometry was identical across all simulations, with $a = 0.2\,\mathrm{m}$ and $b = 0.3\,\mathrm{m}$. The variability in the dataset arises from the material properties— density $\rho$, yield stress $\tau_0$, and viscosity $\mu$—as well as from the inclination angle $\alpha$ of the plane. A characteristic mesh size of $m = 5 \cdot 10^{-3}\,\mathrm{m}$ was adopted, resulting in approximately 2800 particles. PFEM solution snapshots were stored every $\Delta t = 0.005\,\mathrm{s}$, which corresponds to the timestep used by the NeuralPFEM model. The main parameters of the simulations, including the range of variability of the varying material and geometric parameters are summarized in Table~\ref{tab:1}.

\begin{table}[h]
\centering
\caption{Summary of the inclined plane Bingham fluid test case parameters.}
\label{tab:1}
\begin{tabular}{lcc}
\toprule
\textbf{} & \textbf{Symbol and unit} & \textbf{Value} \\
\midrule
\multicolumn{3}{l}{\textbf{Fixed parameters}} \\
\midrule
Block length & $a\,[\mathrm{m}]$ & $0.2$ \\
Block height & $b\,[\mathrm{m}]$ & $0.3$ \\
\midrule
\multicolumn{3}{l}{\textbf{Variable parameters}} \\
\midrule
Inclination angle & $\alpha\,[^\circ]$ & $[10,\,40]$ \\
Density & $\rho\,[\mathrm{kg/m^3}]$ & $[1500,\,2500]$ \\
Yield stress & $\tau_0\,[\mathrm{Pa}]$ & $[15,\,1000]$ \\
Viscosity & $\mu\,[\mathrm{Pa \cdot s}]$ & $[1,\,100]$ \\
\midrule
\multicolumn{3}{l}{\textbf{Numerical parameters}} \\
\midrule
Mesh size & $m\,[\mathrm{m}]$ & $5 \times 10^{-3}$ \\
Number of nodes & $N[-]$ & 2813 \\
Time step & $\Delta t\,[\mathrm{s}]$ & $5 \times 10^{-3}$ \\
\bottomrule
\end{tabular}
\end{table}

\begin{figure}[h]
    \centering
    \subfloat[Case I1, $t=0.1 \, s$]{\includegraphics[width=0.33\linewidth]{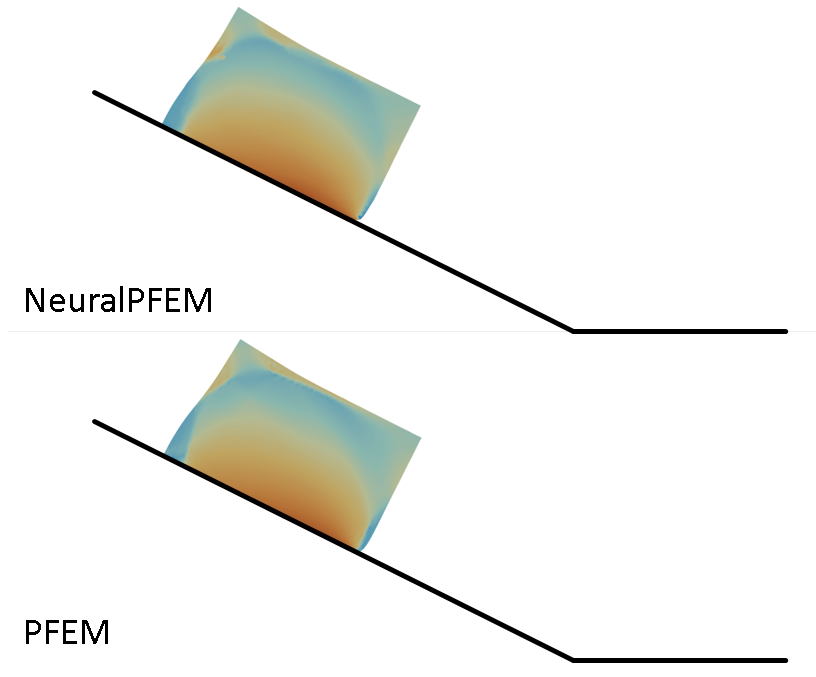} \label{fig:6a}}
    \subfloat[Case I1, $t=0.5 \, s$]{\includegraphics[width=0.33\linewidth]{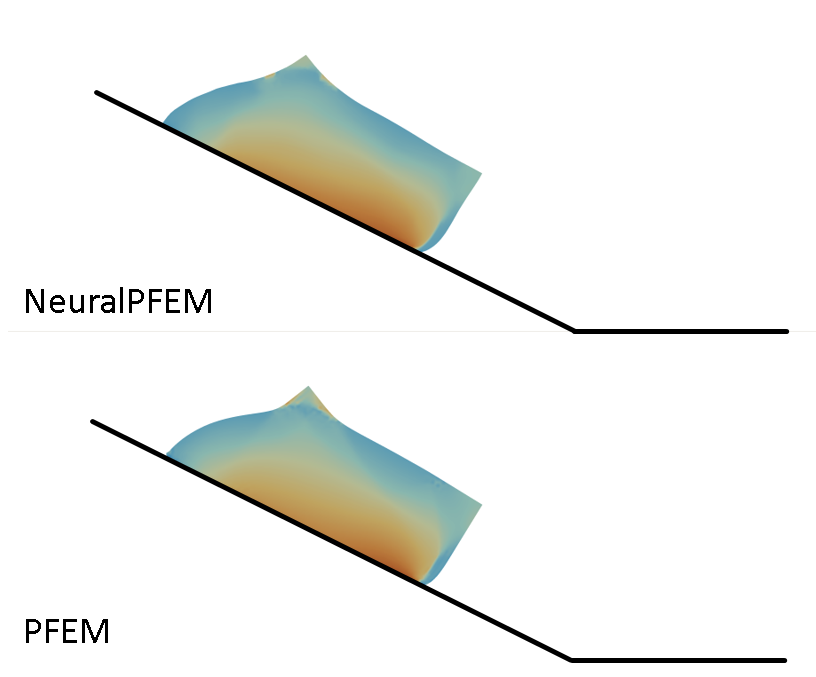} \label{fig:6b}}
    \subfloat[Case I1, $t=2 \, s$]{\includegraphics[width=0.33\linewidth]{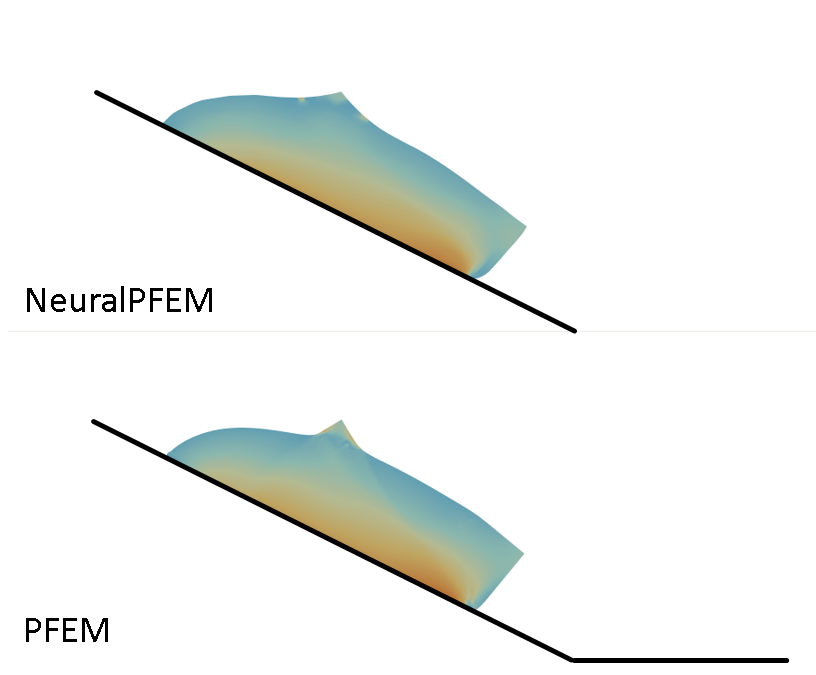} \label{fig:6c}}\\
    \subfloat[Case I2, $t=0.1 \, s$]{\includegraphics[width=0.33\linewidth]{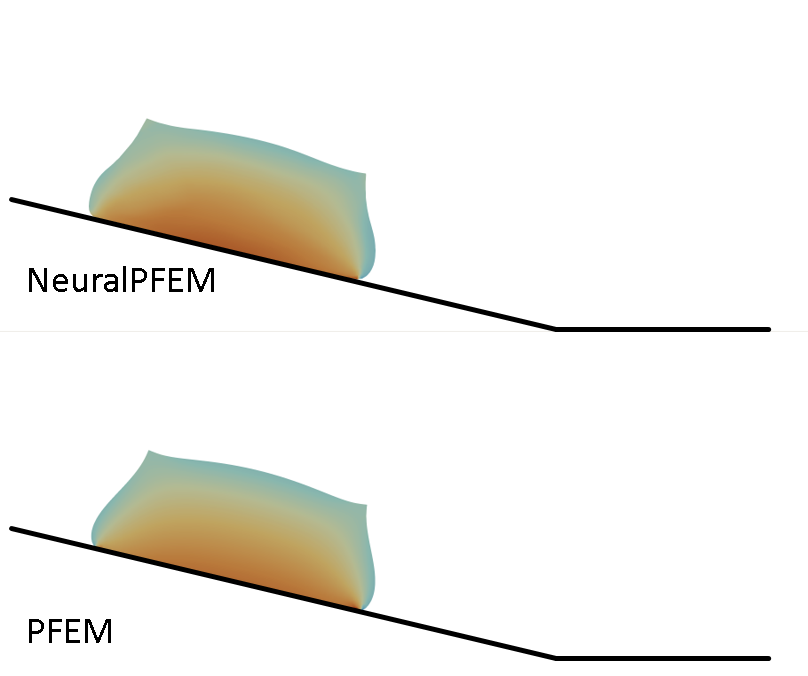} \label{fig:6d}}
    \subfloat[Case I2, $t=0.5 \, s$]{\includegraphics[width=0.33\linewidth]{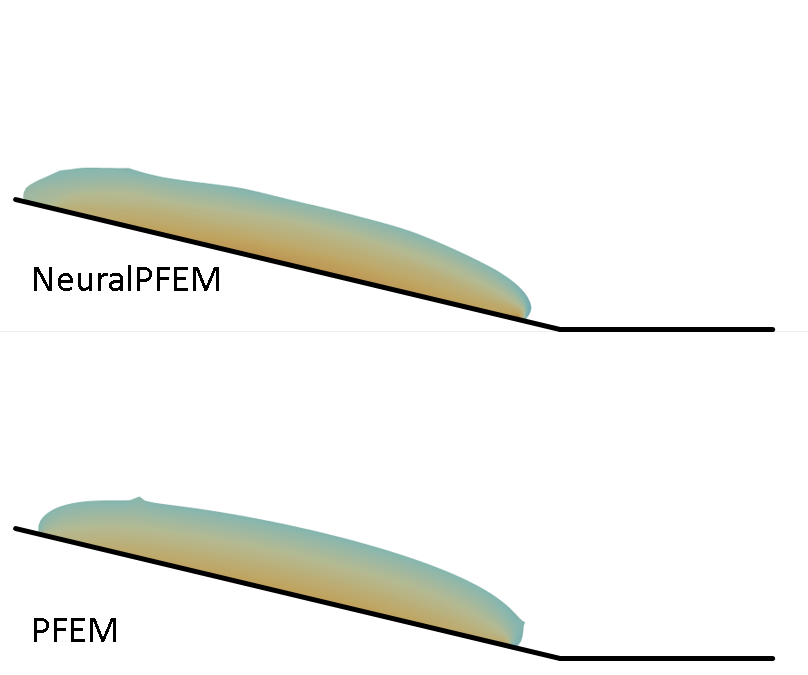} \label{fig:6e}}
    \subfloat[Case I2, $t=2 \, s$]{\includegraphics[width=0.33\linewidth]{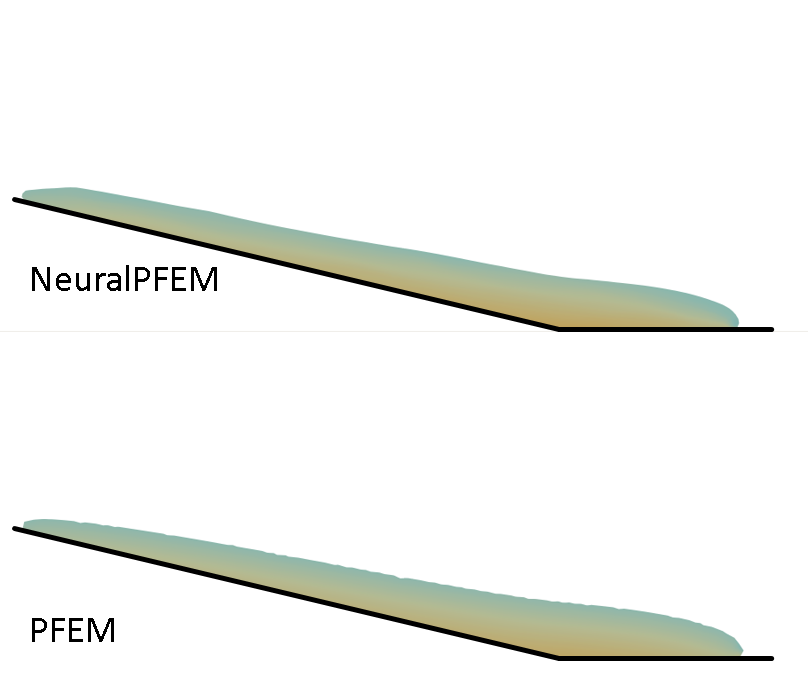} \label{fig:6f}}\\
    \subfloat[Case I3, $t=0.1 \, s$]{\includegraphics[width=0.33\linewidth]{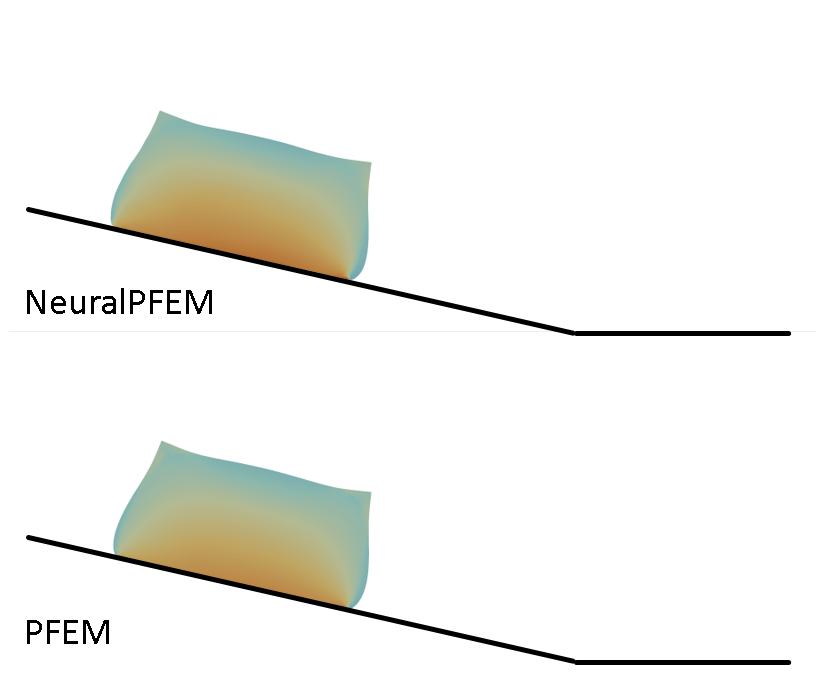} \label{fig:6g}}
    \subfloat[Case I3, $t=0.5 \, s$]{\includegraphics[width=0.33\linewidth]{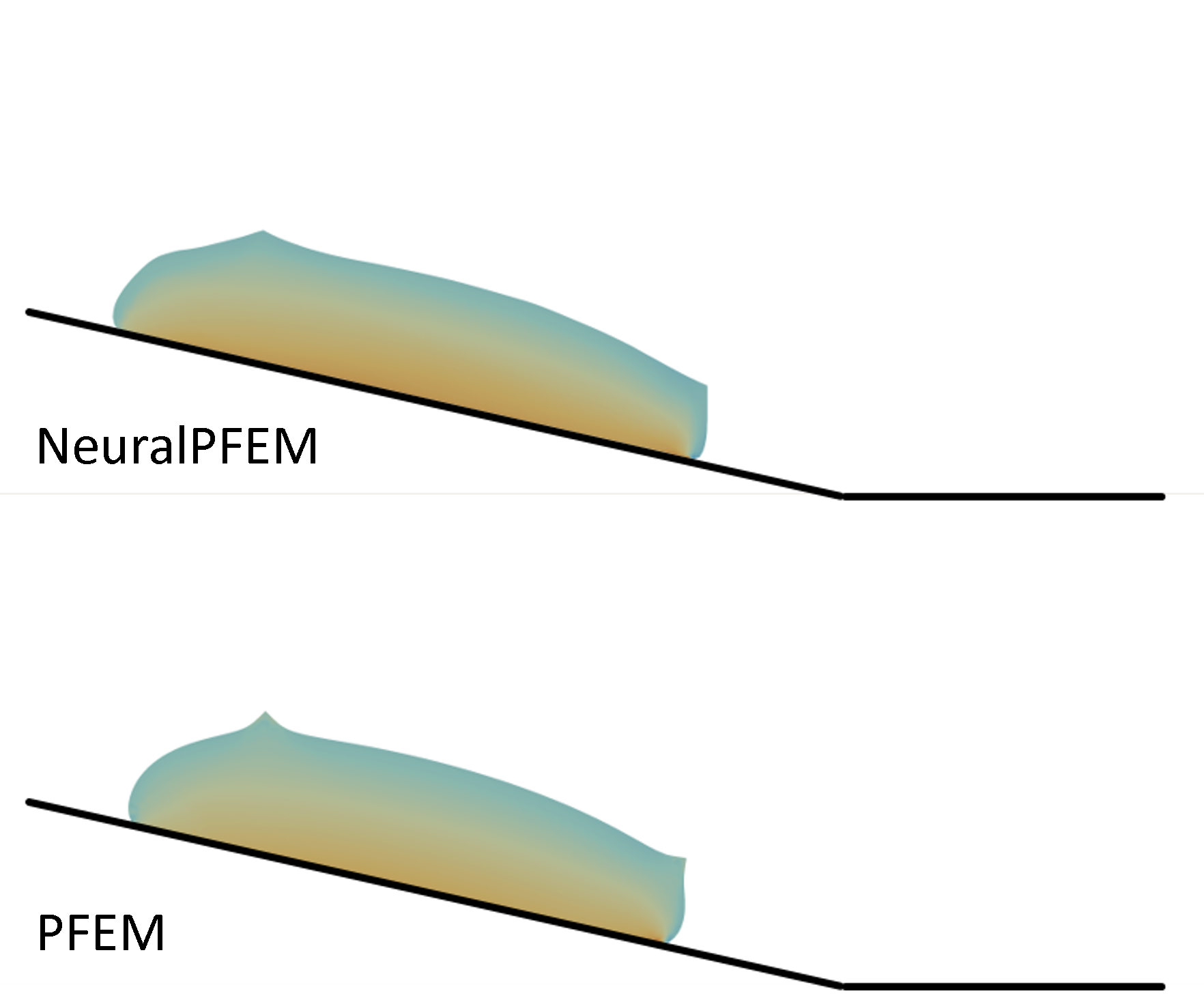} \label{fig:6h}}
    \subfloat[Case I3, $t=2 \, s$]{\includegraphics[width=0.33\linewidth]{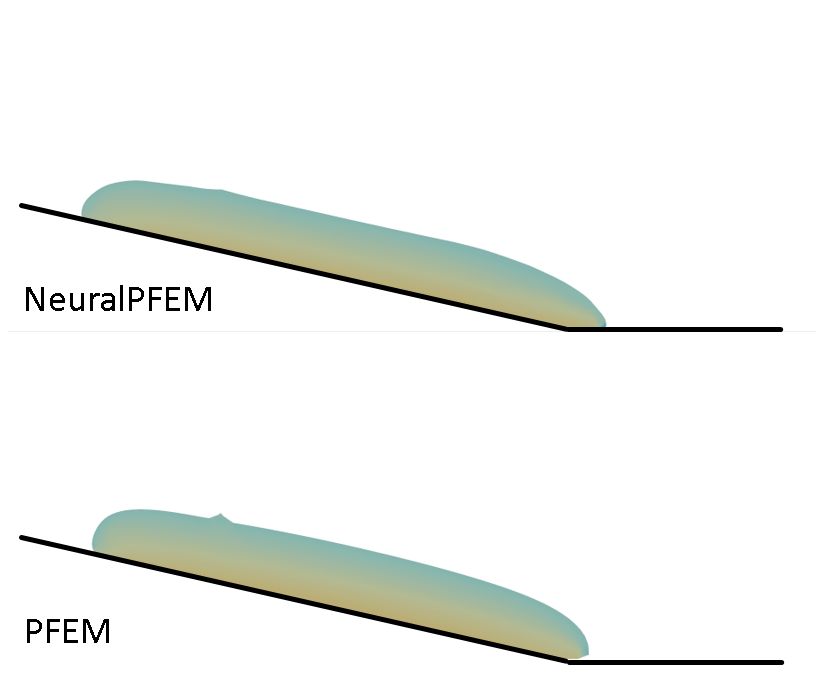} \label{fig:6i}}\\
    \caption{Temporal snapshots of NeuralPFEM with standard self-attention predictions (top row of each pair) and reference PFEM solutions (bottom row) for test cases I1 (\ref{fig:6a}-\ref{fig:6c}), I2 (\ref{fig:6d}-\ref{fig:6f}), and I3 (\ref{fig:6g}-\ref{fig:6i}) at selected time steps. The pressure is visualised as a scalar field on the deformed mesh.}
    \label{fig:6}
\end{figure}

\begin{figure}[h]
    \centering
    \subfloat[Case I1]{\includegraphics[width=0.33\linewidth]{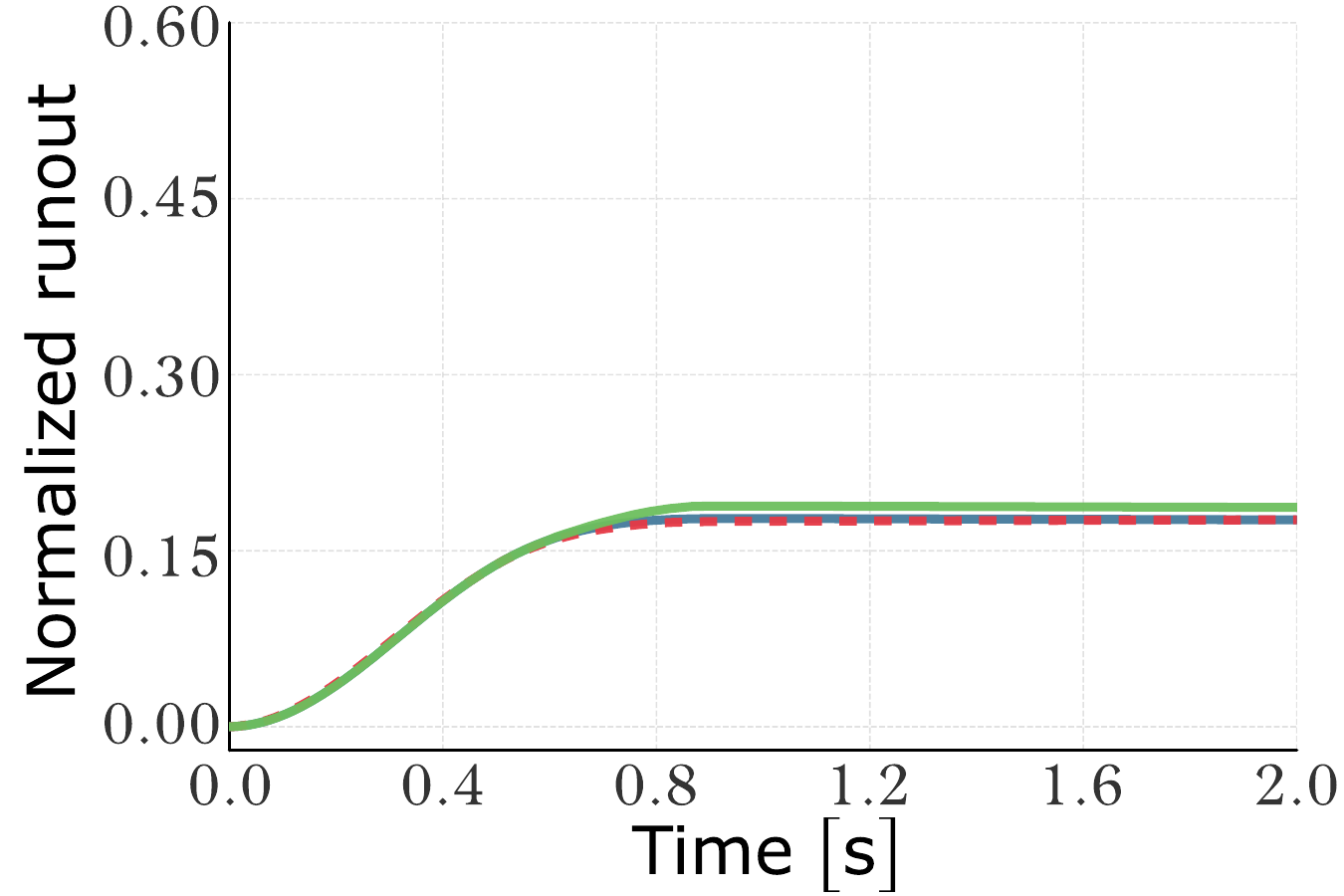}}
    \subfloat[Case I2]{\includegraphics[width=0.33\linewidth]{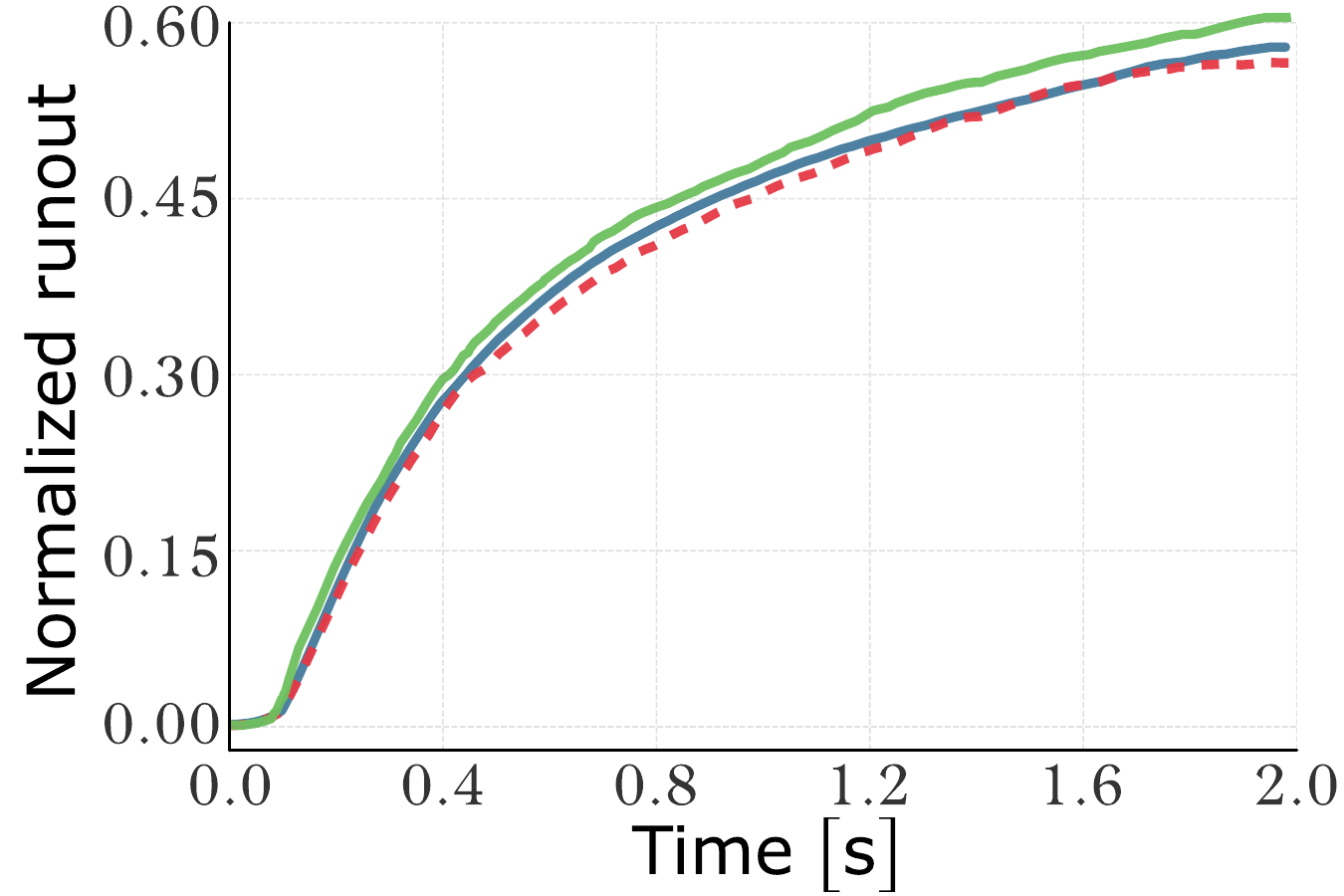}}
    \subfloat[Case I3]{\includegraphics[width=0.33\linewidth]{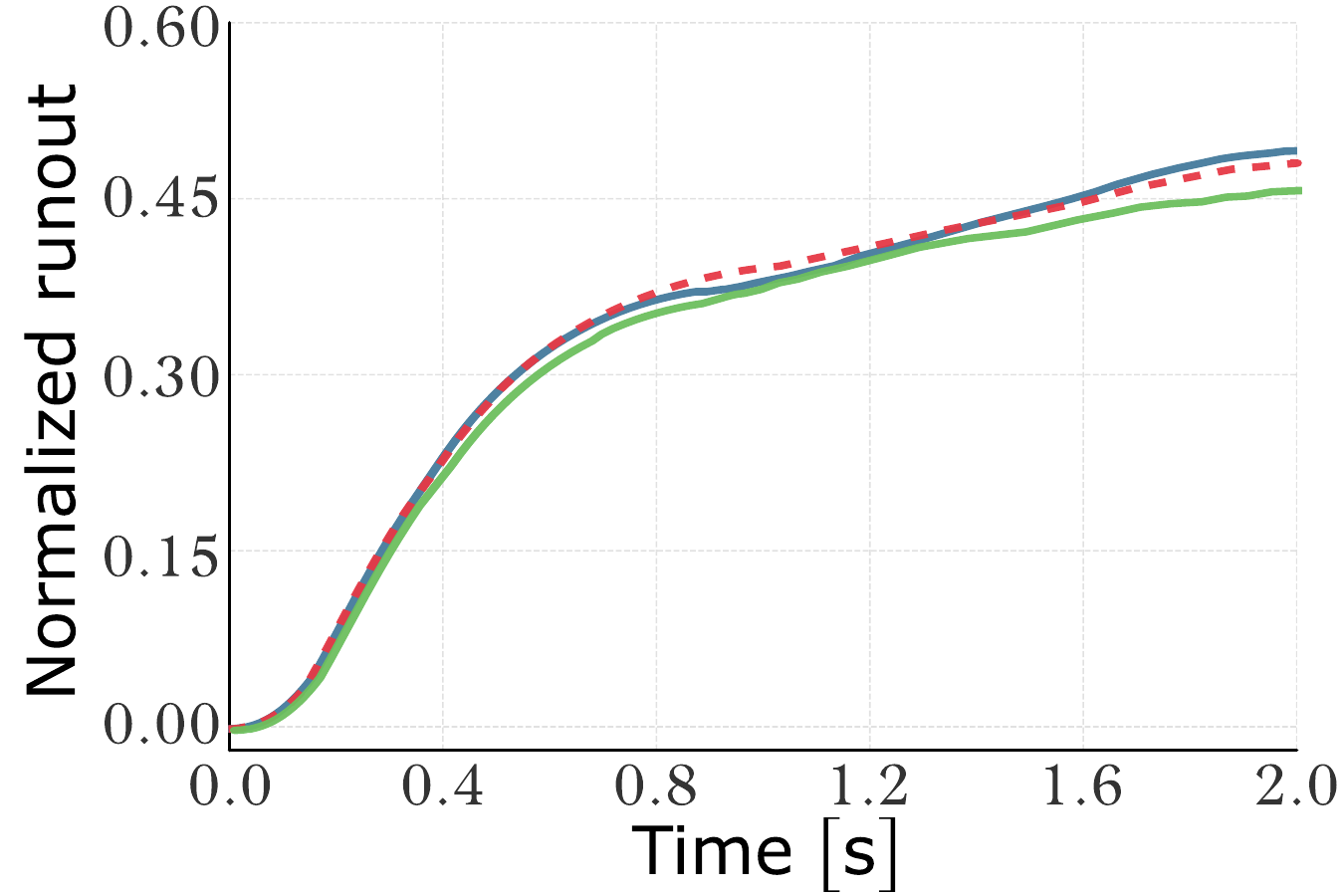}}\\
    \caption{Comparison of the normalized runout evolution over time obtained with PFEM (\textcolor[HTML]{E63946}{\rule[0.5ex]{1.3ex}{0.5ex} \rule[0.5ex]{1.3ex}{0.5ex}}), NeuralPFEM with standard (\textcolor[HTML]{457B9D}{\rule[0.5ex]{3ex}{0.5ex}}) and linear (\textcolor[HTML]{6ec05f}{\rule[0.5ex]{3ex}{0.5ex}}) self-attention for the test cases shown in Figure~\ref{fig:6}.}
    \label{fig:7}
\end{figure}

\begin{table}[h]
\centering
\caption{Material parameters of the test cases shown in Figure \ref{fig:6}.}
\label{tab:2}
\begin{tabular}{lcccc}
\toprule
\textbf{ID} & $\alpha$ [rad] & $\rho$ [kg/m$^3]$ & $\mu$ [Pa$\cdot$s] & $\tau_0$ [Pa] \\
\midrule
I1 & 0.472 & 1717   & 74 & 942  \\
I2 & 0.237 & 1876   & 47 & 18 \\
I3 & 0.223 & 1410   & 61 & 179 \\
\bottomrule
\end{tabular}
\end{table}

Once training is complete, the best-performing model, selected via validation performance, is then evaluated against reference PFEM solutions on the 20 unseen test simulations.

Figure~\ref{fig:6} presents three temporal snapshots for representative test cases whose material parameters are reported in Table~\ref{tab:2}, with pressure visualised as a scalar field on the deformed mesh. The three cases exhibit markedly different physical behaviours as a direct consequence of their distinct material properties and inclination angles. In Case I1 (Figures ~\ref{fig:6a}-~\ref{fig:6c}), the high yield stress causes the fluid to arrest well before reaching the end of the incline, despite the steep inclination angle. In Cases I2 (Figures ~\ref{fig:6d}-~\ref{fig:6f}) and I3 (Figures ~\ref{fig:6g}-~\ref{fig:6i}), the reduced yield stress facilitates flow even at shallower inclinations. Although the inclination angles in these two cases are nearly identical, the influence of material properties is clearly discernible: I3 barely reaches the transition point before settling into equilibrium (Figure ~\ref{fig:6i}), whereas I2 traverses the entire inclined section and continues flowing along the horizontal plane (Figure ~\ref{fig:6f}). These three scenarios collectively span a wide range of dynamic and equilibrium behaviours, and also involve geometrically distinct configurations: varying the inclination angle alters not only the flow dynamics but also the geometry of the boundary on which conditions must be imposed. Despite this diversity, NeuralPFEM consistently reproduces the correct qualitative and quantitative behaviour across all cases. The model accurately captures the different responses associated with distinct material properties and adapts effectively to changes in geometry and boundary conditions. In addition, the pressure field is also well reproduced in all configurations. The predicted free-surface shapes closely match the reference PFEM results at all reported time steps, including the final equilibrium configurations. The normalised runout curves in Figure~\ref{fig:7} confirm this accuracy quantitatively: the predicted runout evolution closely tracks the reference trajectory throughout the simulation, capturing the transient spreading phase, the deceleration regime, and the final arrested state.

\begin{table}
\centering
\caption{Comparison of error, memory footprint, and computational cost for GNN, standard self-attention, and linear self-attention variants of NeuralPFEM on the 2D Bingham flow on an inclined plane.}
\label{tab:3}
\begin{tabular}{l cccc}
\toprule
\textbf{Model} & \textbf{MP steps} & \textbf{Mean Chamfer distance [m]} & \textbf{Peak memory [MB]} & \textbf{Time / step [s]} \\
\midrule
\multirow{3}{*}{\textbf{GNN}}
& 1 & $1.7 \cdot 10^{-1}$ & 470 & 0.02 \\ 
& 10 & $8.5 \cdot 10^{-3}$ & 808 & 0.04 \\
& 20 & $3.5 \cdot 10^{-3}$ &  1356 & 0.06 \\
& 60 & $3.6 \times 10^{-3}$ & 4682 & 0.13 \\
\addlinespace[0.5em]
\textbf{Self-attention} &  &  &  &  \\
Standard & -- & $3.3 \cdot 10^{-3}$ & 493 & 0.04 \\
Linear & -- & $3.5 \cdot 10^{-3}$ & 572 & 0.05 \\
\bottomrule
\end{tabular}
\end{table}

Table~\ref{tab:3} reports the predictive accuracy and peak memory footprint of the different NeuralPFEM variants on the 2D Bingham flow benchmark. GNN, standard self-attention, and linear self-attention achieve comparable levels of accuracy on this task. In all cases, the mean Chamfer distance over the test dataset remains below the characteristic mesh size adopted in the simulation, indicating that the predicted node positions are, on average, within the discretization error of the reference solution.

For the GNN processor, the results reflect the behavior discussed in Section~\ref{sec:memory}. Achieving adequate predictive accuracy requires a sufficiently large number of message-passing steps: when $M$ is too small, information remains too localized and cannot propagate across the domain, leading to poor predictions. Conversely, increasing $M$ improves accuracy, but at the cost of a rapid growth in memory consumption. This trade-off is already evident in this relatively small 2D problem, highlighting a limitation in scalability. In the present configuration, the theoretical condition $M = \frac{L}{h}$, where $L$ is the characteristic domain length, yields $M = 60$ message-passing steps. The empirical results indicate that increasing $M$ beyond $20$ does not lead to further improvements, probably due to the presence of oversmoothing phenomena. Nonetheless, even satisfying this empirical depth requirement carries a substantial memory cost compared to the problem size. This exposes the major limitation of GNN-based processors: the need of sufficiently deep architectures to ensure adequate information propagation makes high memory consumption unavoidable, ultimately restricting their applicability to larger and higher-resolution problems.

Both the standard and linear attention variants achieve an accuracy comparable to the $M=20$ GNN while requiring substantially lower peak memory. This is a direct consequence of the architectural properties described in Section~\ref{sec:attn-npfem}: self-attention enables global information exchange in a single layer, thereby satisfying the depth condition of~\cite{tesan2026} by construction, without the need for iterative message passing or explicit edge storage. As a result, the dominant memory term associated with edge embeddings is entirely eliminated.

Comparing the two attention mechanisms, the standard softmax formulation provides a marginally lower error, reflecting its higher expressiveness due to exact pairwise interactions. The linear attention variant, while slightly less accurate, remains within the same error regime. Its primary benefit lies in its improved scalability, as it reduces the quadratic complexity in the number of nodes to linear, which becomes critical for larger problem sizes. In the present setting, where the number of nodes is limited, this advantage is not reflected in the computational cost, which is instead slightly higher. At the same time, the comparable memory consumption between the two variants suggests that the FlashAttention implementation effectively mitigates the theoretic memory overhead of standard attention.

A key advantage of the proposed approach with respect to GNN-based formulations is its ability to generalize across mesh resolutions at inference time, even when they differ from those used during training. In GNNs, edge features are typically constructed with inter-nodal distances, which tie the model to the characteristic element size of the training meshes. As a result, changing the mesh resolution in prediction alters these geometric features and degrade accuracy. In contrast, the self-attention processor removes the need for explicit geometric edge features altogether. Spatial information is instead encoded via nodal positional embeddings, which represent each node’s location independently of the mesh density. This design allows the same trained model to be applied to discretizations that differ from those used during training. Unlike Eulerian problems with fixed geometries, where increasing resolution at inference time merely leads to an oversampling of the solution learned at the resolution of the training meshes, this property is particularly beneficial in Lagrangian free-surface settings. Higher resolution enhances geometric representation of the evolving boundary, yielding a more accurate reconstruction of the free surface.

To quantify this generalization capability, we evaluate NeuralPFEM on the test dataset using meshes with characteristic sizes different from the training resolution of $h = 0.005\,\mathrm{m}$ spanning from a coarser mesh at $h = 0.01\,\mathrm{m}$ and a finer mesh at $h = 0.002\,\mathrm{m}$. As shown in Table~\ref{tab:4}, the accuracy metrics obtained with the standard attention model remain consistent across these resolutions, indicating a robust generalization without performance degradation. In contrast, the GNN model exhibits increasing performance deterioration as the mesh resolution deviates further from the training configuration.

It is worth remarking that the practical value of resolution invariance is particularly pronounced in our Lagrangian free-surface setting. In standard fixed-mesh applications, using a model on a finer grid at inference time simply oversamples the originally learned dynamics, offering no actual increase in physical accuracy. In our framework, however, discretization resolution plays an additional geometrical role. The domain shape emerges directly from the distribution of the points, where the boundary of the node set defines the free surface. Thus, a finer mesh can more accurately resolve the shape, capturing more complex shape features and reducing geometric error. Resolution invariance is therefore practically valuable here: a single trained model can be paired with a finer mesh at inference time to obtain a more faithful geometric representation of the free surface, without any retraining cost.

\begin{table}
\centering
\caption{Chamfer distance comparison as a function of mesh size.}
\label{tab:4}
\begin{tabular}{ccc}
\toprule
\textbf{Mesh size $[m]$} & \multicolumn{2}{c}{\textbf{Mean Chamfer distance $[m]$}} \\
\cmidrule(lr){2-3}
 & \textbf{GNN} & \textbf{Attention} \\
\midrule
0.002 & $9.9 \cdot 10^{-3}$ & $3.5 \cdot 10^{-3}$ \\
0.004 & $4.2 \cdot 10^{-3}$ & $3.1 \cdot 10^{-3}$ \\
0.005 & $3.5 \cdot 10^{-3}$ & $3.3 \cdot 10^{-3}$ \\
0.006 & $3.8 \cdot 10^{-3}$ & $3.5 \cdot 10^{-3}$ \\
0.009 & $5.4 \cdot 10^{-2}$ & $3.8 \cdot 10^{-3}$ \\
0.010 & $6.7 \cdot 10^{-2}$ & $4.0 \cdot 10^{-3}$ \\
\bottomrule
\end{tabular}
\end{table}

\subsection{3D Bingham cone slump test}
\begin{figure}[h]
    \centering
    \subfloat[\label{fig:8a}]{\includegraphics[width=0.5\linewidth]{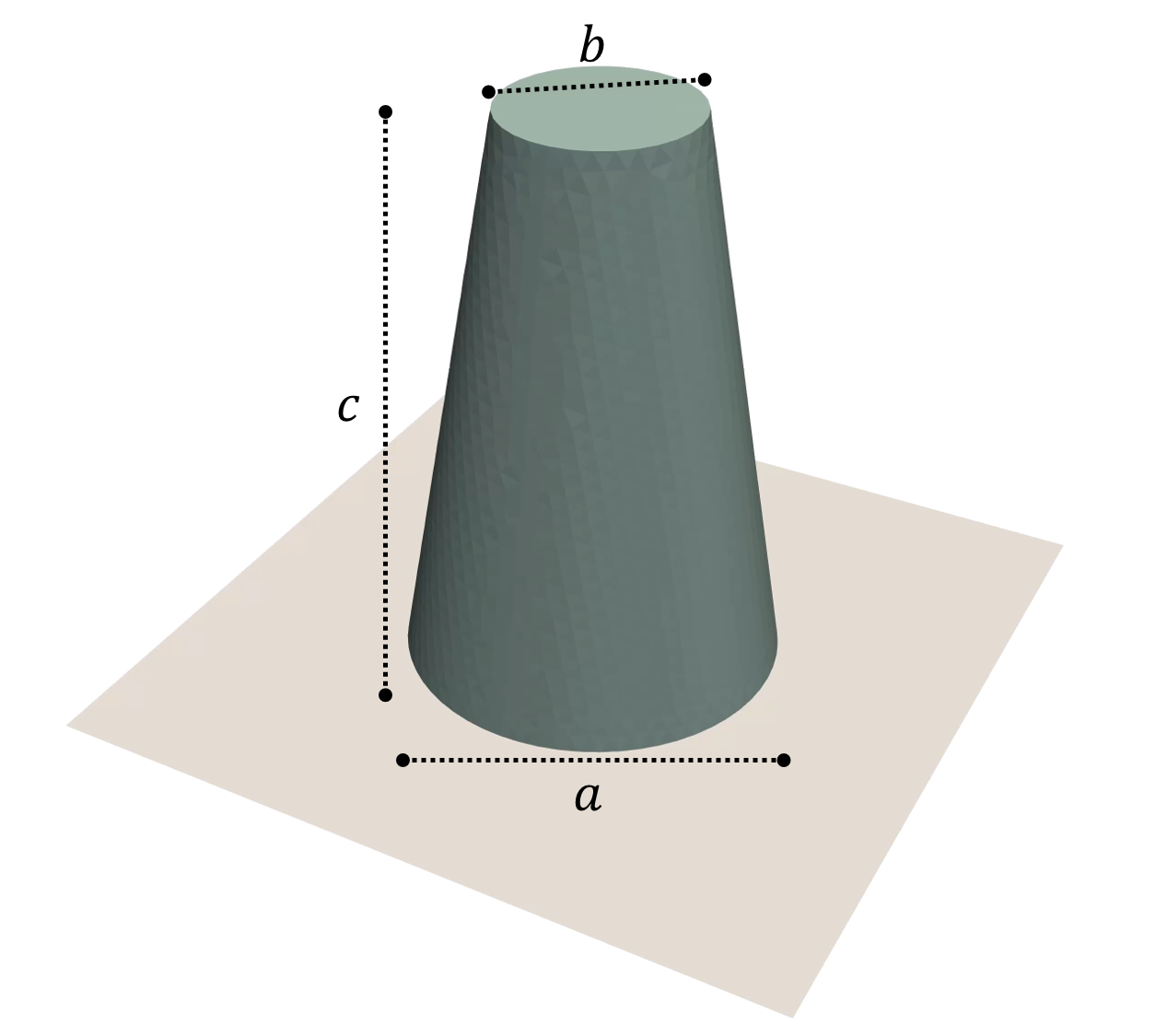}} 
    \subfloat[\label{fig:8b}]{\includegraphics[width=0.5\linewidth]{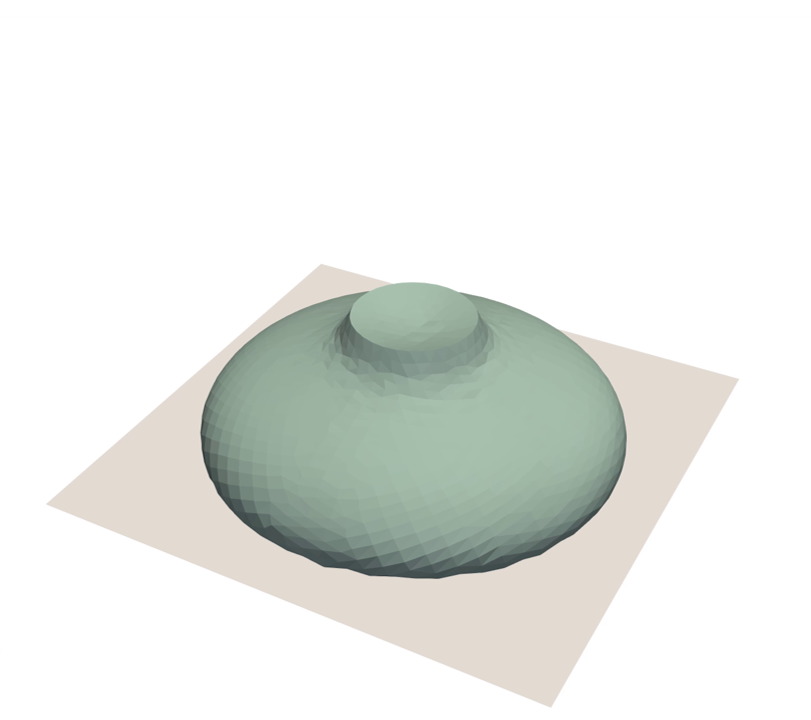}}
    \caption{3D bingham cone slump experiment: initial configuration (left) and final equilibrium state (right).}
    \label{fig:8}
\end{figure}

The second benchmark extends the evaluation to three dimensions through a Bingham fluid cone slump test. The initial configuration is a truncated cone whose dimensions match those of the standard Abrams cone, a benchmark widely used in civil engineering practice to characterize the workability and rheological properties of cementitious materials such as fresh concrete (Figure~\ref{fig:8a}). The base diameter is $a = 0.2\,\mathrm{m}$, the top diameter $b = 0.1\,\mathrm{m}$, and the height $c = 0.3\,\mathrm{m}$. At the initial instant, the conical box is removed and the fluid spreads freely under gravity until the driving stress falls below the yield threshold and the material arrests (Figure~\ref{fig:8b}). The final spread diameter is a standard metric for assessing material workability in practice and serves here as a validation target alongside the full transient evolution of the flow.

This test case serves a dual purpose. In addition to assessing the model's performances in a three-dimensional setting, it is used to evaluate the ability of NeuralPFEM to reproduce derived mechanical quantities — specifically, the stress field within the flowing material. The mesh structure of NeuralPFEM enables the computation of spatial derivatives of the predicted velocity and pressure fields through finite element shape functions. For a Bingham fluid, the deviatoric stress tensor $\bm{\tau}$ is related to the strain rate tensor by the constitutive relation \ref{eq:bingham}. The strain rate field is recovered by differentiating the predicted velocity field through the finite-element shape functions defined on the current mesh, and the stress field is then recovered elementwise through the constitutive law. This post-processing step is performed at inference time and relies entirely on the accuracy of the predicted velocity field and on the geometric information encoded in the mesh. Accurate stress reconstruction therefore serves as an indirect but stringent assessment of the quality of the velocity predictions, since errors in velocity gradients are amplified by differentiation.

The dataset was generated by varying the material parameters (density $\rho$, yield stress $\tau_0$, and viscosity $\mu$) over the ranges reported in Table~\ref{tab:5}, while keeping the initial geometry fixed. A characteristic mesh size of $h = 10^{-2}\,\mathrm{m}$ was adopted, yielding approximately 5000 nodes per simulation. Solution snapshots were stored every $\Delta t = 0.005\,\mathrm{s}$. The dataset consists of 60 training simulations, 10 for validation, and 20 for testing.

\begin{table}
\centering
\caption{Summary of the 3D Bingham cone slump test case parameters.}
\label{tab:5}
\begin{tabular}{lcc}
\toprule
\textbf{} & \textbf{Symbol and unit} & \textbf{Value} \\
\midrule
\multicolumn{3}{l}{\textbf{Fixed parameters}} \\
\midrule
Base diameter & $a\,[\mathrm{m}]$ & $0.2$ \\
Top diameter & $b\,[\mathrm{m}]$ & $0.1$ \\
Height & $c\,[\mathrm{m}]$ & $0.3$ \\
\midrule
\multicolumn{3}{l}{\textbf{Variable parameters}} \\
\midrule
Density & $\rho\,[\mathrm{kg/m^3}]$ & $[2000,\,2500]$ \\
Yield stress & $\tau_0\,[\mathrm{Pa}]$ & $[100,\,1800]$ \\
Viscosity & $\mu\,[\mathrm{Pa \cdot s}]$ & $[10,\,100]$ \\
\midrule
\multicolumn{3}{l}{\textbf{Numerical parameters}} \\
\midrule
Mesh size & $m\,[\mathrm{m}]$ & $10^{-2}$ \\
Number of nodes & $N[-]$ & $4768$ \\
Time step & $\Delta t\,[\mathrm{s}]$ & $0.005$ \\
\bottomrule
\end{tabular}
\end{table}

\begin{figure}[p]
    \centering
    \subfloat[Case S1, $t=0.1 \, s$]{\includegraphics[width=0.33\linewidth]{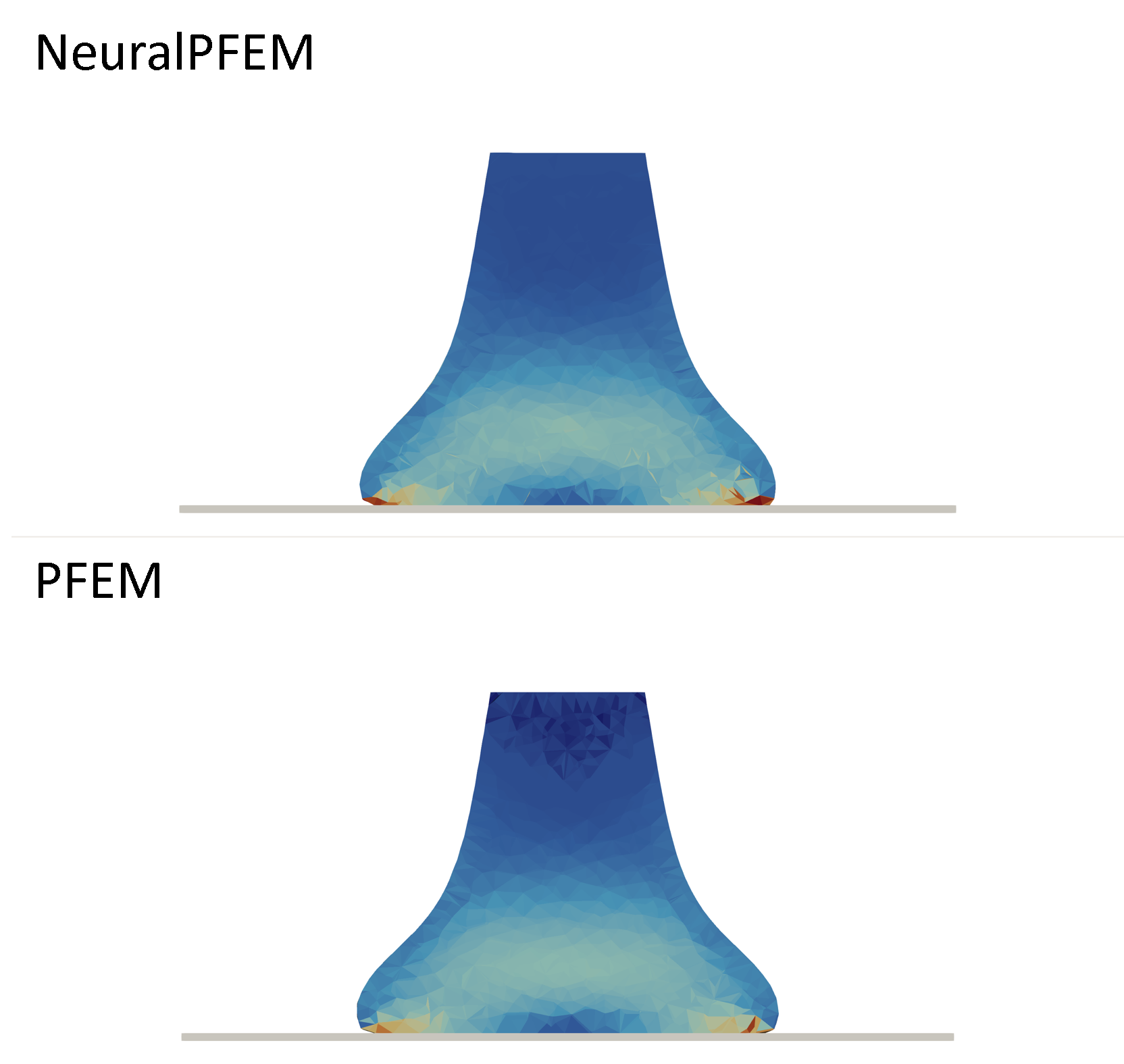}\label{fig:9a}}
    \subfloat[Case S1, $t=0.25 \, s$]{\includegraphics[width=0.33\linewidth]{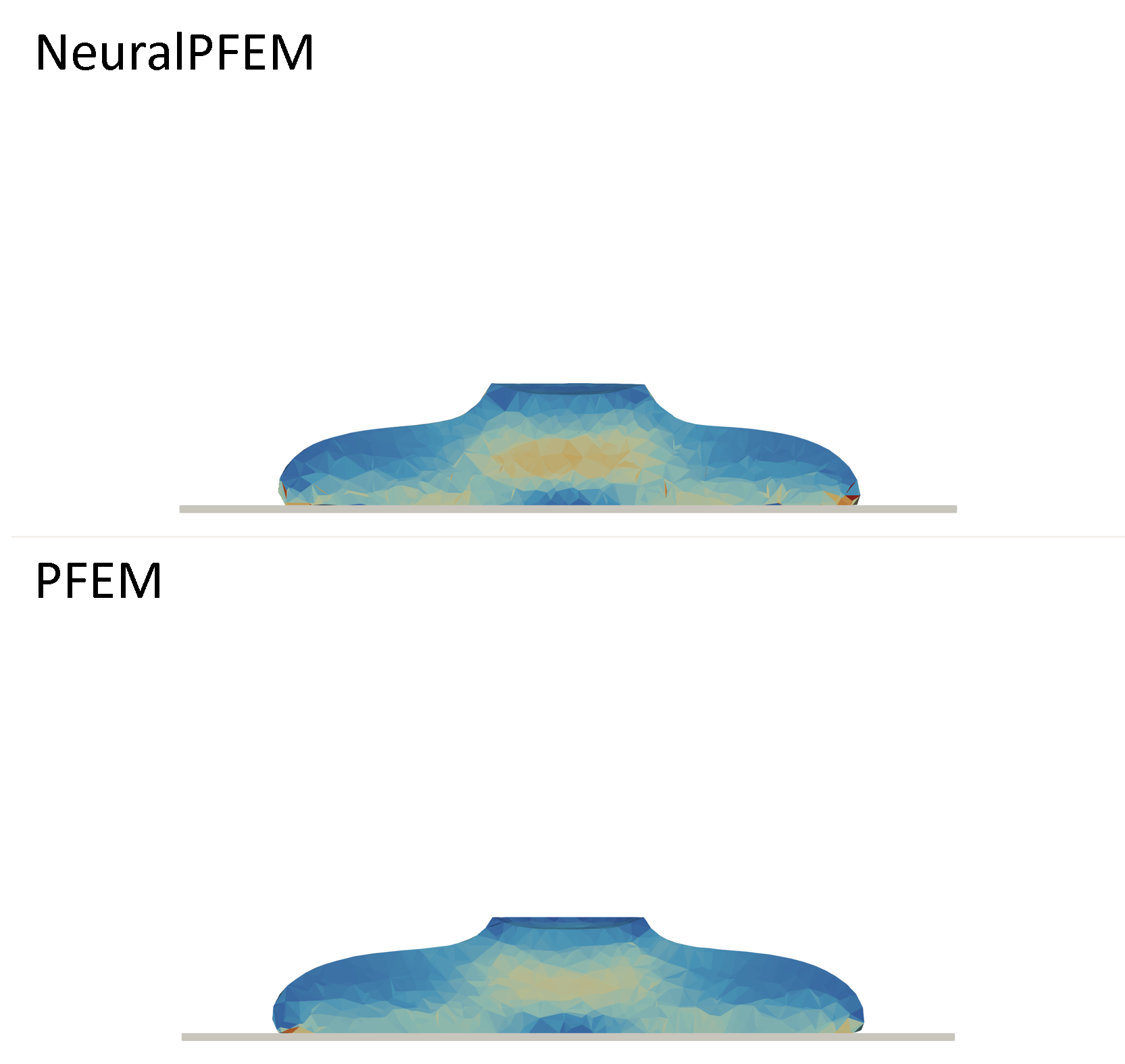}\label{fig:9b}}
    \subfloat[Case S1, $t=0.5 \, s$]{\includegraphics[width=0.33\linewidth]{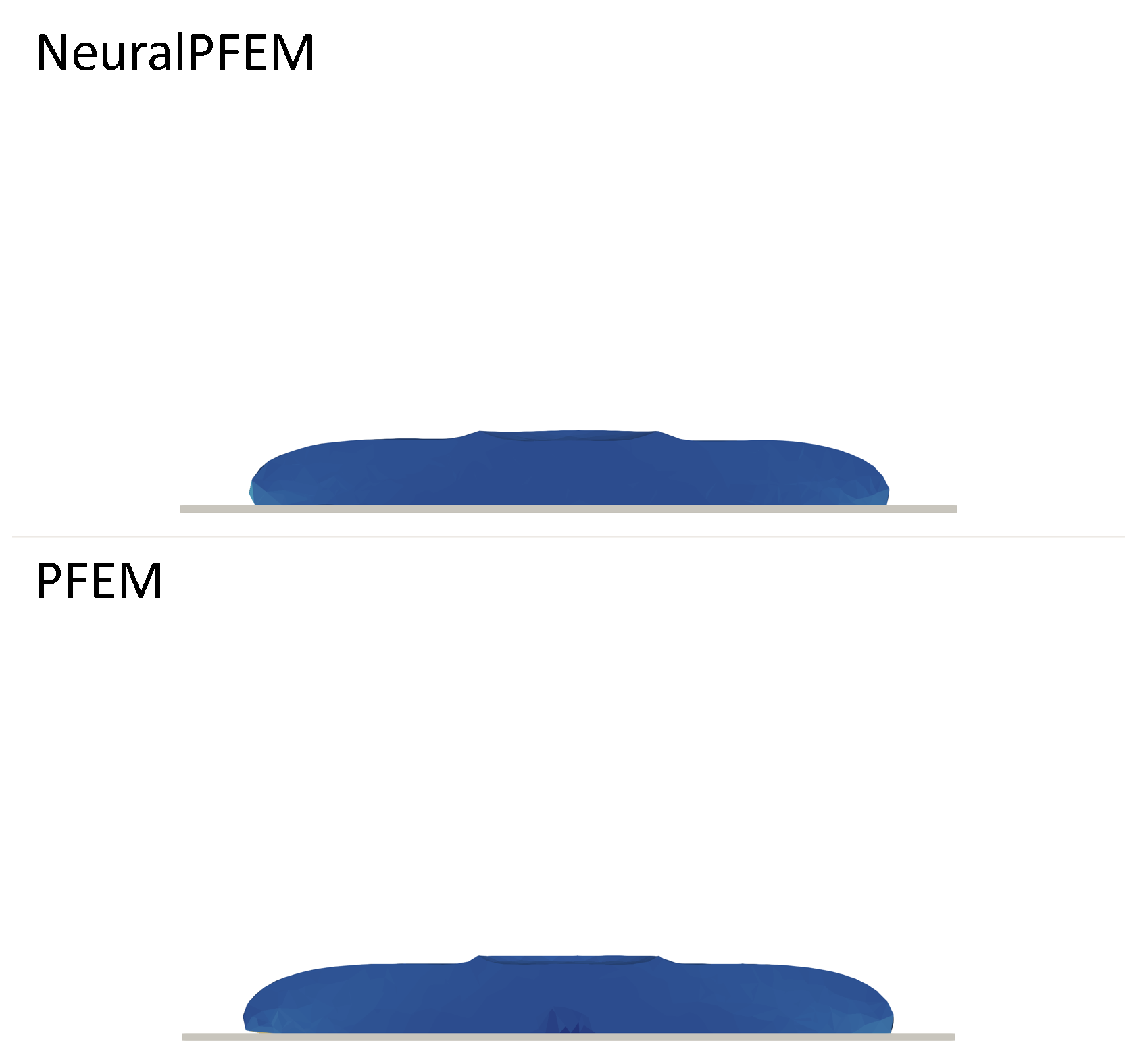}\label{fig:9c}}\\
    \subfloat[Case S2, $t=0.1 \, s$]{\includegraphics[width=0.33\linewidth]{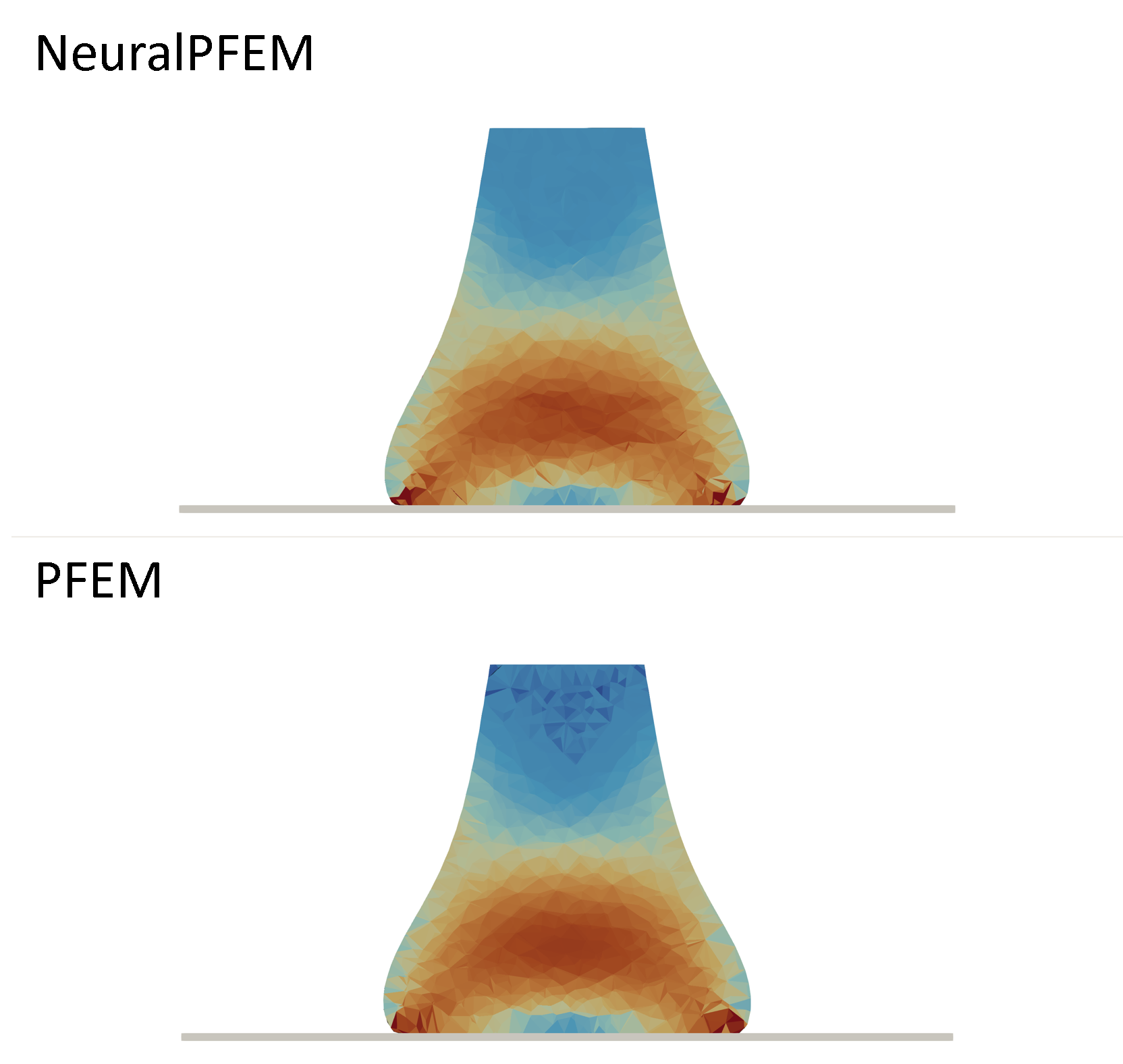}\label{fig:9d}}
    \subfloat[Case S2, $t=0.25 \, s$]{\includegraphics[width=0.33\linewidth]{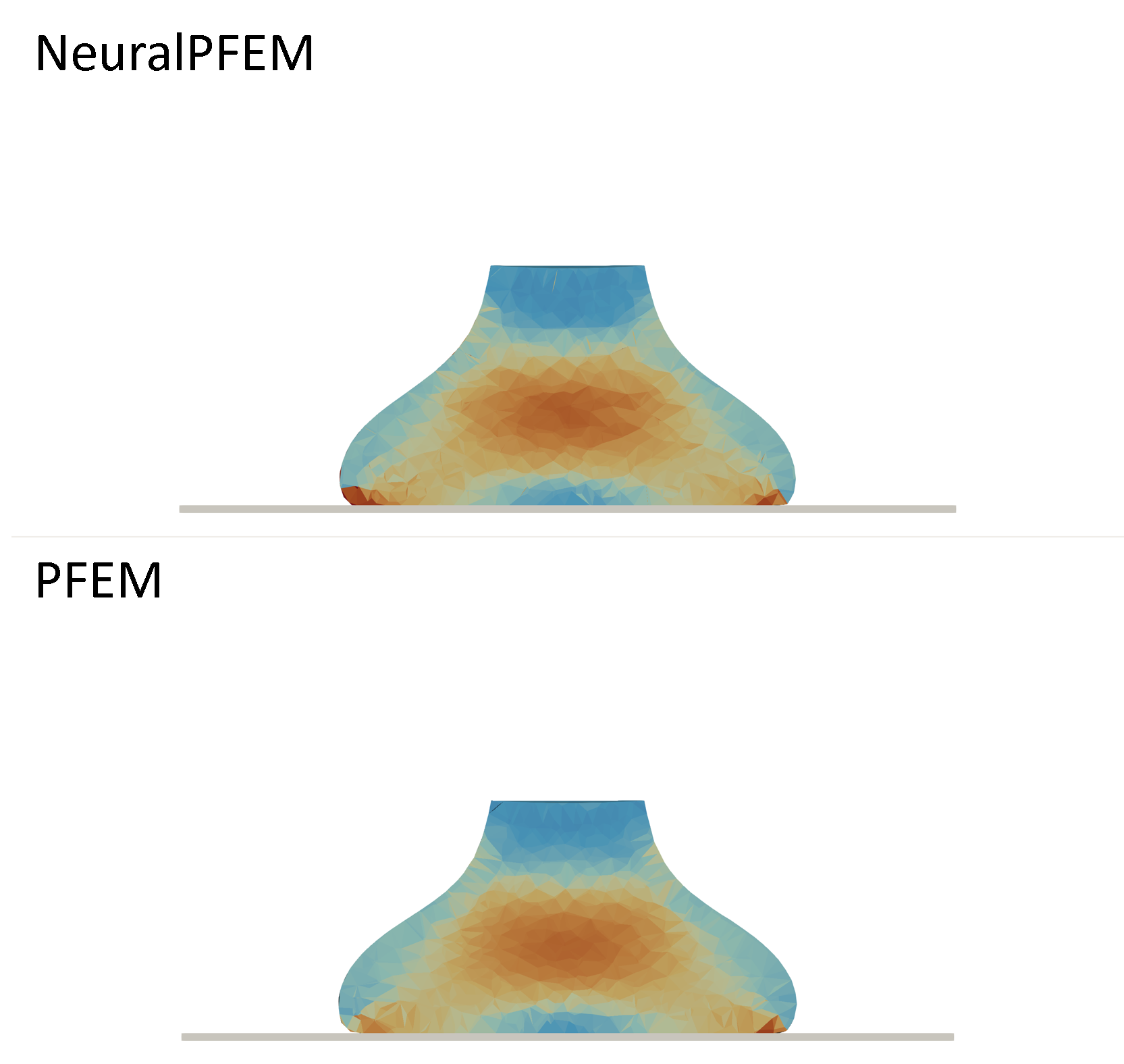}\label{fig:9e}}
    \subfloat[Case S2, $t=0.5 \, s$]{\includegraphics[width=0.33\linewidth]{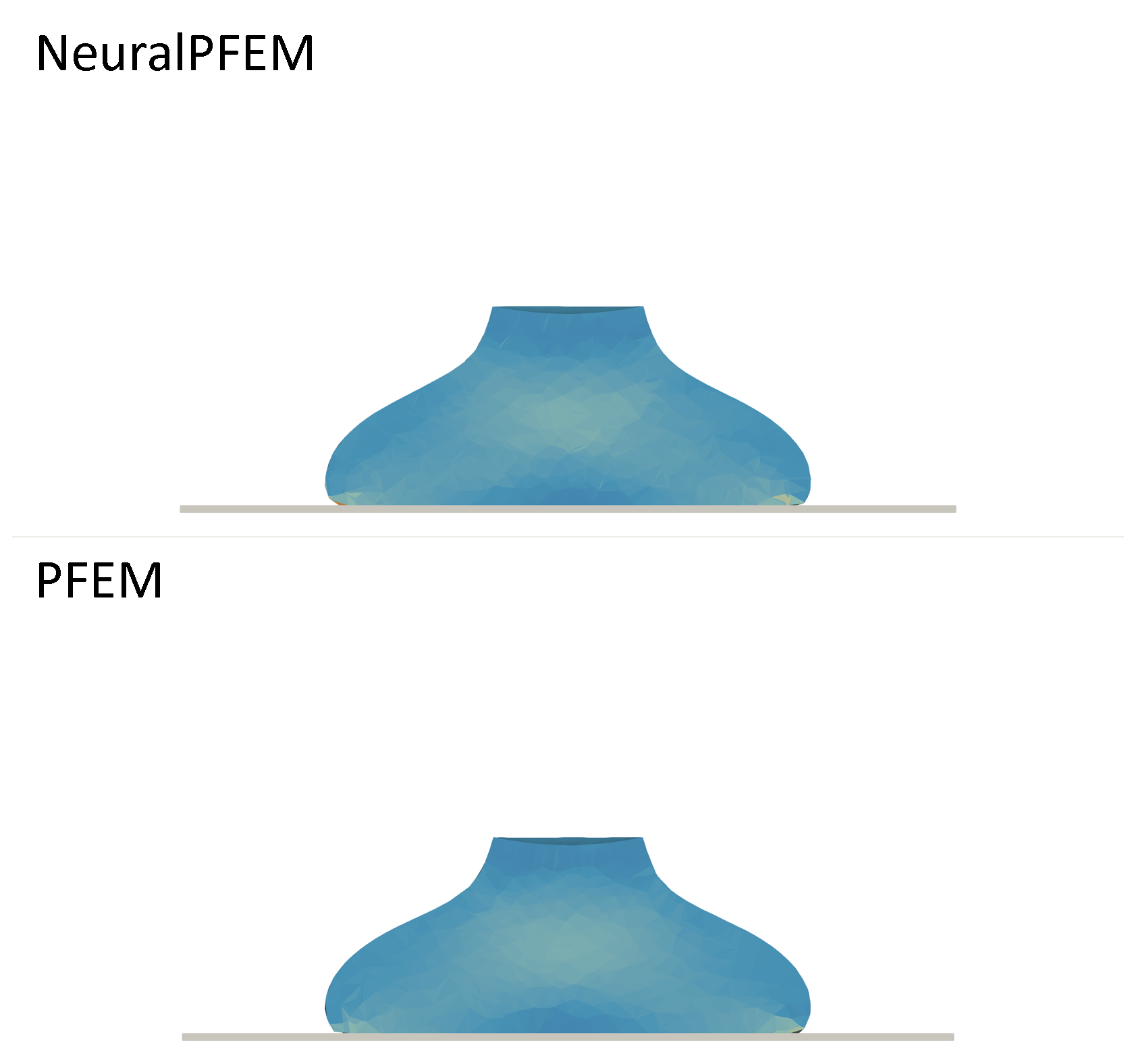}\label{fig:9f}}\\
    \subfloat[Case S3, $t=0.1 \, s$]{\includegraphics[width=0.33\linewidth]{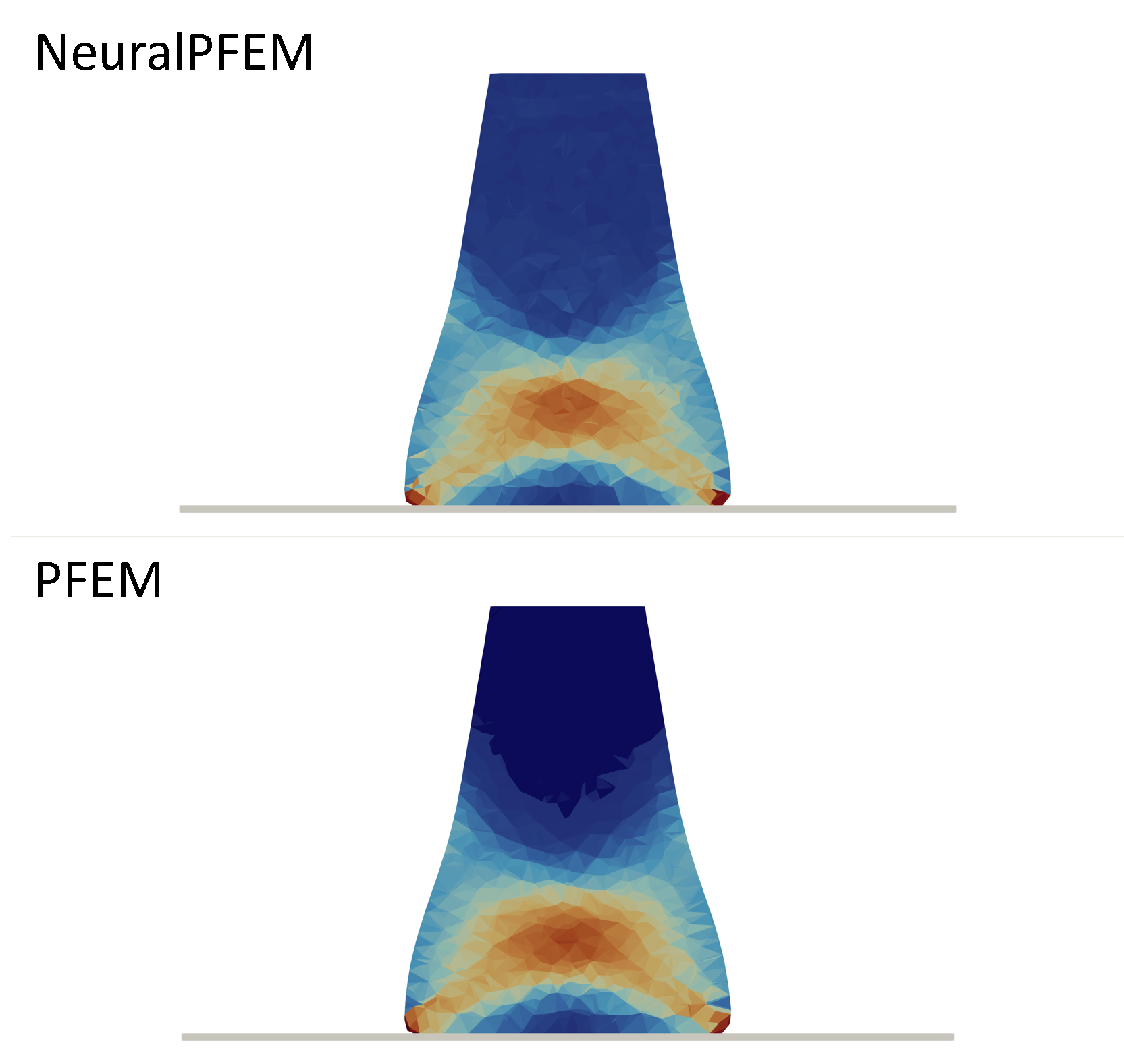}\label{fig:9g}}
    \subfloat[Case S3, $t=0.25 \, s$]{\includegraphics[width=0.33\linewidth]{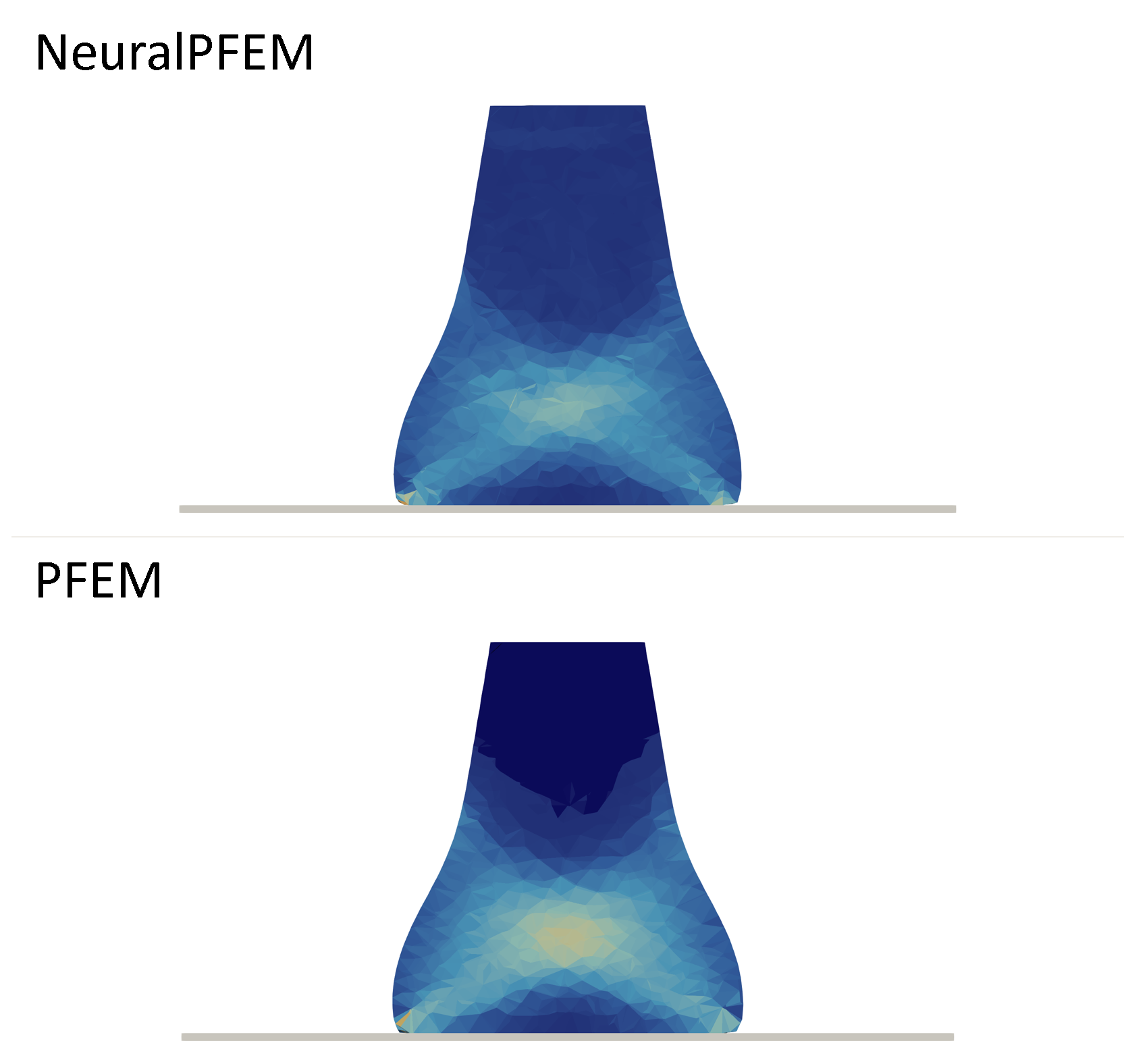}\label{fig:9h}}
    \subfloat[Case S3, $t=0.5 \, s$]{\includegraphics[width=0.33\linewidth]{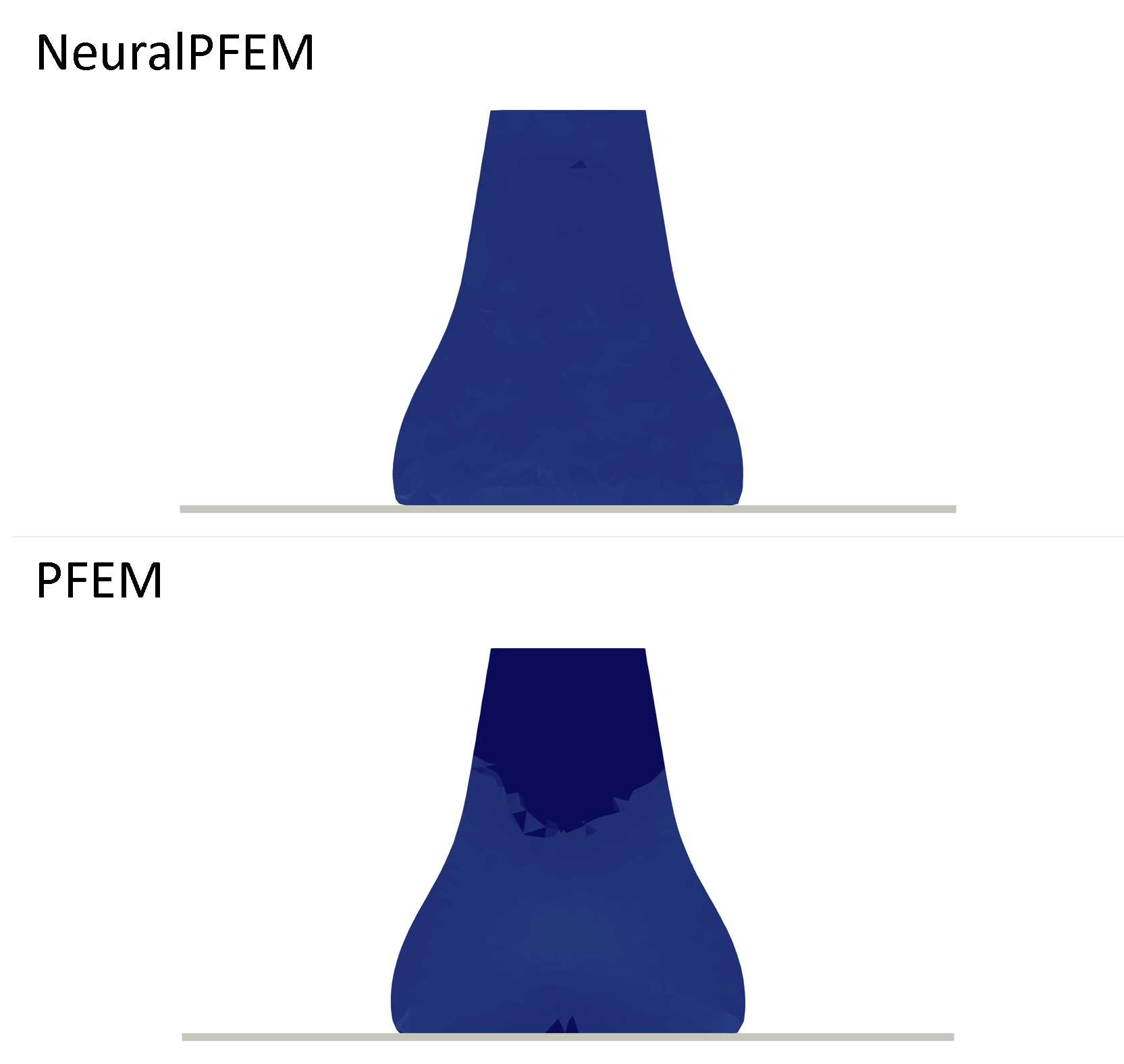}\label{fig:9i}}\\
    \caption{Temporal snapshots of NeuralPFEM with standard self-attention predictions (top row of each pair) and reference PFEM solutions (bottom row) for test cases S1 (\ref{fig:9a}-\ref{fig:9c}), S2 (\ref{fig:9d}-\ref{fig:9f}), and S3  (\ref{fig:9g}-\ref{fig:9i}) at selected time steps. The von Mises stress is visualised as a scalar field on the deformed mesh, recovered from the predicted nodal velocities and pressures through finite element shape functions and the Bingham constitutive relation.}
    \label{fig:9}
\end{figure}

As reported in Table ~\ref{tab:7}, linear and standard attention yield Chamfer distances of $8.3 \cdot 10^{-3}$ and $7.8 \cdot 10^{-3}$, respectively. Consistent with previous observations, the standard attention model achieves slightly better accuracy, although both remain below the characteristic mesh size adopted in the simulations. At this higher node count, the computational advantage of the linear attention formulation becomes apparent in terms of runtime, while memory usage remains comparable between the two approaches, as expected. Figure~\ref{fig:9} shows three temporal snapshots obtained with standard attention model for the three representative test cases S1, S2, and S3, whose material parameters are reported in Table~\ref{tab:6}. The three cases span a wide range of rheological behaviors: Case S1, characterized by a relatively low yield stress, exhibits pronounced spreading and reaches large final diameter. Case S2, with higher viscosity and yield stress, exhibits slower and more constrained spreading. Case S3 has the highest yield stress among the three, which arrests the flow at an early stage and produces the smallest slump diameter. Across all cases, NeuralPFEM accurately reproduces both the transient spreading dynamics and the final equilibrium configuration, capturing the correct three-dimensional free-surface geometry throughout the simulation. Importantly, this agreement extends to the reconstructed stress fields. The model is able to recover physically consistent stress distributions. In particular, high-stress zones are concentrated near the base and along regions of significant deformation, while stress magnitudes decrease as the material approaches arrest. This reflects the model’s ability to reproduce the yield behavior intrinsic to Bingham fluids and demonstrates that NeuralPFEM not only predicts geometrically accurate flow evolution but also preserves the underlying mechanical structure of the problem, validating the quality of the learned velocity gradients. This result highlights one of the main advantages of retaining the PFEM discretization within the surrogate framework: derived mechanical quantities can be reconstructed directly from the predicted fields using standard finite-element operators.

\begin{table}[h]
\centering
\caption{Material parameters of the test cases shown in Figure \ref{fig:9}.}
\label{tab:6}
\begin{tabular}{lccc}
\toprule
\textbf{ID} & $\rho$ (kg/m$^3$) & $\mu$ (Pa$\cdot$s) & $\tau_0$ (Pa) \\
\midrule
S1 & 2469 & 17 & 238  \\
S2 & 2486 & 67 & 468 \\
S3 & 2292 & 24 & 1416 \\
\bottomrule
\end{tabular}
\end{table}

\begin{table}[h]
\centering
\caption{Comparison of error, memory footprint, and computational cost for standard self-attention and linear self-attention variants of NeuralPFEM on the 3D Bingham slump.}
\label{tab:7}
\begin{tabular}{l ccc}
\toprule
\textbf{Model} & \textbf{Mean Chamfer distance [m]} & \textbf{Peak memory [MB]} & \textbf{Time / step [s]}  \\
\midrule
Standard attention & $8.3 \cdot 10^{-3}$ & 745 & 0.09 \\
Linear attention & $7.8 \cdot 10^{-3}$ & 891 & 0.06 \\
\bottomrule
\end{tabular}
\end{table}

\clearpage

\subsection{3D casting test}

\begin{figure}[h]
    \centering
    \subfloat[\label{fig:10a}]{\includegraphics[width=0.33\linewidth]{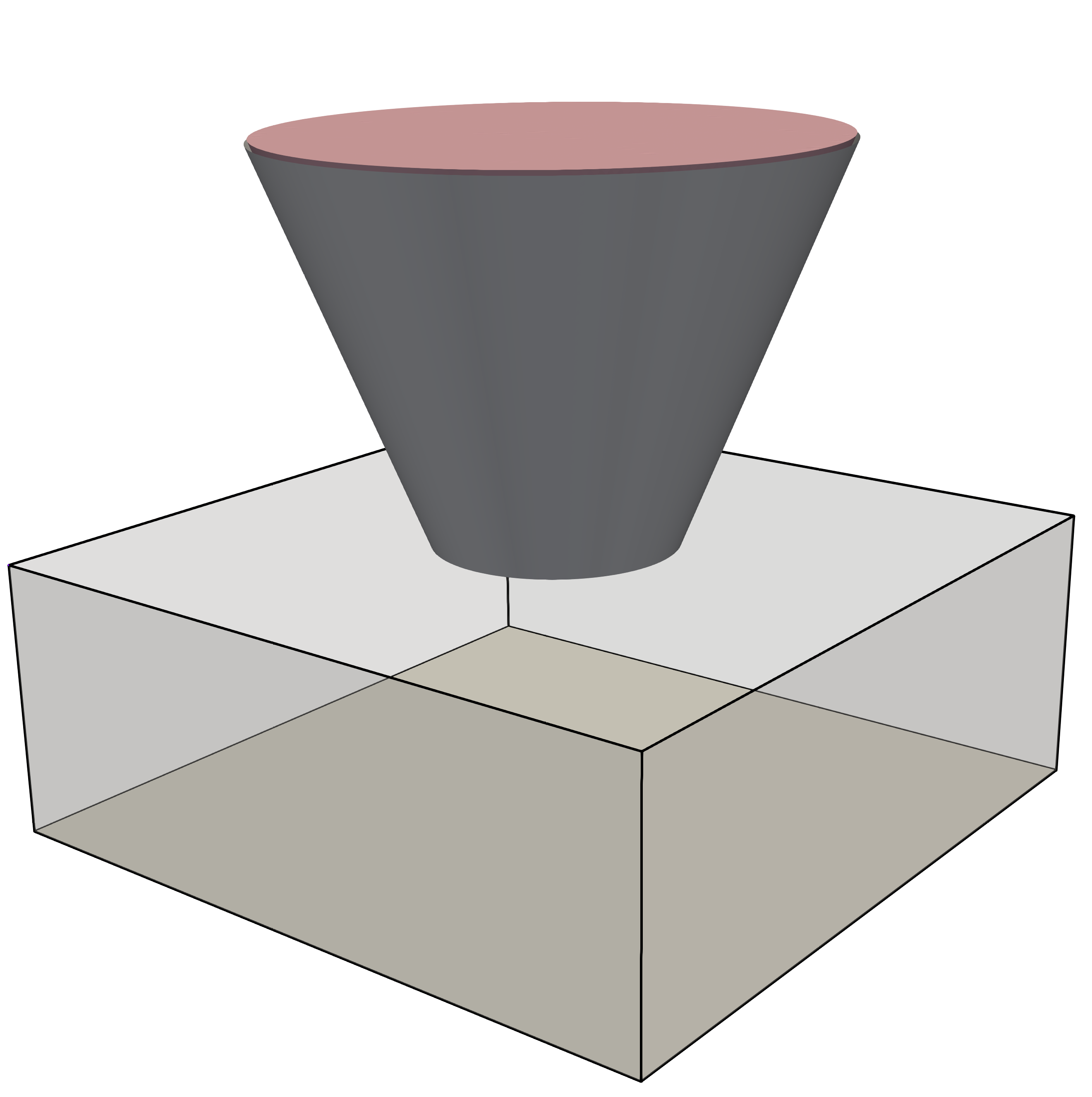}} 
    \subfloat[\label{fig:10b}]{\includegraphics[width=0.33\linewidth]{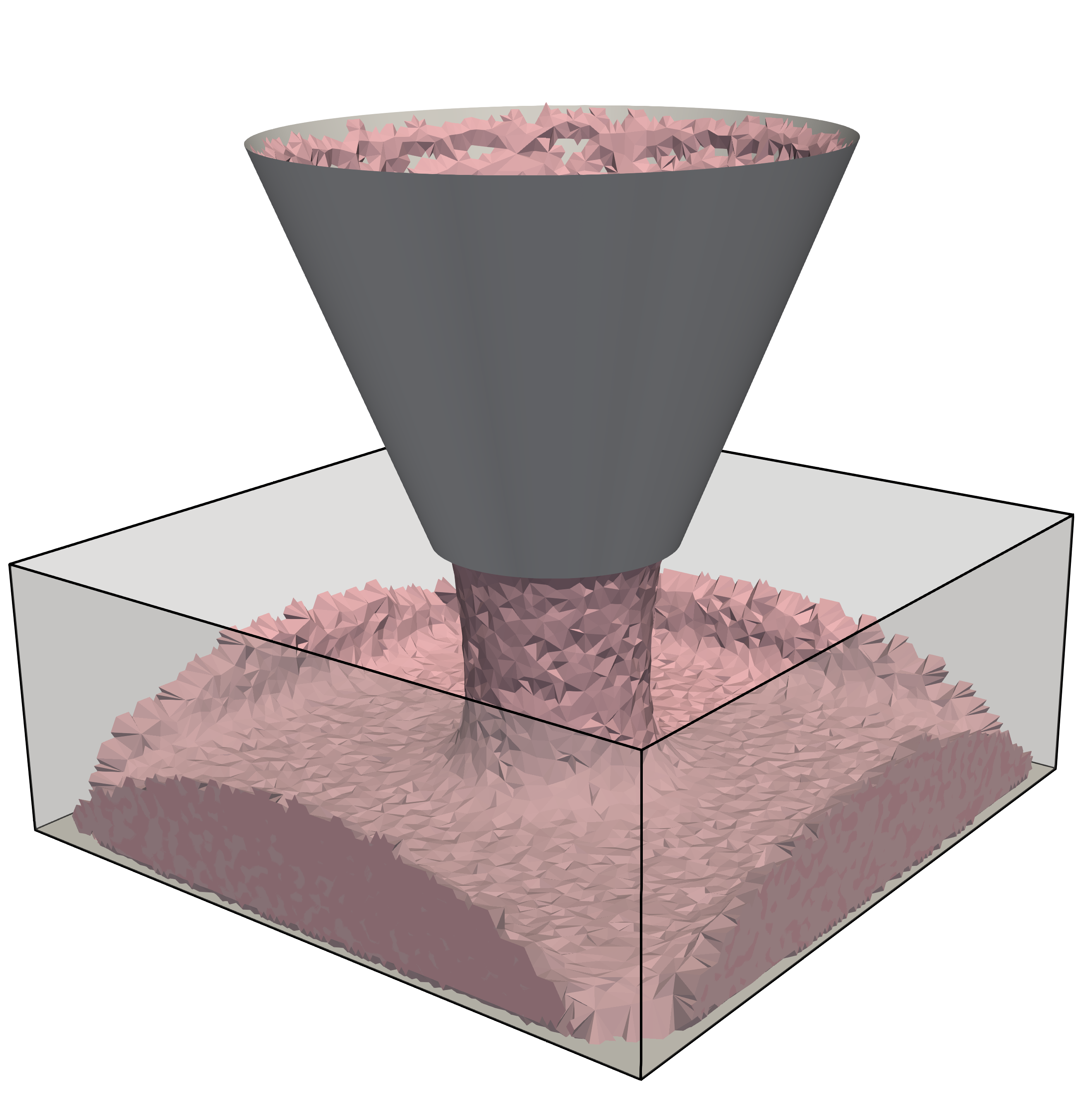}}
    \subfloat[\label{fig:10c}]{\includegraphics[width=0.33\linewidth]{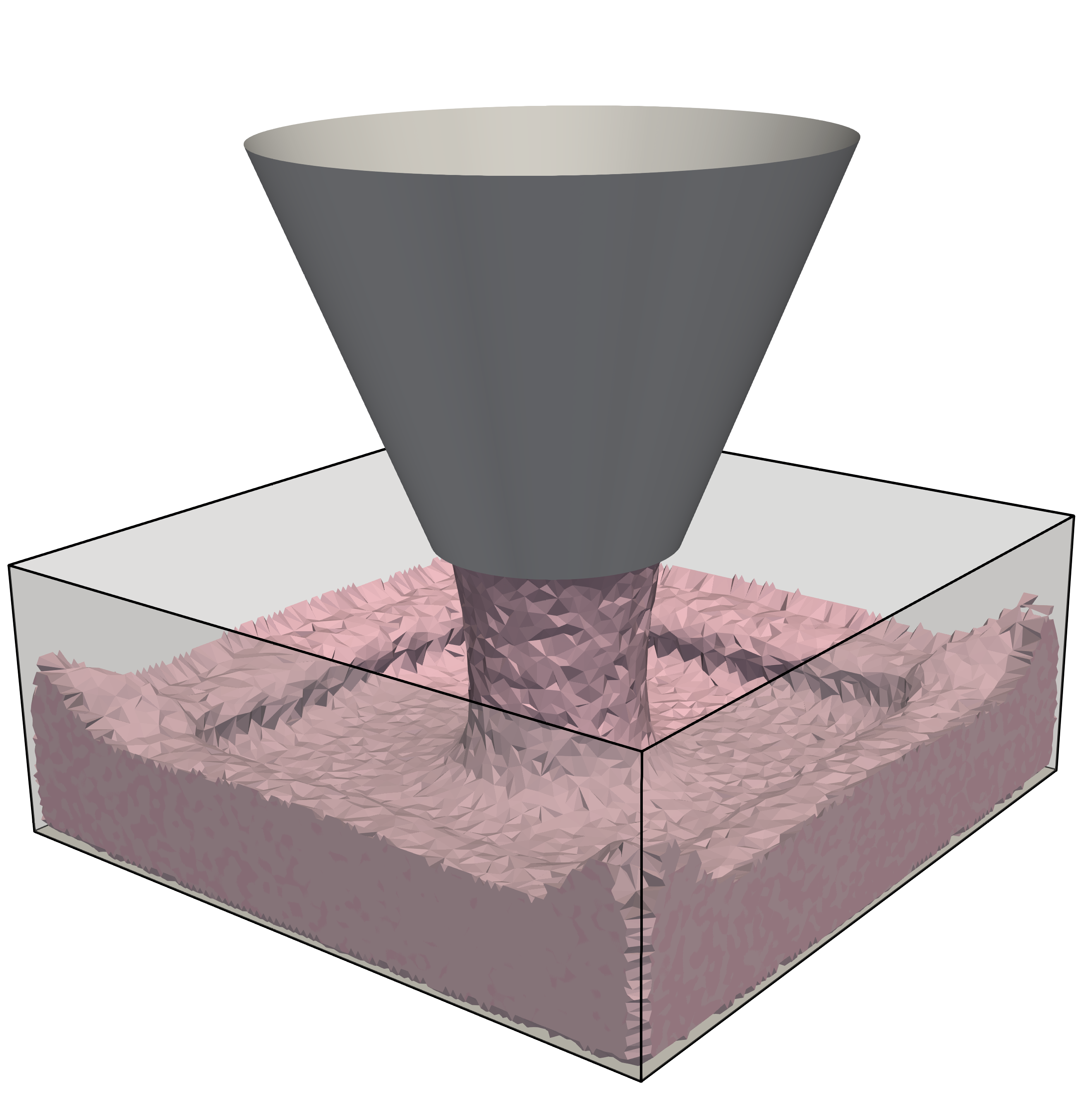}}
    \caption{3D casting experiment: the fluid, initially suspended above the container, is released and flows under gravity through the bottom opening, progressively filling the box.}
    \label{fig:10}
\end{figure}

As a final test case, we consider the casting of a Newtonian fluid into a rigid parallelepipedal container. The initial configuration consists of a truncated cone of fluid suspended above the box (Figure~\ref{fig:10a}). At the initial time, the constraint at the bottom base is removed, allowing the fluid to flow freely under gravity through the resulting opening. Following an initial free-fall phase, the fluid first impacts the bottom surface and subsequently the lateral walls, progressively filling the container (Figures~\ref{fig:10b}--\ref{fig:10c}). Compared to the two previous benchmarks, this problem exhibits substantially more complex dynamics, with the fluid undergoing large free-surface deformation as it impacts the box walls. Moreover, in this test the domain is discretized with approximately $N = 10^5$ nodes, deliberately chosen to test the scalability of the proposed architectures and to probe the practical limits of the standard softmax attention formulation. The simulation parameters are summarized in Table~\ref{tab:8}.

Table~\ref{tab:9} reports the memory consumption and computational cost of the two attention variants. As shown in the first column, the FlashAttention implementation is highly effective at limiting memory usage. However, it does not reduce the computational complexity, which remains $\mathcal{O}(N^2)$ in terms of floating-point operations. Consequently, at this scale in the number of nodes, standard attention incurs a prohibitive computational cost. Linear attention, on the other hand, reduces the computational complexity to $\mathcal{O}(N d^2)$, resulting in a substantial decrease in training time per step. Interestingly, it exhibits a higher memory consumption compared to standard attention, which is not expected from a theoretical standpoint. This discrepancy is likely attributable to the efficiency of the FlashAttention implementation, which is highly optimized, in contrast to our more naive implementation of linear attention. All results presented in this section were obtained with the linear attention variant.

Due to the high computational cost of three-dimensional PFEM simulations at this resolution, and since the test was performed mainly to assess the computational scalability of the neural surrogate, the dataset size was intentionally kept limited: 30 simulations were generated, varying only the fluid viscosity $\mu$ into a limited range. As a result, the evaluation is qualitative, assessing whether NeuralPFEM with linear attention is capable of reproducing the correct flow morphology, free-surface dynamics, and final filling configuration across the range of viscosities considered.

\begin{figure}
    \centering
    \subfloat[Case C1, $t=0.2 \, s$]{\includegraphics[width=0.5\linewidth]{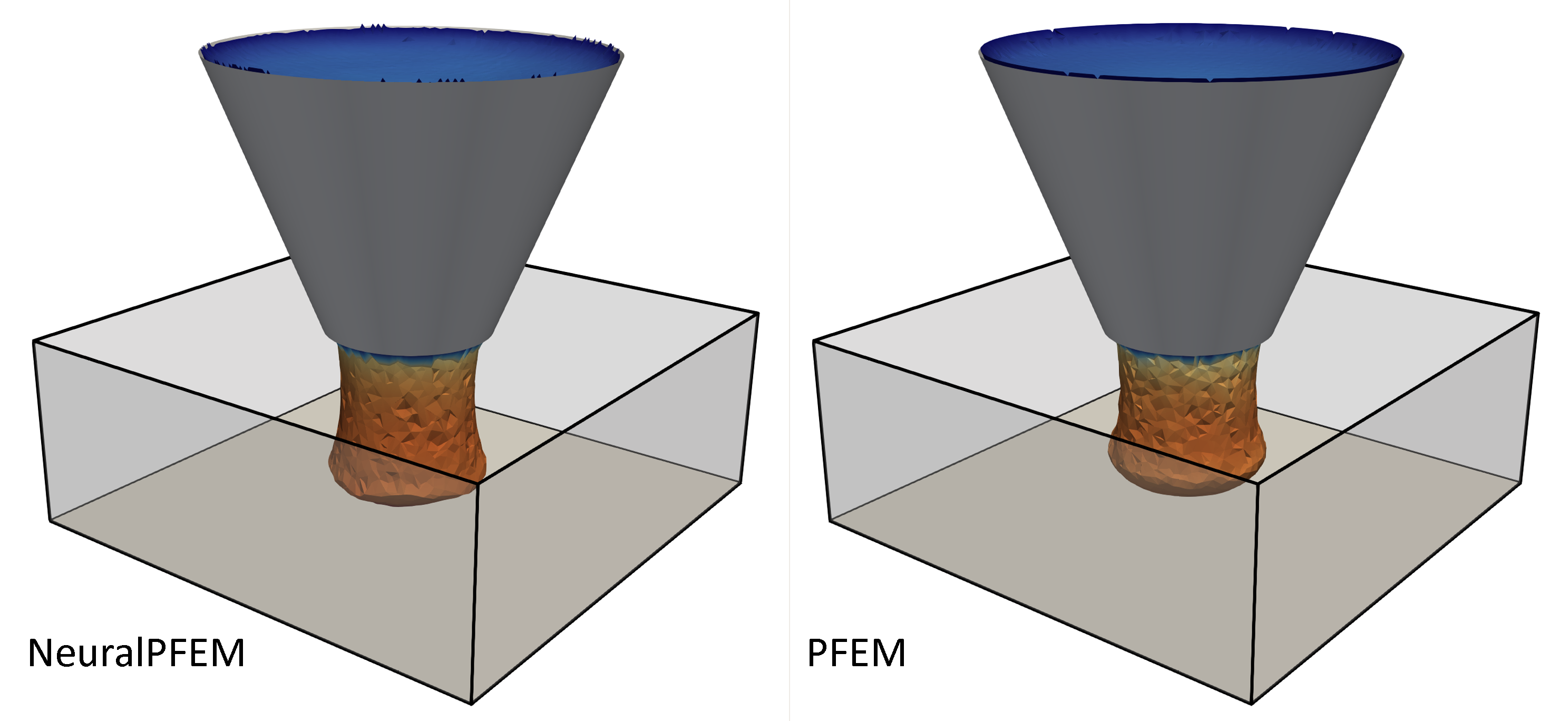}\label{fig:11a}}
    \subfloat[Case C1, $t=0.5 \, s$]{\includegraphics[width=0.5\linewidth]{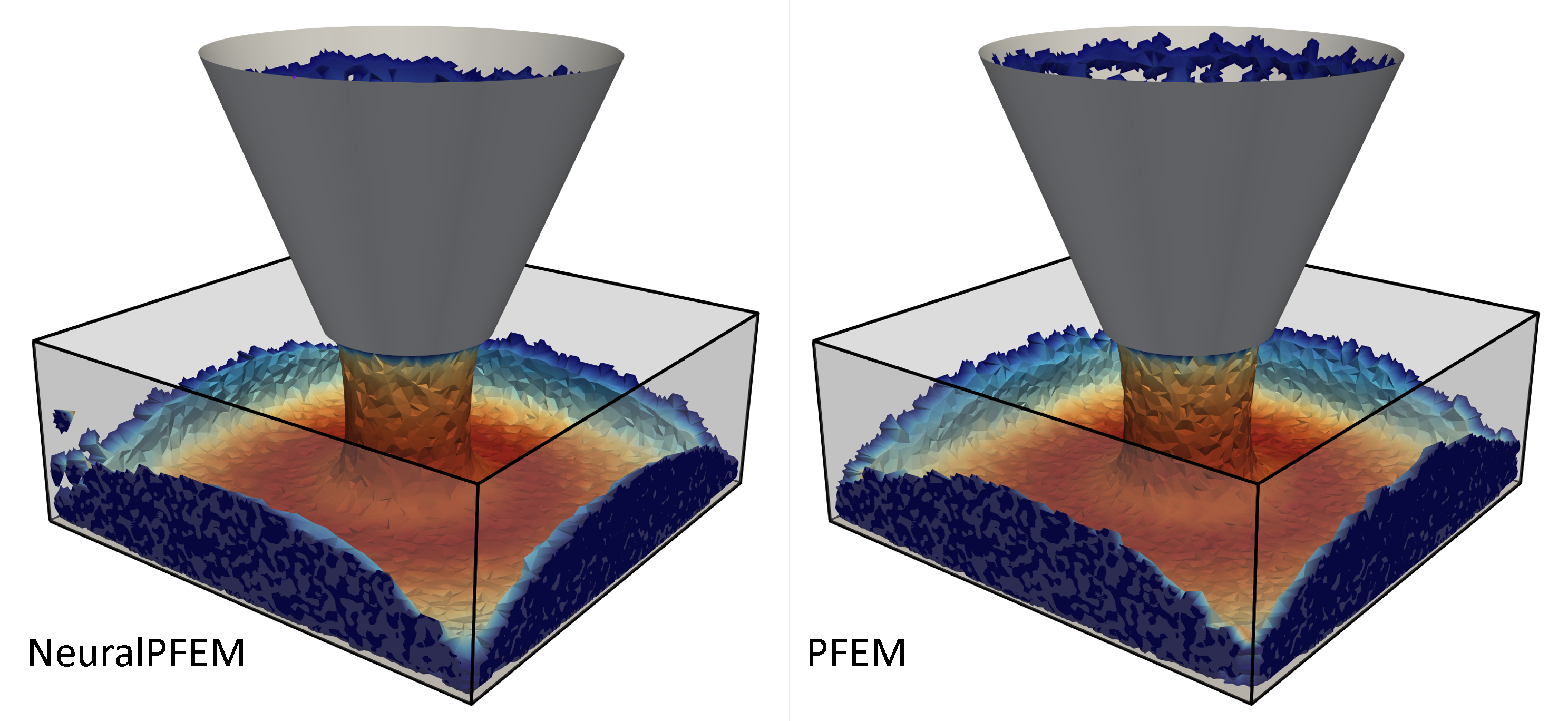}\label{fig:11b}}\\
    \subfloat[Case C1, $t=0.8 \, s$]{\includegraphics[width=0.5\linewidth]{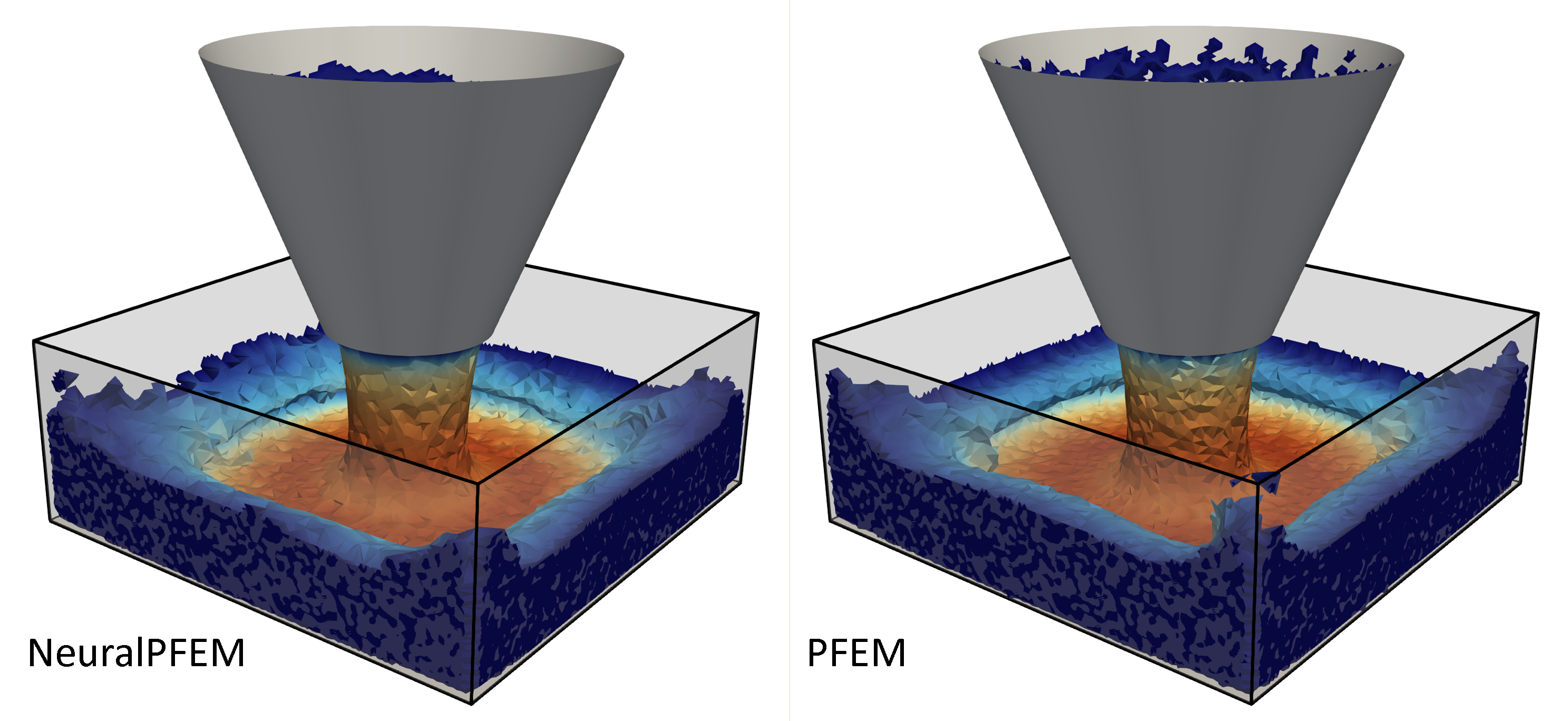}\label{fig:11c}}
    \subfloat[Case C1, $t=1.5 \, s$]{\includegraphics[width=0.5\linewidth]{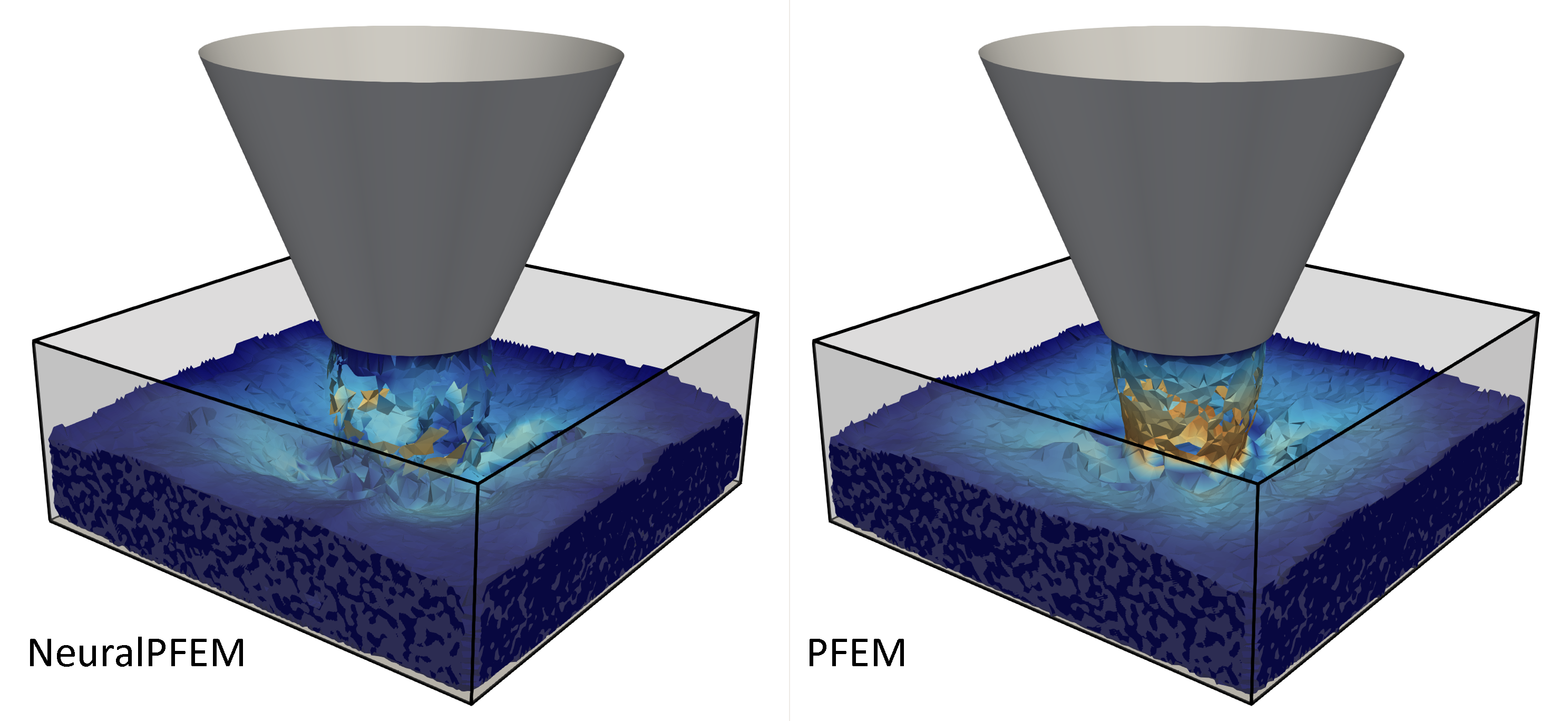}\label{fig:11d}}\\
    \subfloat[Case C2, $t=0.2 \, s$]{\includegraphics[width=0.5\linewidth]{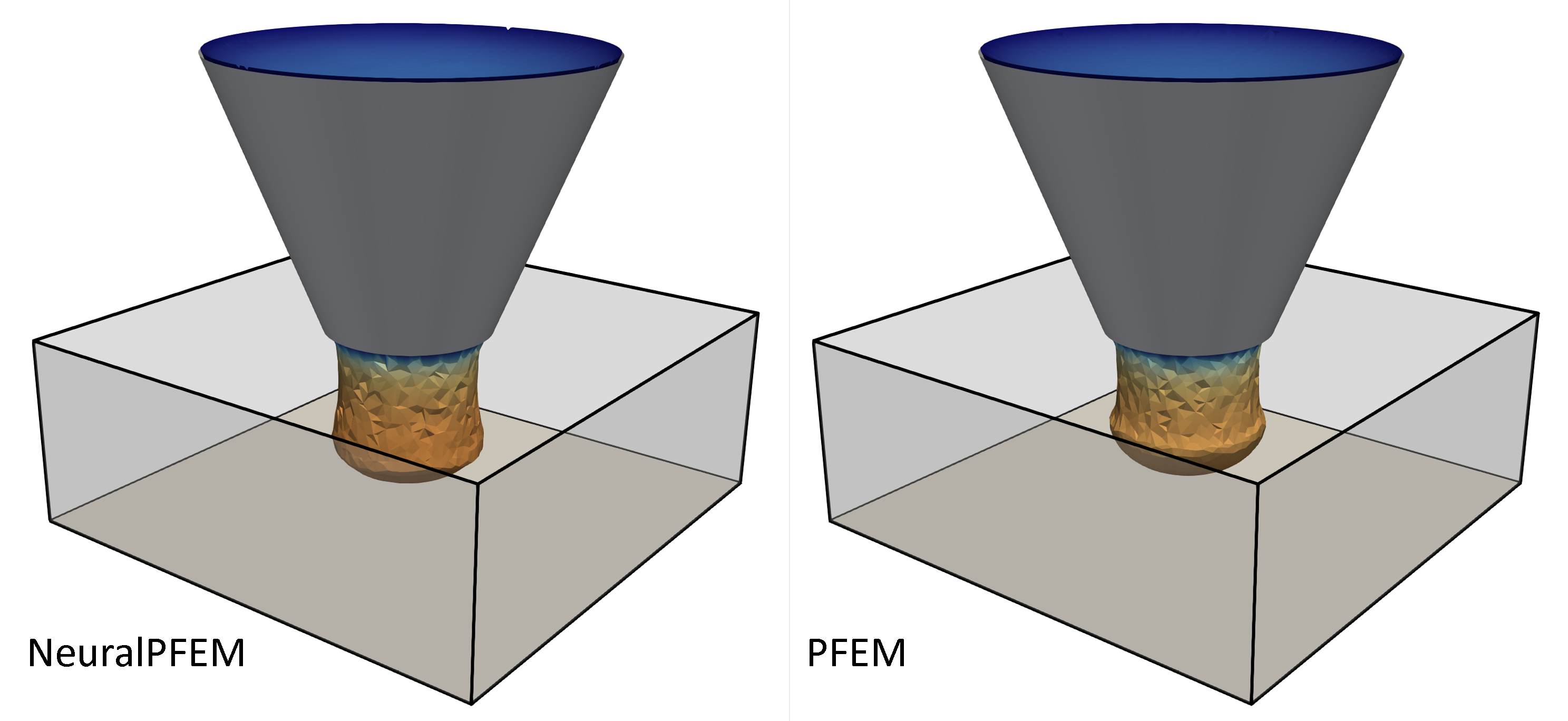}\label{fig:11e}}
    \subfloat[Case C2, $t=0.5 \, s$]{\includegraphics[width=0.5\linewidth]{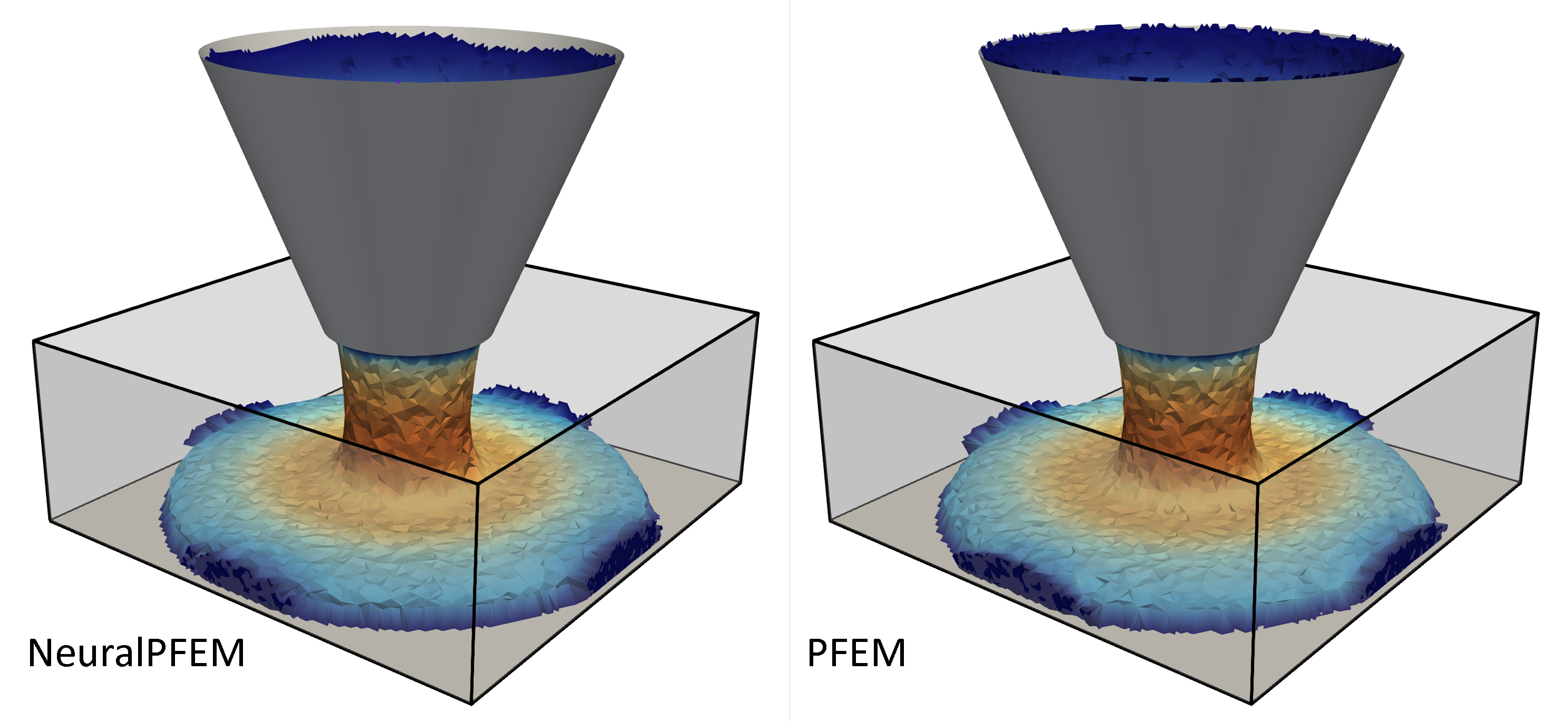}\label{fig:11f}}\\
    \subfloat[Case C2, $t=0.8 \, s$]{\includegraphics[width=0.5\linewidth]{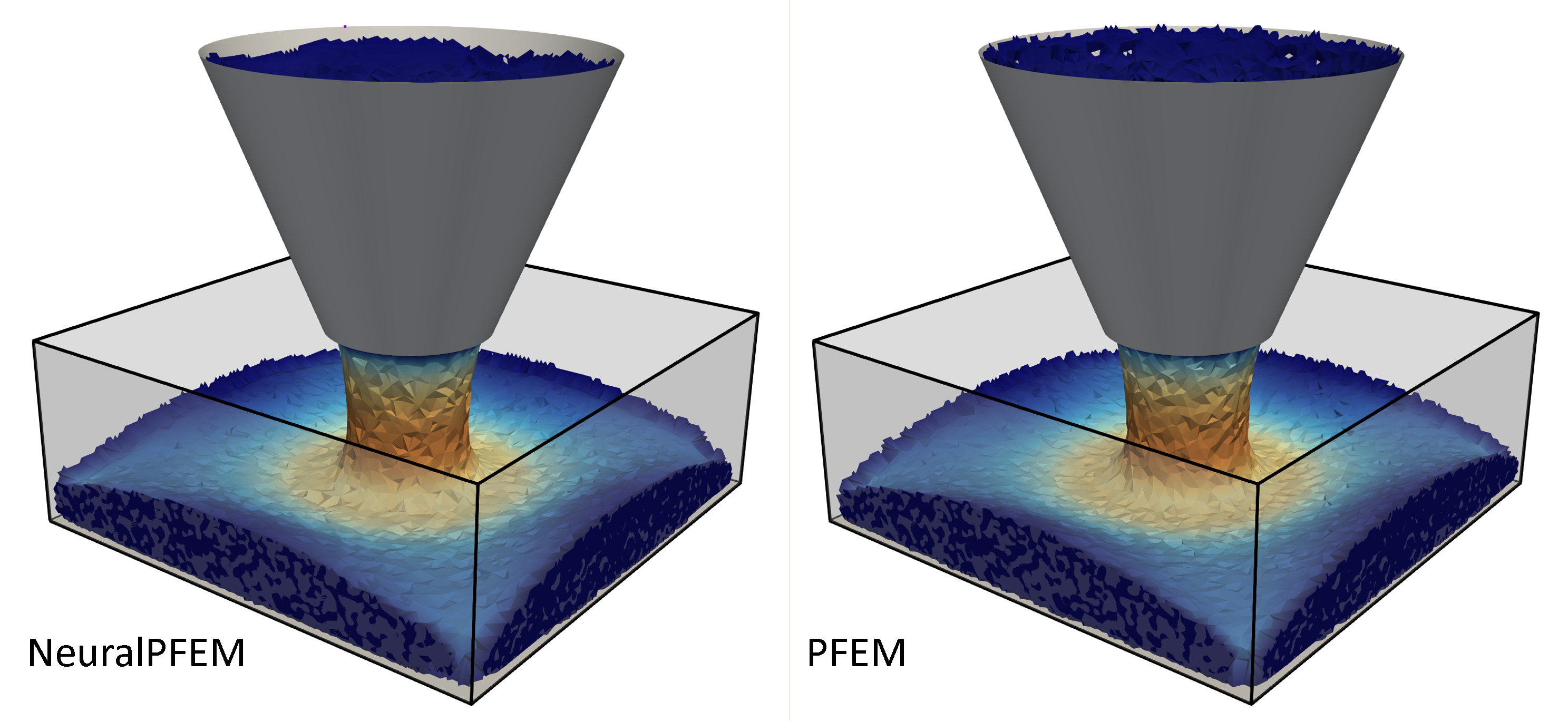}\label{fig:11g}}
    \subfloat[Case C2, $t=1.5 \, s$]{\includegraphics[width=0.5\linewidth]{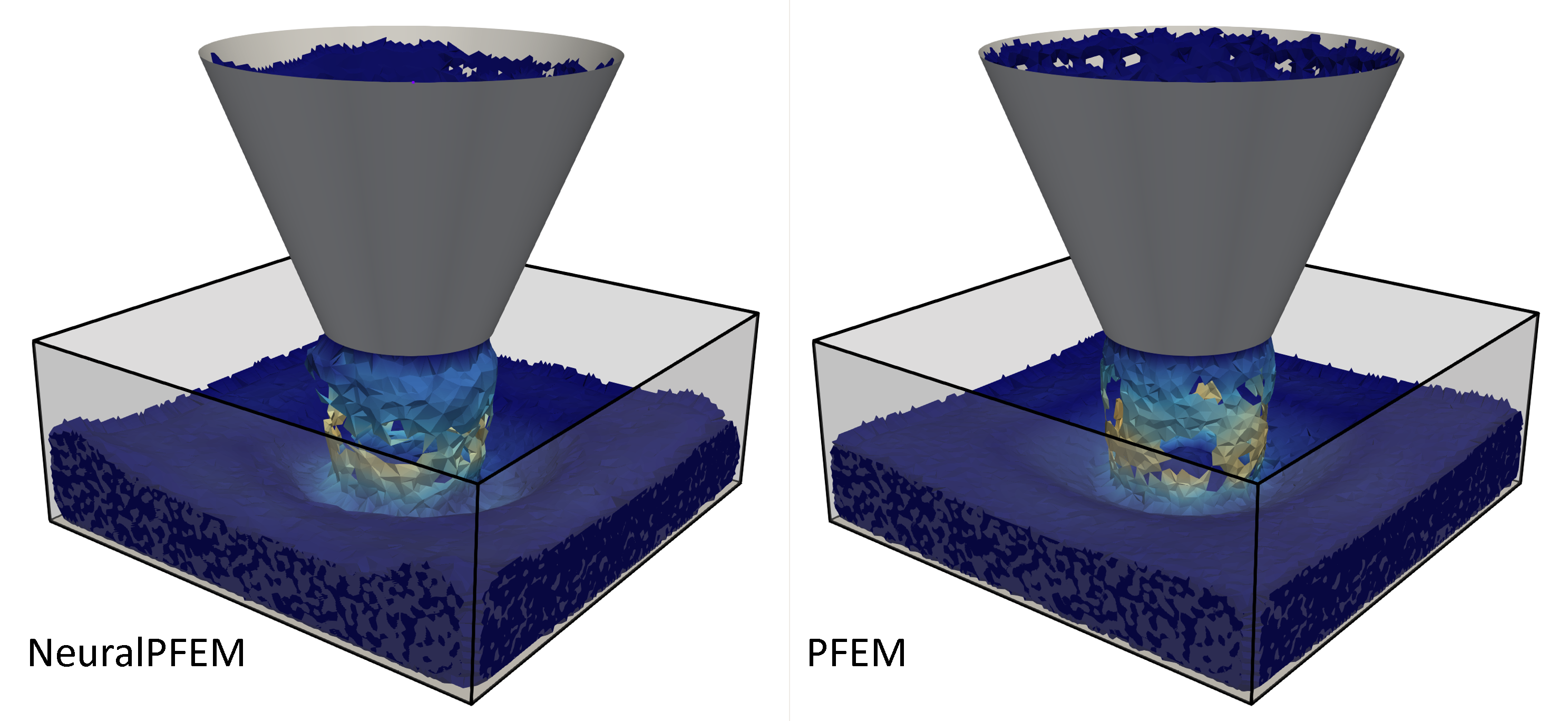}\label{fig:11h}}\\
    \caption{Temporal snapshots of NeuralPFEM with linear self-attention predictions and reference PFEM solutions for test cases C1 ($\mu = 10 \, \mathrm{Pa \cdot s}$, \ref{fig:11a}-\ref{fig:11d}) and C2 ($\mu = 60 \, \mathrm{Pa \cdot s}, \ref{fig:11e}-\ref{fig:11h}$) at selected time steps. The velocity magnitude is visualised as a scalar field on the deformed mesh.}
    \label{fig:11}
\end{figure}

Figure~\ref{fig:11} presents representative snapshots of the predicted and reference PFEM solutions at selected time steps for two test cases, C1 and C2, corresponding to viscosity values of $\mu = 10 \, \mathrm{Pa \cdot s}$ and $\mu = 60 \, \mathrm{Pa \cdot s}$, respectively. These cases display clearly distinct flow regimes. In the low-viscosity case, the fluid undergoes a rapid free fall followed by an energetic spreading after impact with the bottom surface, with pronounced inertial effects producing sharper free-surface deformations and more complex interactions with the container walls (Figures~\ref{fig:11a}--\ref{fig:11d}). Conversely, the higher-viscosity case exhibits a slower and more damped evolution, characterized by reduced spreading, smoother interfaces, and a more gradual filling process dominated by viscous dissipation (Figures~\ref{fig:11e}--\ref{fig:11h}). Despite these substantial differences in flow behavior, the surrogate model accurately reproduces the dynamics in both regimes. Close agreement with PFEM reference solutions across all time steps highlights the ability of the model to generalize effectively to previously unseen viscosity values within the training range, consistently capturing both inertia-dominated and viscosity-dominated flow features.

This benchmark demonstrates that NeuralPFEM with linear attention is capable of scaling to large, complex three-dimensional problems where the computational cost of standard attention becomes prohibitive. Linear attention therefore provides a practical solution for large-scale surrogate modeling within the NeuralPFEM framework. 

\begin{table}
\centering
\caption{Summary of the 3D casting test case parameters.}
\label{tab:8}
\begin{tabular}{lcc}
\toprule
\textbf{} & \textbf{Symbol and unit} & \textbf{Value} \\
\midrule
\multicolumn{3}{l}{\textbf{Fixed parameters}} \\
\midrule
Base diameter & $a\,[\mathrm{m}]$ & $0.55$ \\
Top diameter & $b\,[\mathrm{m}]$ & $1.2$ \\
Height & $c\,[\mathrm{m}]$ & $0.85$ \\
Density & $\rho\,[\mathrm{kg/m^3}]$ & $ 2000 $ \\
\midrule
\multicolumn{3}{l}{\textbf{Variable parameters}} \\
\midrule
Viscosity & $\mu\,[\mathrm{Pa \cdot s}]$ & $[5,\,70]$ \\
\midrule
\multicolumn{3}{l}{\textbf{Numerical parameters}} \\
\midrule
Mesh size & $m\,[\mathrm{m}]$ & $1.8 \cdot 10^{-2}$ \\
Number of nodes & $N[-]$ & $93618$ \\
Time step & $\Delta t\,[\mathrm{s}]$ & $0.005$ \\
\bottomrule
\end{tabular}
\end{table}

\begin{table}
\centering
\caption{Comparison between standard and linear attention in memory and computational time for 1 step training in the 3D casting example.}
\label{tab:9}
\begin{tabular}{l cc}
\toprule
\textbf{Model} & \textbf{Peak memory [MB]} & \textbf{Time / step [s]} \\
\midrule
Standard attention & 2876 & 1.06\\
Linear attention & 4522 & 0.16\\
\bottomrule
\end{tabular}
\end{table}

\clearpage    
\clearpage

\section{Conclusions}\label{sec:conclusions}
In this work, we introduced a novel self-attention-based architecture for mesh-based Lagrangian fluid simulations within the NeuralPFEM framework. To overcome the memory bottlenecks of message-passing graph neural networks, where the explicit storage and updating of edge embeddings become prohibitive in large-scale three-dimensional scenarios, we adopted a framework that learns spatial correlations adaptively, without imposing predefined structures. We demonstrated that self-attention provides a principled and memory-efficient alternative to explicit message passing, enabling domain-wide information exchange without relying on graph edges.

Two variants of the attention mechanism were developed and evaluated. Standard softmax attention achieves full expressiveness through learned, data-driven interaction weights, and its memory footprint is kept tractable via the FlashAttention implementation. To further enhance scalability, a linear attention variant was also investigated. By discarding the softmax operation and exploiting the associativity of matrix multiplication, this formulation avoids the explicit construction of the pairwise interaction matrix, significantly reducing both memory and computational complexity. This enables training on large-scale datasets where standard attention would otherwise be computationally intractable.

The proposed architectures were validated on free-surface flow benchmarks involving non-Newtonian fluids. By training and testing on datasets with varying material parameters, we demonstrated that the models accurately capture a wide diversity of rheological behaviors. The attention-based formulations match the predictive accuracy of the message-passing baseline while delivering substantial reductions in memory consumption. Furthermore, by removing explicit geometric edge features, the model generalizes effectively across mesh resolutions not encountered during training. Crucially, by retaining the underlying mesh topology, the framework preserves the ability to reconstruct derived mechanical quantities, such as stress fields, from the predicted kinematics using standard finite element shape functions.

In particular, the linear attention formulation demonstrates the capacity to scale to complex, three-dimensional problems, providing a highly practical solution for large-scale surrogate modeling. However, since linear attention remains less expressive than full softmax attention exploring architectures that bridge this gap represents a natural direction for future research. Recent approaches such as Transolver~\cite{zhou2026} and UPT~\cite{alkin2025a} have shown that it is possible to efficiently apply standard attention to problems with millions of degrees of freedom. These methods rely on reducing the effective sequence length processed by the attention mechanism while preserving global interaction information, thereby retaining the expressive power of softmax attention without incurring prohibitive computational costs. Another particularly relevant direction is represented by hybrid architectures such as GAOT \cite{wen2026}, which combine graph neural networks (GNNs) with attention mechanisms. In this approach, the GNN is first used to encode the unstructured mesh into a lower-dimensional latent representation, effectively reducing the number of tokens on which attention must be applied. Attention is then performed in this compressed space, enabling significantly improved scalability. This strategy is especially well suited for NeuralPFEM, where the data is naturally defined on an unstructured mesh, making it straightforward to leverage GNN-based encodings.

\section*{Declarations}
The authors declare no potential conflict of interest.

\bibliographystyle{elsarticle-num.bst}
\bibliography{Bibliography}  

@misc{adams2025,
  title = {{{GeoTransolver}}: {{Learning Physics}} on {{Irregular Domains Using Multi-scale Geometry Aware Physics Attention Transformer}}},
  shorttitle = {{{GeoTransolver}}},
  author = {Adams, Corey and Ranade, Rishikesh and Cherukuri, Ram and Choudhry, Sanjay},
  year = 2025,
  month = dec,
  number = {arXiv:2512.20399},
  eprint = {2512.20399},
  primaryclass = {cs},
  publisher = {arXiv},
  doi = {10.48550/arXiv.2512.20399},
  urldate = {2026-02-10},
  archiveprefix = {arXiv}
}

@misc{alkin2025,
  title = {{{AB-UPT}}: {{Scaling Neural CFD Surrogates}} for {{High-Fidelity Automotive Aerodynamics Simulations}} via {{Anchored-Branched Universal Physics Transformers}}},
  shorttitle = {{{AB-UPT}}},
  author = {Alkin, Benedikt and Bleeker, Maurits and Kurle, Richard and Kronlachner, Tobias and Sonnleitner, Reinhard and Dorfer, Matthias and Brandstetter, Johannes},
  year = 2025,
  month = oct,
  number = {arXiv:2502.09692},
  eprint = {2502.09692},
  primaryclass = {cs},
  publisher = {arXiv},
  doi = {10.48550/arXiv.2502.09692},
  urldate = {2026-02-10},
  archiveprefix = {arXiv}
}

@misc{alkin2025a,
  title = {{{NeuralDEM}} -- {{Real-time Simulation}} of {{Industrial Particulate Flows}}},
  author = {Alkin, Benedikt and Kronlachner, Tobias and Papa, Samuele and Pirker, Stefan and Lichtenegger, Thomas and Brandstetter, Johannes},
  year = 2025,
  month = feb,
  number = {arXiv:2411.09678},
  eprint = {2411.09678},
  primaryclass = {cs},
  publisher = {arXiv},
  doi = {10.48550/arXiv.2411.09678},
  urldate = {2025-03-06},
  archiveprefix = {arXiv}
}

@misc{battaglia2018,
  title = {Relational Inductive Biases, Deep Learning, and Graph Networks},
  author = {Battaglia, Peter W. and Hamrick, Jessica B. and Bapst, Victor and {Sanchez-Gonzalez}, Alvaro and Zambaldi, Vinicius and Malinowski, Mateusz and Tacchetti, Andrea and Raposo, David and Santoro, Adam and Faulkner, Ryan and Gulcehre, Caglar and Song, Francis and Ballard, Andrew and Gilmer, Justin and Dahl, George and Vaswani, Ashish and Allen, Kelsey and Nash, Charles and Langston, Victoria and Dyer, Chris and Heess, Nicolas and Wierstra, Daan and Kohli, Pushmeet and Botvinick, Matt and Vinyals, Oriol and Li, Yujia and Pascanu, Razvan},
  year = 2018,
  month = oct,
  number = {arXiv:1806.01261},
  eprint = {1806.01261},
  primaryclass = {cs},
  publisher = {arXiv},
  doi = {10.48550/arXiv.1806.01261},
  urldate = {2026-01-29},
  archiveprefix = {arXiv}
}

@article{beckermann2025,
  title = {A New Strategy Using the {{Proper Generalized Decomposition}} to Model Time Evolving Spatial Domains},
  author = {Beckermann, Max and Scanff, Ronan and Cremonesi, Massimiliano and Barbarulo, Andrea},
  year = 2025,
  month = sep,
  journal = {Computers \& Structures},
  volume = {316},
  pages = {107860},
  issn = {0045-7949},
  doi = {10.1016/j.compstruc.2025.107860},
  urldate = {2026-01-29}
}

@article{brivio2025,
  title = {Handling Geometrical Variability in Nonlinear Reduced Order Modeling through {{Continuous Geometry-Aware DL-ROMs}}},
  author = {Brivio, Simone and Fresca, Stefania and Manzoni, Andrea},
  year = 2025,
  month = jul,
  journal = {Computer Methods in Applied Mechanics and Engineering},
  volume = {442},
  pages = {117989},
  issn = {0045-7825},
  doi = {10.1016/j.cma.2025.117989},
  urldate = {2025-10-24}
}

@inproceedings{cao2021,
  title = {Choose a {{Transformer}}: {{Fourier}} or {{Galerkin}}},
  shorttitle = {Choose a {{Transformer}}},
  booktitle = {Advances in {{Neural Information Processing Systems}}},
  author = {Cao, Shuhao},
  year = 2021,
  volume = {34},
  pages = {24924--24940},
  publisher = {Curran Associates, Inc.},
  urldate = {2026-01-29}
}

@article{carbonell2022,
  title = {Geotechnical Particle Finite Element Method for Modeling of Soil-Structure Interaction under Large Deformation Conditions},
  author = {Carbonell, Josep Maria and Monforte, Llu{\'i}s and Ciantia, Matteo O. and Arroyo, Marcos and Gens, Antonio},
  year = 2022,
  month = jun,
  journal = {Journal of Rock Mechanics and Geotechnical Engineering},
  volume = {14},
  number = {3},
  pages = {967--983},
  issn = {1674-7755},
  doi = {10.1016/j.jrmge.2021.12.006},
  urldate = {2026-01-29}
}

@article{cerquaglia2019,
  title = {A Fully Partitioned {{Lagrangian}} Framework for {{FSI}} Problems Characterized by Free Surfaces, Large Solid Deformations and Displacements, and Strong Added-Mass Effects},
  author = {Cerquaglia, M. L. and Thomas, D. and Boman, R. and Terrapon, V. and Ponthot, J. -P.},
  year = 2019,
  month = may,
  journal = {Computer Methods in Applied Mechanics and Engineering},
  volume = {348},
  pages = {409--442},
  issn = {0045-7825},
  doi = {10.1016/j.cma.2019.01.021},
  urldate = {2025-01-23}
}

@article{choi2024,
  title = {Graph {{Neural Network-based}} Surrogate Model for Granular Flows},
  author = {Choi, Yongjin and Kumar, Krishna},
  year = 2024,
  month = feb,
  journal = {Computers and Geotechnics},
  volume = {166},
  pages = {106015},
  issn = {0266-352X},
  doi = {10.1016/j.compgeo.2023.106015},
  urldate = {2026-02-10}
}

@article{cremonesi2020,
  title = {A {{State}} of the {{Art Review}} of the {{Particle Finite Element Method}} ({{PFEM}})},
  author = {Cremonesi, Massimiliano and Franci, Alessandro and Idelsohn, Sergio and O{\~n}ate, Eugenio},
  year = 2020,
  month = nov,
  journal = {Arch Computat Methods Eng},
  volume = {27},
  number = {5},
  pages = {1709--1735},
  issn = {1886-1784},
  doi = {10.1007/s11831-020-09468-4},
  urldate = {2025-01-23},
  langid = {english}
}

@article{dao2022,
  title = {{{FlashAttention}}: {{Fast}} and {{Memory-Efficient Exact Attention}} with {{IO-Awareness}}},
  shorttitle = {{{FlashAttention}}},
  author = {Dao, Tri and Fu, Dan and Ermon, Stefano and Rudra, Atri and R{\'e}, Christopher},
  year = 2022,
  month = dec,
  journal = {Advances in Neural Information Processing Systems},
  volume = {35},
  pages = {16344--16359},
  urldate = {2026-01-29},
  langid = {english}
}

@misc{dosovitskiy2021,
  title = {An {{Image}} Is {{Worth}} 16x16 {{Words}}: {{Transformers}} for {{Image Recognition}} at {{Scale}}},
  shorttitle = {An {{Image}} Is {{Worth}} 16x16 {{Words}}},
  author = {Dosovitskiy, Alexey and Beyer, Lucas and Kolesnikov, Alexander and Weissenborn, Dirk and Zhai, Xiaohua and Unterthiner, Thomas and Dehghani, Mostafa and Minderer, Matthias and Heigold, Georg and Gelly, Sylvain and Uszkoreit, Jakob and Houlsby, Neil},
  year = 2021,
  month = jun,
  number = {arXiv:2010.11929},
  eprint = {2010.11929},
  primaryclass = {cs},
  publisher = {arXiv},
  doi = {10.48550/arXiv.2010.11929},
  urldate = {2026-01-29},
  archiveprefix = {arXiv}
}

@misc{duan2023,
  title = {A {{Comprehensive Study}} on {{Large-Scale Graph Training}}: {{Benchmarking}} and {{Rethinking}}},
  shorttitle = {A {{Comprehensive Study}} on {{Large-Scale Graph Training}}},
  author = {Duan, Keyu and Liu, Zirui and Wang, Peihao and Zheng, Wenqing and Zhou, Kaixiong and Chen, Tianlong and Hu, Xia and Wang, Zhangyang},
  year = 2023,
  month = mar,
  number = {arXiv:2210.07494},
  eprint = {2210.07494},
  primaryclass = {cs},
  publisher = {arXiv},
  doi = {10.48550/arXiv.2210.07494},
  urldate = {2026-01-29},
  archiveprefix = {arXiv}
}

@inproceedings{edelsbrunner1992,
  title = {Three-Dimensional Alpha Shapes},
  booktitle = {Proceedings of the 1992 Workshop on {{Volume}} Visualization},
  author = {Edelsbrunner, Herbert and M{\"u}cke, Ernst P.},
  year = 1992,
  month = dec,
  series = {{{VVS}} '92},
  pages = {75--82},
  publisher = {Association for Computing Machinery},
  address = {New York, NY, USA},
  doi = {10.1145/147130.147153},
  urldate = {2026-01-29},
  isbn = {978-0-89791-527-4}
}

@article{franci2020,
  title = {{{3D}} Simulation of {{Vajont}} Disaster. {{Part}} 1: {{Numerical}} Formulation and Validation},
  shorttitle = {{{3D}} Simulation of {{Vajont}} Disaster. {{Part}} 1},
  author = {Franci, Alessandro and Cremonesi, Massimiliano and Perego, Umberto and Crosta, Giovanni and O{\~n}ate, Eugenio},
  year = 2020,
  month = dec,
  journal = {Engineering Geology},
  volume = {279},
  pages = {105854},
  issn = {0013-7952},
  doi = {10.1016/j.enggeo.2020.105854},
  urldate = {2025-01-23}
}

@article{fu2024,
  title = {A Hybrid {{Lagrangian}}--{{Eulerian}} Particle Finite Element Method for Free-Surface and Fluid--Structure Interaction Problems},
  author = {Fu, Cheng and Cremonesi, Massimiliano and Perego, Umberto},
  year = 2024,
  journal = {International Journal for Numerical Methods in Engineering},
  volume = {125},
  number = {5},
  pages = {e7402},
  issn = {1097-0207},
  doi = {10.1002/nme.7402},
  urldate = {2026-01-29},
  langid = {english}
}

@article{hughes1986,
  title = {A New Finite Element Formulation for Computational Fluid Dynamics: {{V}}. {{Circumventing}} the Babu\v ska-Brezzi Condition: A Stable {{Petrov-Galerkin}} Formulation of the Stokes Problem Accommodating Equal-Order Interpolations},
  shorttitle = {A New Finite Element Formulation for Computational Fluid Dynamics},
  author = {Hughes, Thomas J. R. and Franca, Leopoldo P. and Balestra, Marc},
  year = 1986,
  month = nov,
  journal = {Computer Methods in Applied Mechanics and Engineering},
  volume = {59},
  number = {1},
  pages = {85--99},
  issn = {0045-7825},
  doi = {10.1016/0045-7825(86)90025-3},
  urldate = {2026-01-29}
}

@article{idelsohn2004,
  title = {The Particle Finite Element Method: A Powerful Tool to Solve Incompressible Flows with Free-Surfaces and Breaking Waves},
  shorttitle = {The Particle Finite Element Method},
  author = {Idelsohn, S.r. and O{\~n}ate, E. and Pin, F. Del},
  year = 2004,
  journal = {International Journal for Numerical Methods in Engineering},
  volume = {61},
  number = {7},
  pages = {964--989},
  issn = {1097-0207},
  doi = {10.1002/nme.1096},
  urldate = {2025-01-23},
  copyright = {Copyright \copyright{} 2004 John Wiley \& Sons, Ltd.},
  langid = {english}
}

@article{idelsohn2009,
  title = {Multi-Fluid Flows with the {{Particle Finite Element Method}}},
  author = {Idelsohn, Sergio and {Mier-Torrecilla}, Monica and O{\~n}ate, Eugenio},
  year = 2009,
  month = jul,
  journal = {Computer Methods in Applied Mechanics and Engineering},
  volume = {198},
  number = {33},
  pages = {2750--2767},
  issn = {0045-7825},
  doi = {10.1016/j.cma.2009.04.002},
  urldate = {2025-01-23}
}

@misc{iparraguirre2026,
  title = {{{MeshGraphNet-Transformer}}: {{Scalable Mesh-based Learned Simulation}} for {{Solid Mechanics}}},
  shorttitle = {{{MeshGraphNet-Transformer}}},
  author = {Iparraguirre, Mikel M. and Alfaro, Iciar and Gonzalez, David and Cueto, Elias},
  year = 2026,
  month = feb,
  number = {arXiv:2601.23177},
  eprint = {2601.23177},
  primaryclass = {cs},
  publisher = {arXiv},
  doi = {10.48550/arXiv.2601.23177},
  urldate = {2026-02-10},
  archiveprefix = {arXiv}
}

@misc{joshi2025,
  title = {Transformers Are {{Graph Neural Networks}}},
  author = {Joshi, Chaitanya K.},
  year = 2025,
  month = jun,
  number = {arXiv:2506.22084},
  eprint = {2506.22084},
  primaryclass = {cs},
  publisher = {arXiv},
  doi = {10.48550/arXiv.2506.22084},
  urldate = {2025-10-27},
  archiveprefix = {arXiv}
}

@article{jumper2021,
  title = {Highly Accurate Protein Structure Prediction with {{AlphaFold}}},
  author = {Jumper, John and Evans, Richard and Pritzel, Alexander and Green, Tim and Figurnov, Michael and Ronneberger, Olaf and Tunyasuvunakool, Kathryn and Bates, Russ and {\v Z}{\'i}dek, Augustin and Potapenko, Anna and Bridgland, Alex and Meyer, Clemens and Kohl, Simon A. A. and Ballard, Andrew J. and Cowie, Andrew and {Romera-Paredes}, Bernardino and Nikolov, Stanislav and Jain, Rishub and Adler, Jonas and Back, Trevor and Petersen, Stig and Reiman, David and Clancy, Ellen and Zielinski, Michal and Steinegger, Martin and Pacholska, Michalina and Berghammer, Tamas and Bodenstein, Sebastian and Silver, David and Vinyals, Oriol and Senior, Andrew W. and Kavukcuoglu, Koray and Kohli, Pushmeet and Hassabis, Demis},
  year = 2021,
  month = aug,
  journal = {Nature},
  volume = {596},
  number = {7873},
  pages = {583--589},
  publisher = {Nature Publishing Group},
  issn = {1476-4687},
  doi = {10.1038/s41586-021-03819-2},
  urldate = {2026-01-29},
  copyright = {2021 The Author(s)},
  langid = {english}
}

@inproceedings{katharopoulos2020,
  title = {Transformers Are {{RNNs}}: Fast Autoregressive Transformers with Linear Attention},
  shorttitle = {Transformers Are {{RNNs}}},
  booktitle = {Proceedings of the 37th {{International Conference}} on {{Machine Learning}}},
  author = {Katharopoulos, Angelos and Vyas, Apoorv and Pappas, Nikolaos and Fleuret, Fran{\c c}ois},
  year = 2020,
  month = jul,
  series = {{{ICML}}'20},
  volume = {119},
  pages = {5156--5165},
  publisher = {JMLR.org},
  urldate = {2026-05-29}
}

@article{lanteri2025,
  title = {A Mesh-Based {{Graph Neural Network}} Approach for Surrogate Modeling of {{Lagrangian}} Free Surface Fluid Flows},
  author = {Lanteri, Federico and Cremonesi, Massimiliano},
  year = 2025,
  month = oct,
  journal = {Computers \& Fluids},
  volume = {301},
  pages = {106773},
  issn = {0045-7930},
  doi = {10.1016/j.compfluid.2025.106773},
  urldate = {2025-09-10}
}

@article{leyssens2024,
  title = {A {{Delaunay}} Refinement Algorithm for the Particle Finite Element Method Applied to Free Surface Flows},
  author = {Leyssens, Thomas and Henry, Michel and Lambrechts, Jonathan and Remacle, Jean-Fran{\c c}ois},
  year = 2024,
  journal = {International Journal for Numerical Methods in Engineering},
  volume = {125},
  number = {18},
  pages = {e7554},
  issn = {1097-0207},
  doi = {10.1002/nme.7554},
  urldate = {2025-01-23},
  copyright = {\copyright{} 2024 John Wiley \& Sons Ltd.},
  langid = {english}
}

@article{li2022,
  title = {Graph Neural Network-Accelerated {{Lagrangian}} Fluid Simulation},
  author = {Li, Zijie and Farimani, Amir Barati},
  year = 2022,
  month = apr,
  journal = {Computers \& Graphics},
  volume = {103},
  pages = {201--211},
  issn = {00978493},
  doi = {10.1016/j.cag.2022.02.004},
  urldate = {2025-01-21},
  langid = {english}
}

@article{li2022a,
  title = {Graph Neural Networks Accelerated Molecular Dynamics},
  author = {Li, Zijie and Meidani, Kazem and Yadav, Prakarsh and Barati Farimani, Amir},
  year = 2022,
  month = apr,
  journal = {J. Chem. Phys.},
  volume = {156},
  number = {14},
  pages = {144103},
  issn = {0021-9606},
  doi = {10.1063/5.0083060},
  urldate = {2026-02-10}
}

@misc{li2023,
  title = {Transformer for {{Partial Differential Equations}}' {{Operator Learning}}},
  author = {Li, Zijie and Meidani, Kazem and Farimani, Amir Barati},
  year = 2023,
  month = apr,
  number = {arXiv:2205.13671},
  eprint = {2205.13671},
  primaryclass = {cs},
  publisher = {arXiv},
  doi = {10.48550/arXiv.2205.13671},
  urldate = {2025-10-24},
  archiveprefix = {arXiv}
}

@inproceedings{li2023a,
  title = {Geometry-Informed Neural Operator for Large-Scale {{3D PDEs}}},
  booktitle = {Proceedings of the 37th {{International Conference}} on {{Neural Information Processing Systems}}},
  author = {Li, Zongyi and Kovachki, Nikola Borislavov and Choy, Chris and Li, Boyi and Kossaifi, Jean and Otta, Shourya Prakash and Nabian, Mohammad Amin and Stadler, Maximilian and Hundt, Christian and Azizzadenesheli, Kamyar and Anandkumar, Anima},
  year = 2023,
  month = dec,
  series = {{{NIPS}} '23},
  pages = {35836--35854},
  publisher = {Curran Associates Inc.},
  address = {Red Hook, NY, USA},
  urldate = {2026-05-12}
}

@article{mccabe2023,
  title = {Towards {{Stability}} of {{Autoregressive Neural Operators}}},
  author = {McCabe, Michael and Harrington, Peter and Subramanian, Shashank and Brown, Jed},
  year = 2023,
  month = jun,
  journal = {Transactions on Machine Learning Research},
  issn = {2835-8856},
  urldate = {2026-05-27},
  langid = {english}
}

@article{meduri2018,
  title = {A Partitioned Fully Explicit {{Lagrangian}} Finite Element Method for Highly Nonlinear Fluid-Structure Interaction Problems},
  author = {Meduri, S. and Cremonesi, M. and Perego, U. and Bettinotti, O. and Kurkchubasche, A. and Oancea, V.},
  year = 2018,
  journal = {International Journal for Numerical Methods in Engineering},
  volume = {113},
  number = {1},
  pages = {43--64},
  issn = {1097-0207},
  doi = {10.1002/nme.5602},
  urldate = {2025-01-23},
  copyright = {Copyright \copyright{} 2017 John Wiley \& Sons, Ltd.},
  langid = {english}
}

@article{meduri2019,
  title = {An Efficient Runtime Mesh Smoothing Technique for {{3D}} Explicit {{Lagrangian}} Free-Surface Fluid Flow Simulations},
  author = {Meduri, S. and Cremonesi, M. and Perego, U.},
  year = 2019,
  journal = {International Journal for Numerical Methods in Engineering},
  volume = {117},
  number = {4},
  pages = {430--452},
  issn = {1097-0207},
  doi = {10.1002/nme.5962},
  urldate = {2026-05-29},
  copyright = {\copyright{} 2018 John Wiley \& Sons, Ltd.},
  langid = {english}
}

@article{onate2004,
  title = {The Particle Finite Element Method --- an Overview},
  author = {O{\~n}ate, E. and Idelsohn, S. R. and Del Pin, F. and Aubry, R.},
  year = 2004,
  month = sep,
  journal = {Int. J. Comput. Methods},
  volume = {01},
  number = {02},
  pages = {267--307},
  publisher = {World Scientific Publishing Co.},
  issn = {0219-8762},
  doi = {10.1142/S0219876204000204},
  urldate = {2026-01-29}
}

@article{papanastasiou1987,
  title = {Flows of Materials with Yield},
  author = {Papanastasiou, Tasos C.},
  year = 1987,
  month = jul,
  journal = {J. Rheol.},
  volume = {31},
  number = {5},
  pages = {385--404},
  publisher = {Society of Rheology},
  issn = {0148-6055},
  doi = {10.1122/1.549926}
}

@misc{perez2017,
  title = {{{FiLM}}: {{Visual Reasoning}} with a {{General Conditioning Layer}}},
  shorttitle = {{{FiLM}}},
  author = {Perez, Ethan and Strub, Florian and de Vries, Harm and Dumoulin, Vincent and Courville, Aaron},
  year = 2017,
  month = dec,
  number = {arXiv:1709.07871},
  eprint = {1709.07871},
  primaryclass = {cs},
  publisher = {arXiv},
  doi = {10.48550/arXiv.1709.07871},
  urldate = {2026-04-16},
  archiveprefix = {arXiv}
}

@book{quarteroni2016,
  title = {Reduced {{Basis Methods}} for {{Partial Differential Equations}}},
  author = {Quarteroni, Alfio and Manzoni, Andrea and Negri, Federico},
  year = 2016,
  series = {{{UNITEXT}}},
  volume = {92},
  publisher = {Springer International Publishing},
  address = {Cham},
  doi = {10.1007/978-3-319-15431-2},
  urldate = {2026-01-29},
  copyright = {http://www.springer.com/tdm},
  isbn = {978-3-319-15430-5 978-3-319-15431-2}
}

@article{rizzieri2024,
  title = {Numerical Simulation of the Extrusion and Layer Deposition Processes in {{3D}} Concrete Printing with the {{Particle Finite Element Method}}},
  author = {Rizzieri, Giacomo and Ferrara, Liberato and Cremonesi, Massimiliano},
  year = 2024,
  month = feb,
  journal = {Comput Mech},
  volume = {73},
  number = {2},
  pages = {277--295},
  issn = {1432-0924},
  doi = {10.1007/s00466-023-02367-y},
  urldate = {2025-01-23},
  langid = {english}
}

@article{rizzieri2024a,
  title = {Simulation of Viscoelastic Free-Surface Flows with the {{Particle Finite Element Method}}},
  author = {Rizzieri, Giacomo and Ferrara, Liberato and Cremonesi, Massimiliano},
  year = 2024,
  month = oct,
  journal = {Comp. Part. Mech.},
  volume = {11},
  number = {5},
  pages = {2043--2067},
  issn = {2196-4386},
  doi = {10.1007/s40571-024-00730-1},
  urldate = {2025-01-23},
  langid = {english}
}

@article{rizzieri2025,
  title = {A Partitioned {{Lagrangian}} Finite Element Approach for the Simulation of Viscoelastic and Elasto-Viscoplastic Free-Surface Flows},
  author = {Rizzieri, Giacomo and Ferrara, Liberato and Cremonesi, Massimiliano},
  year = 2025,
  month = aug,
  journal = {Computer Methods in Applied Mechanics and Engineering},
  volume = {443},
  pages = {118071},
  issn = {0045-7825},
  doi = {10.1016/j.cma.2025.118071},
  urldate = {2025-10-27}
}

@article{rizzieri2026,
  title = {{{{\emph{ShapeGen3DCP}}}}: {{A}} Deep Learning Framework for Layer Shape Prediction in {{3D}} Concrete Printing},
  shorttitle = {{{{\emph{ShapeGen3DCP}}}}},
  author = {Rizzieri, Giacomo and Lanteri, Federico and Ferrara, Liberato and Cremonesi, Massimiliano},
  year = 2026,
  month = mar,
  journal = {Computers \& Structures},
  volume = {323},
  pages = {108142},
  issn = {0045-7949},
  doi = {10.1016/j.compstruc.2026.108142},
  urldate = {2026-04-16}
}

@misc{rusch2023,
  title = {A {{Survey}} on {{Oversmoothing}} in {{Graph Neural Networks}}},
  author = {Rusch, T. Konstantin and Bronstein, Michael M. and Mishra, Siddhartha},
  year = 2023,
  month = mar,
  number = {arXiv:2303.10993},
  eprint = {2303.10993},
  primaryclass = {cs},
  publisher = {arXiv},
  doi = {10.48550/arXiv.2303.10993},
  urldate = {2026-05-19},
  archiveprefix = {arXiv}
}

@article{ryzhakov2016,
  title = {Lagrangian Finite Element Model for the {{3D}} Simulation of Glass Forming Processes},
  author = {Ryzhakov, P. B. and Garc{\'i}a, J. and O{\~n}ate, E.},
  year = 2016,
  month = dec,
  journal = {Computers \& Structures},
  volume = {177},
  pages = {126--140},
  issn = {0045-7949},
  doi = {10.1016/j.compstruc.2016.09.007},
  urldate = {2025-01-23}
}

@misc{saberi2025,
  title = {{{RheOFormer}}: {{A}} Generative Transformer Model for Simulation of Complex Fluids and Flows},
  shorttitle = {{{RheOFormer}}},
  author = {Saberi, Maedeh and Farimani, Amir Barati and Jamali, Safa},
  year = 2025,
  month = oct,
  number = {arXiv:2510.01365},
  eprint = {2510.01365},
  primaryclass = {cs},
  publisher = {arXiv},
  doi = {10.48550/arXiv.2510.01365},
  urldate = {2026-04-16},
  archiveprefix = {arXiv}
}

@misc{sanchez-gonzalez2020,
  title = {Learning to {{Simulate Complex Physics}} with {{Graph Networks}}},
  author = {{Sanchez-Gonzalez}, Alvaro and Godwin, Jonathan and Pfaff, Tobias and Ying, Rex and Leskovec, Jure and Battaglia, Peter W.},
  year = 2020,
  month = sep,
  number = {arXiv:2002.09405},
  eprint = {2002.09405},
  primaryclass = {cs},
  publisher = {arXiv},
  doi = {10.48550/arXiv.2002.09405},
  urldate = {2025-01-21},
  archiveprefix = {arXiv}
}

@article{scarselli2009,
  title = {The {{Graph Neural Network Model}}},
  author = {Scarselli, Franco and Gori, Marco and Tsoi, Ah Chung and Hagenbuchner, Markus and Monfardini, Gabriele},
  year = 2009,
  month = jan,
  journal = {IEEE Transactions on Neural Networks},
  volume = {20},
  number = {1},
  pages = {61--80},
  issn = {1941-0093},
  doi = {10.1109/TNN.2008.2005605},
  urldate = {2026-01-29}
}

@article{sharma2026,
  title = {A Physics-Informed Graph Neural Network Conserving Linear and Angular Momentum for Dynamical Systems},
  author = {Sharma, Vinay and Fink, Olga},
  year = 2026,
  month = jan,
  journal = {Nat Commun},
  volume = {17},
  number = {1},
  pages = {1045},
  publisher = {Nature Publishing Group},
  issn = {2041-1723},
  doi = {10.1038/s41467-025-67802-5},
  urldate = {2026-02-10},
  copyright = {2026 The Author(s)},
  langid = {english}
}

@article{su2024,
  title = {{{RoFormer}}: {{Enhanced}} Transformer with {{Rotary Position Embedding}}},
  shorttitle = {{{RoFormer}}},
  author = {Su, Jianlin and Ahmed, Murtadha and Lu, Yu and Pan, Shengfeng and Bo, Wen and Liu, Yunfeng},
  year = 2024,
  month = feb,
  journal = {Neurocomputing},
  volume = {568},
  pages = {127063},
  issn = {0925-2312},
  doi = {10.1016/j.neucom.2023.127063},
  urldate = {2026-01-29}
}

@article{tesan2026,
  title = {On the Under-Reaching Phenomenon in Message Passing Neural {{PDE}} Solvers: {{Revisiting}} the {{CFL}} Condition},
  shorttitle = {On the Under-Reaching Phenomenon in Message Passing Neural {{PDE}} Solvers},
  author = {Tes{\'a}n, Lucas and Iparraguirre, Mikel M. and Gonz{\'a}lez, David and Martins, Pedro and Cueto, El{\'i}as},
  year = 2026,
  month = feb,
  journal = {Computer Methods in Applied Mechanics and Engineering},
  volume = {449},
  pages = {118476},
  issn = {0045-7825},
  doi = {10.1016/j.cma.2025.118476},
  urldate = {2026-02-10}
}

@article{tierz2025,
  title = {Graph Neural Networks Informed Locally by Thermodynamics},
  author = {Tierz, Alicia and Alfaro, Ic{\'i}ar and Gonz{\'a}lez, David and Chinesta, Francisco and Cueto, El{\'i}as},
  year = 2025,
  month = mar,
  journal = {Engineering Applications of Artificial Intelligence},
  volume = {144},
  pages = {110108},
  issn = {0952-1976},
  doi = {10.1016/j.engappai.2025.110108},
  urldate = {2025-07-17}
}

@inproceedings{vaswani2017,
  title = {Attention Is {{All}} You {{Need}}},
  booktitle = {Advances in {{Neural Information Processing Systems}}},
  author = {Vaswani, Ashish and Shazeer, Noam and Parmar, Niki and Uszkoreit, Jakob and Jones, Llion and Gomez, Aidan N and ukasz Kaiser, {\L} and Polosukhin, Illia},
  year = 2017,
  volume = {30},
  publisher = {Curran Associates, Inc.},
  urldate = {2026-01-29}
}

@article{wang2026,
  title = {{{FluidFormer}} : {{Transformer}} with Continuous Convolution for Particle-Based Fluid Simulation},
  shorttitle = {{{FluidFormer}}},
  author = {Wang, Nianyi and Zheng, Shuai and Chen, Yu and Zhao, Hai and Fang, Zhou},
  year = 2026,
  month = jun,
  journal = {Neural Networks},
  volume = {198},
  pages = {108631},
  issn = {0893-6080},
  doi = {10.1016/j.neunet.2026.108631},
  urldate = {2026-02-10}
}

@article{wen2026,
  title = {Geometry {{Aware Operator Transformer}} as an Efficient and Accurate Neural Surrogate for {{PDEs}} on Arbitrary Domains},
  author = {Wen, Shizheng and Kumbhat, Arsh and Lingsch, Levi and Mousavi, Sepehr and Zhao, Yizhou and Chandrashekar, Praveen and Mishra, Siddhartha},
  year = 2026,
  month = apr,
  journal = {Advances in Neural Information Processing Systems},
  volume = {38},
  pages = {155423--155501},
  urldate = {2026-05-14},
  langid = {english}
}

@misc{wu2024,
  title = {Transolver: {{A Fast Transformer Solver}} for {{PDEs}} on {{General Geometries}}},
  shorttitle = {Transolver},
  author = {Wu, Haixu and Luo, Huakun and Wang, Haowen and Wang, Jianmin and Long, Mingsheng},
  year = 2024,
  month = jun,
  number = {arXiv:2402.02366},
  eprint = {2402.02366},
  primaryclass = {cs},
  publisher = {arXiv},
  doi = {10.48550/arXiv.2402.02366},
  urldate = {2026-02-10},
  archiveprefix = {arXiv}
}

@article{zhao2025,
  title = {A Physical-Information-Flow-Constrained Temporal Graph Neural Network-Based Simulator for Granular Materials},
  author = {Zhao, Shiwei and Chen, Hao and Zhao, Jidong},
  year = 2025,
  month = jan,
  journal = {Computer Methods in Applied Mechanics and Engineering},
  volume = {433},
  pages = {117536},
  issn = {0045-7825},
  doi = {10.1016/j.cma.2024.117536},
  urldate = {2025-10-24}
}

@article{zhou2025,
  title = {Improving Long-Term Autoregressive Spatiotemporal Predictions: {{A}} Proof of Concept with Fluid Dynamics},
  shorttitle = {Improving Long-Term Autoregressive Spatiotemporal Predictions},
  author = {Zhou, Hao and Cheng, Sibo},
  year = 2025,
  month = dec,
  journal = {Computer Methods in Applied Mechanics and Engineering},
  volume = {447},
  pages = {118332},
  issn = {0045-7825},
  doi = {10.1016/j.cma.2025.118332},
  urldate = {2026-05-27}
}

@misc{zhou2026,
  title = {Transolver-3: {{Scaling Up Transformer Solvers}} to {{Industrial-Scale Geometries}}},
  shorttitle = {Transolver-3},
  author = {Zhou, Hang and Wu, Haixu and Shangguan, Haonan and Ma, Yuezhou and Weng, Huikun and Wang, Jianmin and Long, Mingsheng},
  year = 2026,
  month = feb,
  number = {arXiv:2602.04940},
  eprint = {2602.04940},
  primaryclass = {cs},
  publisher = {arXiv},
  doi = {10.48550/arXiv.2602.04940},
  urldate = {2026-02-10},
  archiveprefix = {arXiv}
}

@inproceedings{zhuoran2021,
  title = {Efficient {{Attention}}: {{Attention}} with {{Linear Complexities}}},
  shorttitle = {Efficient {{Attention}}},
  booktitle = {2021 {{IEEE Winter Conference}} on {{Applications}} of {{Computer Vision}} ({{WACV}})},
  author = {Zhuoran, Shen and Mingyuan, Zhang and Haiyu, Zhao and Shuai, Yi and Hongsheng, Li},
  year = 2021,
  month = jan,
  pages = {3530--3538},
  issn = {2642-9381},
  doi = {10.1109/WACV48630.2021.00357},
  urldate = {2026-05-29}
}






\end{document}